\pdfoutput=1

\documentclass[11pt,twoside,a4paper,cmspaper,final,collab]{cms-tdr}
\def\svnVersion{418614P}\def\cmsCernNoTag{CERN-EP/2016-293}\def\cmsCernDate{\today}\def\cmsMessage{Published in Physical Review D as \href{http://dx.doi.org/10.1103/PhysRevD.96.012004}{\doi{10.1103/PhysRevD.96.012004.}}}

\begin{document}\cmsNoteHeader{SUS-16-009}

\hyphenation{had-ron-i-za-tion}
\hyphenation{cal-or-i-me-ter}
\hyphenation{de-vices}
\RCS$Revision: 416552 $
\RCS$HeadURL: svn+ssh://svn.cern.ch/reps/tdr2/papers/SUS-16-009/trunk/SUS-16-009.tex $
\RCS$Id: SUS-16-009.tex 416552 2017-07-17 15:31:54Z pastika $
\newlength\cmsFigWidth
\ifthenelse{\boolean{cms@external}}{\setlength\cmsFigWidth{0.98\columnwidth}}{\setlength\cmsFigWidth{0.8\textwidth}}
\ifthenelse{\boolean{cms@external}}{\providecommand{\cmsLeft}{top\xspace}}{\providecommand{\cmsLeft}{left\xspace}}
\ifthenelse{\boolean{cms@external}}{\providecommand{\cmsRight}{bottom\xspace}}{\providecommand{\cmsRight}{right\xspace}}
\ifthenelse{\boolean{cms@external}}{\providecommand{\cmsTable}[1]{#1}}{\providecommand{\cmsTable}[1]{\resizebox{\textwidth}{!}{#1}}}
\ifthenelse{\boolean{cms@external}}{\providecommand{\CL}{C.L.\xspace}}{\providecommand{\CL}{CL\xspace}}

\newcommand{\znunu}{\ensuremath{\cPZ\to\cPgn\cPagn}\xspace}
\newcommand{\zll}{\ensuremath{\cPZ\to\ell\ell}\xspace}
\newcommand{\W}{\PW\xspace}
\newcommand{\ttbarW}{\ensuremath{\ttbar\PW}\xspace}
\newcommand{\ttbarZ}{\ensuremath{\ttbar\cPZ}\xspace}
\newcommand{\ttZ}{\ttbarZ}
\newcommand{\ttW}{\ttbarW}
\newcommand{\stopq}{\PSQt}
\newcommand{\sbottomq}{\PSQb}
\newcommand{\gluino}{\PSg}
\newcommand{\topq}{\PQt}
\newcommand{\bq}{\PQb}
\newcommand{\lsp}{\PSGczDo}
\newcommand{\chipmone}{\PSGcpmDo}
\newcommand{\mgluino}{\ensuremath{m_{\gluino}}\xspace}
\newcommand{\mstop}{\ensuremath{m_{\PSQt}}\xspace}
\newcommand{\MTTwo}{\ensuremath{M_{\mathrm{T2}}}\xspace}
\newcommand{\njets}{\ensuremath{N_{\mathrm{j}}}\xspace}
\newcommand{\ntops}{\ensuremath{N_{\PQt}}\xspace}
\newcommand{\nbjets}{\ensuremath{N_{\PQb}}\xspace}
\newcommand{\mt}{\ensuremath{m_{\mathrm{T}}}\xspace}
\newcommand{\zjets}{{{\cPZ}\text{+jets}}\xspace}
\newcommand{\wjets}{{{\PW}\text{+jets}}\xspace}

\ifthenelse{\boolean{cms@external}}{\providecommand{\suppMaterial}{\cite{SupplementalMaterial}}}{\providecommand{\suppMaterial}{App.~\ref{app:suppMat}}}

\cmsNoteHeader{SUS-16-009}
\title{\texorpdfstring{Search for supersymmetry in the all-hadronic final state using top quark tagging in $\Pp\Pp$ collisions at $\sqrt{s} = 13\TeV$}{Search for supersymmetry in the all-hadronic final state using top quark tagging in pp collisions at sqrt(s) = 13 TeV}}

\date{\today}

\abstract{A search is presented for supersymmetry in all-hadronic events with missing transverse momentum and tagged top quarks.
The data sample was collected with the CMS detector at the LHC and corresponds to an integrated luminosity of 2.3\fbinv of proton-proton collisions at a center-of-mass energy of 13\TeV. Search regions are defined using the properties of reconstructed jets, the multiplicity of bottom and top quark candidates, and an imbalance in transverse momentum. With no statistically significant excess of events observed beyond the expected contributions from the standard model, we set exclusion limits at 95\% confidence level on the masses of new particles in the context of simplified models of direct and gluino-mediated top squark production.
For direct top squark production with decays to a top quark and a neutralino, top squark masses up to 740\GeV and neutralino masses up to 240\GeV are excluded.
Gluino masses up to 1550\GeV and neutralino masses up to 900\GeV are excluded for a gluino-mediated production case, where each of the pair-produced gluinos decays to a top-antitop quark pair and a neutralino.
}

\hypersetup{%
pdfauthor={CMS Collaboration},%
pdftitle={Search for supersymmetry in the all-hadronic final state using top quark tagging in pp collisions at sqrt(s) = 13 TeV},%
pdfsubject={CMS},%
pdfkeywords={CMS, physics, supersymmetry}}

\maketitle

\section{Introduction}
\label{sec:intro}
The standard model (SM) of fundamental particles and their interactions has been extremely successful in describing phenomena in the atomic and subatomic realms.
The discovery of a boson with properties consistent with the SM Higgs boson~\cite{Aad:2012tfa,Chatrchyan:2012ufa,Chatrchyan:2013lba} at the CERN LHC~\cite{LHC} further strengthened this model.
Assuming that the Higgs boson is a fundamental spin-0 particle, however, the low value of its measured mass, around 125\GeV~\cite{Aad:2015zhl}, implies that there is a fine-tuned cancellation of large quantum corrections to its mass, which is referred to as the hierarchy problem and is currently unexplained~\cite{Dimopoulos:1981au,Witten:1981nf,Dine:1981za,Dimopoulos:1981zb,Kaul:1981hi}.
Supersymmetry (SUSY)~\cite{Barbieri:1982eh,Barbieri:1987fn,Wess:1974tw,Golfand,Volkov,Chamseddine,Kane,Fayet,Hall,Ramond} is one of the most compelling models of new physics as it provides an elegant mechanism to mitigate the hierarchy problem by introducing a symmetry between fermions and bosons.

Supersymmetry proposes a superpartner for each SM particle with the same quantum numbers, except for spin, which differs by a half-integer.
The SM particles and their corresponding superpartners contribute to the loop corrections to the Higgs boson mass with opposite sign~\cite{deCarlos1993320}, and are therefore capable of controlling these corrections.
This behavior can persist despite the breaking of SUSY, which is required to accommodate the lack of observation of superpartners with exactly the same masses as their SM counterparts.
To solve the hierarchy problem in a ``natural'' way, Refs.~\cite{Dimopoulos1995573,Sakai:1981gr,Papucci:2011wy,Brust:2011tb,Feng:2013pwa,Delgado:2012eu} suggest models in which the higgsino mass parameter is of the order of 100\GeV and the masses of the top squark $\stopq$, the bottom squark $\sbottomq$, and the gluino $\gluino$ are near the TeV scale, while the masses of the other sparticles can be beyond the reach of the LHC.
The mass of the top squark is particularly constrained in ``natural'' SUSY models as it is the most important factor in cancelling the top quark contribution to the Higgs boson mass.
In $R$-parity conserving models~\cite{Farrar:1978xj}, superpartners are produced in pairs, and the lightest SUSY particle (LSP) is stable.
Models with a weakly interacting neutralino ($\lsp$) as the LSP are especially attractive because the $\lsp$ can have properties consistent with dark matter~\cite{Feng:2010gw}.

Based on these considerations, we perform a search for top squarks, produced either directly or through gluino decays, with each top squark decaying into a stable $\lsp$ and SM particles.
Previous searches at the LHC in proton-proton collisions at $\sqrt{s}=8\TeV$ have found no evidence for physics beyond the SM, and lower limits have been placed on the top squark mass within the framework of simplified models of the SUSY particle spectrum (SMS)~\cite{Alwall:2008ag,Alwall:2008va,Alves:2011wf,Alves:2011sq,Chatrchyan:2013sza}.
The particle spectra in such models are typically restricted to states that are required for natural SUSY scenarios. Lower limits on the top squark mass, $\mstop$, extend up to $775\GeV$~\cite{Aad:2015pfx,Aad:2014nra,Aad:2014bva,Aad:2013ija,Aad:2014qaa,Aad:2014pda,stop8TeV,Khachatryan:2016oia,Khachatryan:2016pup,Chatrchyan:2013xna,Khachatryan:2015pot}, and those on the gluino mass, $\mgluino$, extend up to 1400\GeV~\cite{Aad:2015iea,Aad:2014wea,Aad:2013wta,Aad:2015zva,Aad:2015mia,Aad:2014lra,Chatrchyan:2013iqa,CMS:2014dpa,Khachatryan:2016zcu,Khachatryan:2015vra,Khachatryan:2015pwa,Chatrchyan:2014lfa}. Lower limits on the neutralino mass, $m_{\lsp}$, extend up to $290\GeV$ for models with direct top squarks production and up to $600\GeV$ for models with gluino-mediated production.
Recent searches in proton-proton collisions at $\sqrt{s}=13\TeV$ have further extended these lower limits, reaching up to $800\GeV$~\cite{Aaboud:2016lwz,Khachatryan:2016xvy,SUS-16-008} for the top squark mass, up to $1760\GeV$ for the gluino mass, and up to $850\GeV$ for the neutralino mass~\cite{Aad:2016eki,Aad:2016tuk,Khachatryan:2016kdk,Khachatryan:2016kod,Khachatryan:2016uwr}.

The search presented in this paper is performed on data collected with the CMS detector at the LHC and corresponding to an integrated luminosity of $2.3\fbinv$ of proton-proton collisions at a center-of-mass energy of 13\TeV.
The search strategy closely follows the one reported in Ref.~\cite{stop8TeV} with several improvements.
We select events containing large missing transverse momentum, at least four jets, at least one jet identified as originating from the hadronization of a $\PQb$ quark (``$\PQb$ jet''), and no identified leptons.
The analysis relies on a highly efficient algorithm to tag groups of jets consistent with top quark decay.
This top quark tagging algorithm is improved relative to the one described in Ref.~\cite{stop8TeV},
to enhance the sensitivity for selecting top quarks with large Lorentz boosts that cause the merging of jets among the top decay products.
The analysis categorizes each event according to the number of identified top quark candidates, in order to both discriminate signal from background and to distinguish among signal hypotheses such as direct top squark production and gluino-mediated top squark production, which contain different multiplicities of top quarks in the final state.
In addition, the kinematic properties of top quark candidates are used as input to the computation of the ``stransverse'' mass (\MTTwo) variable~\cite{Lester:1999tx,Barr:2003rg}, which is used to estimate the mass of pair-produced particles in the presence of invisible particles.
Exclusive search regions are defined using several event properties, including the number of identified $\PQb$ jets, the number of top quark candidates, the missing transverse momentum $\ptvecmiss$, and \MTTwo.

One of the major sources of SM background originates from either top-antitop quark pair ($\ttbar$) or \wjets events in which leptonic \W~boson decay produces a charged lepton that is not reconstructed or identified, and a high momentum neutrino, generating true missing transverse momentum.
Events in which a \Z~boson, produced in association with jets, decays to neutrinos ($\znunu$) also provide a significant contribution to the SM background.
The SM backgrounds are estimated using control samples in the data that are disjoint from the signal regions but have similar kinematic properties and composition.

This paper is structured as follows. Event reconstruction and simulation are described in Sec.~\ref{sec:evtrecoandsim}.
Sec.~\ref{sec:methodAlpha} presents details of the optimization of the analysis, including signal models, the top quark tagging algorithm, and event categorization.
The strategy used to estimate the SM background is detailed in Sec.~\ref{sec:backgroundestimation}.
The results and their interpretation in the context of SUSY are discussed in Sec.~\ref{sec:interpretation}, followed by a summary in Sec.~\ref{sec:summary}.

\section{Detector, event reconstruction, and simulation}
\label{sec:evtrecoandsim}
\subsection{Detector and event reconstruction}
\label{sec:evtsel}

The CMS detector is built around a superconducting solenoid of 6\unit{m} internal diameter,
providing a magnetic field of 3.8\unit{T}. Within the solenoid volume are a silicon pixel and strip tracker, a
lead tungstate crystal electromagnetic calorimeter (ECAL), and a brass and scintillator hadron
calorimeter (HCAL).
The tracking detectors cover $\abs{\eta}< 2.5$.
The ECAL and HCAL, each composed of a barrel and two endcap sections,
extend over a pseudorapidity range $\abs{\eta}< 3.0$. Forward calorimeters on each side of the
interaction point encompass $3.0 < \abs{\eta}< 5.2$.
Muons are identified and measured within $\abs{\eta}< 2.4$ by gas-ionization detectors embedded in the steel flux-return
yoke outside the solenoid.
The first level of the CMS trigger system, composed of custom hardware processors, uses information from the
calorimeters and muon detectors to select the most interesting events in a fixed time interval of less than 4\mus.
The high-level-trigger processor farm further decreases the event rate from around 100\unit{kHz} to less than 1\unit{kHz} before data storage.
A more detailed description of the CMS detector, together with a definition of the
coordinate system used and the relevant kinematic variables, can be found in Ref.~\cite{Chatrchyan:2008zzk}.

The recorded events are reconstructed using the particle-flow (PF) algorithm~\cite{Sirunyan:2017ulk}.
Using the information from the tracker, calorimeters, and muon system, this algorithm reconstructs PF candidates that are classified as charged hadrons, neutral hadrons, photons, muons, or electrons.
The  \ptvecmiss is defined as the negative of the vector sum of the transverse momentum \pt of all PF candidates in the event,
and its magnitude is denoted by \MET.
The PF candidates in an event are clustered into jets using the anti-\kt clustering algorithm~\cite{Cacciari:2008gp}
with size parameter $0.4$ (AK4 jets).
Charged particles from additional $\Pp\Pp$ collisions (``pileup'') from the same or adjacent beam crossing to the one that produced the primary hard-scattering process
 are excluded if they do not originate from the primary interaction vertex, i.e., the vertex with the largest $\sum{\pt^2}$ calculated from all its associated tracks.
The momentum of neutral particles from pileup interactions, and from the underlying event, is subtracted using the \FASTJET technique, which is based on the calculation of the $\eta$-dependent transverse momentum density, evaluated event by event~\cite{PU_JET_AREAS,JET_AREAS}.
The energy and momentum of each jet are corrected using factors derived from simulation, and, for jets in data, an additional residual energy-momentum correction is applied to account for differences in the jet energy-momentum scales~\cite{Khachatryan:2016kdb} between simulations and data.
Only jets with $\pt>30\GeV$ and $\abs{\eta}< 2.4$ or $\abs{\eta}< 5$, depending on the use case, are considered in this search.
The scalar sum of the jet \pt for all jets within $\abs{\eta}< 2.4$ is denoted by \HT in the following.

A jet is considered to be a $\PQb$ jet (``$\PQb$-tagged'') if it passes
the medium operating point requirements of the combined secondary vertex algorithm~\cite{CMS-PAS-BTV-15-001,Chatrchyan:2012jua}, has $\pt>30\GeV$, and is within $\abs{\eta}<2.4$.
The corresponding $\PQb$ quark identification efficiency is 70\% on average per jet in \ttbar events.
The probability of a jet originating from a light quark or gluon to be misidentified as a $\PQb$ quark jet is 1.4\%, averaged over jet \pt in \ttbar events~\cite{CMS-PAS-BTV-15-001}.

Muons are reconstructed by matching tracks in the muon detectors to compatible track segments in the silicon tracker~\cite{Chatrchyan:2012xi} and are required to be within $\abs{\eta}<2.4$.
Electron candidates are reconstructed starting from clusters of energy deposited in the ECAL that are then matched to a track in the silicon tracker~\cite{Khachatryan:2015hwa}.
Electron candidates are required to have $\abs{\eta}<1.44$ or $1.56<\abs{\eta}<2.50$ to avoid the transition region between the ECAL barrel and the endcap.
Muon and electron candidates are required to originate from within 2\unit{mm} of the primary vertex in the transverse plane and within 5\unit{mm} along the $z$ axis.

To obtain a sample of all-hadronic events, events with isolated electrons and muons are vetoed.
The isolation of electron and muon candidates is defined as the $\sum{\pt}$ of all additional PF candidates in a cone around the lepton candidate's trajectory with a radius $\Delta R = \sqrt{\smash[b]{(\Delta\eta)^2 + (\Delta\phi)^2}}$.
The cone size depends on the lepton \pt as follows:
\begin{linenomath}
\begin{equation}
\Delta R =
\begin{cases}
0.2, & \pt \leq 50\GeV \\
\frac{10\GeV}{\pt}, & 50 < \pt < 200\GeV \\
0.05, & \pt \geq 200\GeV .
\end{cases}
\end{equation}
\end{linenomath}
The cone radius for higher-\pt candidates is reduced because highly boosted objects, which may include high-\pt leptons in their decay, are contained in a cone of smaller radius than low-\pt objects.
The isolation sum is corrected for contributions originating from pileup interactions using an estimate of the pileup energy in the cone.
A relative isolation is defined as the ratio of the isolation sum to the candidate \pt, and is required to be less than 0.1 (0.2) for electron (muon) candidates.
Events with isolated electrons (muons) that have $\pt > 10\GeV$ and  $\abs{\eta}< 2.5$ (2.4) are rejected.

In order to further reduce the contribution from background events originating from leptonic \W~boson decays that
feature low-\pt electrons, muons, or hadronically decaying taus ($\Pgt_{\rm h}$),
an additional veto on the presence of isolated tracks is used.
These tracks are required to have $\abs{\eta}< 2.5$, $\pt > 5\,(10)\GeV$, and relative track isolation less than 0.2 (0.1) when they are identified by the PF algorithm as electrons or muons (charged hadrons).
The isolation sum used to compute the relative track isolation is the $\sum{\pt}$ of all additional charged PF candidates within a fixed cone of $\Delta R = 0.3$ around the track.
To preserve signal efficiency, this veto is applied only if the transverse mass ($\mt$) of the isolated track and $\MET$ system is consistent with a \W~boson decay. The $\mt$ is defined as
\begin{linenomath}
\begin{equation}
\mt(\text{track}, \MET) = \sqrt{2 \pt^\text{track} \MET (1 - \cos \Delta\phi)} ,
\label{eq:mt}
\end{equation}
\end{linenomath}
with $\pt^\text{track}$  the \pt of the track and $\Delta\phi$ the azimuthal separation between the track and \ptvecmiss vector.
Specifically, we require $\mt<100\GeV$.

\subsection{Event simulation}
\label{sec:simulation}
Monte Carlo (MC) simulated event samples are used to study the properties of the SM background processes, as well as the signal models.
The \MADGRAPH5\_a\MCATNLO v2.2.2 generator~\cite{Alwall:2014hca} is used in leading-order (LO) mode to simulate events originating from $\ttbar$ production, \wjets with $\PW \to \ell\cPgn$ decays, \zjets with \znunu decays, Drell-Yan (DY)+jets, $\gamma$+jets, quantum chromodynamics (QCD) multijet, gluino pair production, and top squark pair production processes.
The generation of these processes is based on LO parton distribution functions (PDFs) using NNPDF3.0~\cite{Ball:2014uwa}.
Single top quark events produced in the $\PQt\W$~channel are generated with the next-to-leading-order (NLO) \POWHEG v1.0~\cite{Nason:2004rx,Frixione:2007vw,Alioli:2010xd,Re:2010bp} generator.
Rare SM processes, such as $\ttZ$ and $\ttW$, are generated at NLO accuracy with the \MADGRAPH5\_a\MCATNLO v2.2.2 program.
Both the single top quark and rare SM processes are generated using NLO NNPDF3.0 PDFs.
The parton showering and hadronization is simulated with {\PYTHIA} v8.205~\cite{pythia8} using underlying-event tune CUETP8M1~\cite{Khachatryan:2015pea}.

The CMS detector response is simulated using a \GEANTfour-based model~\cite{Agostinelli:2002hh} in the case of SM background processes and a dedicated fast simulation package~\cite{fastsim} for the case of signal processes, where a large number of signal model scenarios are needed.
The fast simulation is tuned to provide results that are consistent with those obtained from the full \GEANTfour-based simulation.
Event reconstruction is performed in the same manner as for collision data.

The signal production cross sections are calculated using NLO plus next-to-leading-logarithm (NLL) calculations~\cite{Borschensky:2014cia}.
The most precise available cross section calculations are used to normalize the SM simulated samples, corresponding to NLO or next-to-NLO accuracy in most cases~\cite{Alwall:2014hca,Czakon:2011xx,Kant:2014oha,Aliev:2010zk,Gehrmann:2014fva,Campbell:1999ah,Campbell:2011bn,Li:2012wna}.

The simulation is corrected to account for discrepancies between data and simulation in the lepton selection efficiency and the $\PQb$ tagging efficiency.
The uncertainties corresponding to these corrections are propagated to the predicted SM yields in the search regions.
Differences in the efficiencies for selecting isolated electrons and muons are measured in \zll~events.
Correction factors and their uncertainties for the $\PQb$ tagging efficiency are derived using multijet- and \ttbar-enriched event samples and are parametrized by the jet kinematics~\cite{CMS-PAS-BTV-15-001}.

\section{Analysis strategy}
\label{sec:methodAlpha}

The analysis is designed for maximum sensitivity to models in which top quarks are produced in the SUSY decay chains discussed in Sec.~\ref{sec:intro}.
The data are first divided into regions based upon the numbers of tagged top quarks (\ntops) and $\PQb$ jets (\nbjets) found in each event.
The search regions are defined by further subdivision of each \ntops, \nbjets bin in several \MET and \MTTwo bins.

\subsection{Benchmark signal models}
\label{sec:signal_models}

\begin{figure}[tb]
  \centering
    \includegraphics[width=0.48\textwidth]{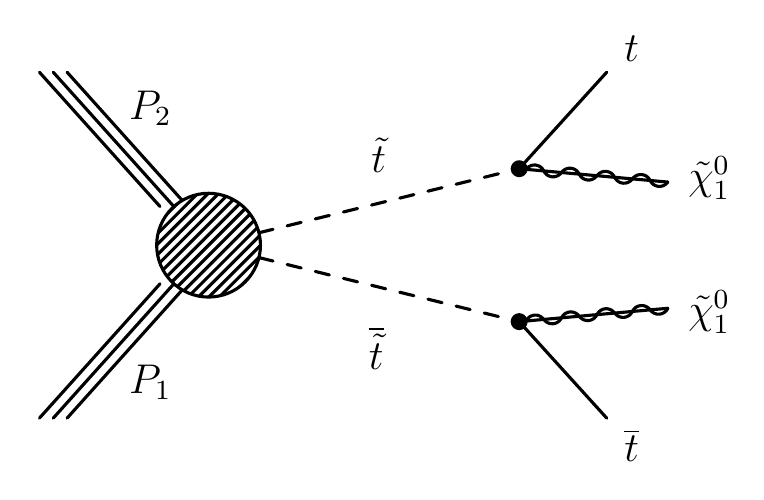}
    \includegraphics[width=0.48\textwidth]{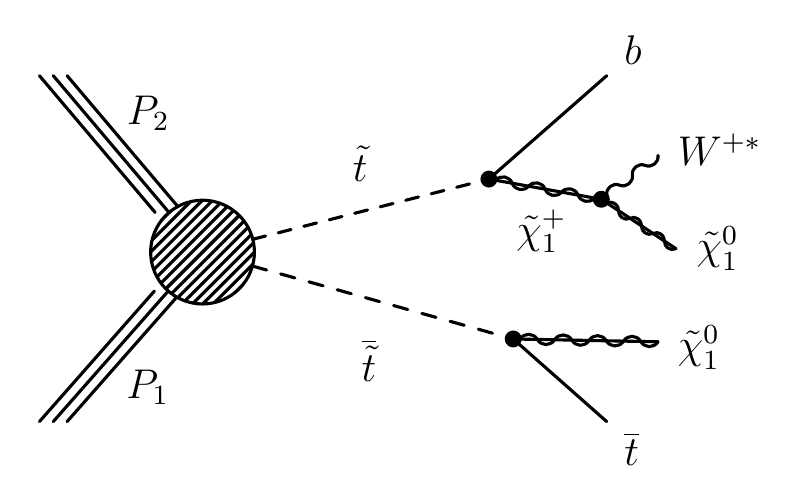}
   \caption{Diagrams representing two cases of the simplified models of direct top squark pair production and decay considered in this study:
the T2tt model with top squark decay via a top quark (\cmsLeft), and the T2tb model with the top squark decaying either via a top quark or via an intermediate chargino (\cmsRight).}
  \label{fig:sms_stop}
\end{figure}

For direct top squark pair production, we consider two decay scenarios within the SMS framework.
In the scenario denoted by ``T2tt,'' each $\stopq$ decays via a top quark: $\stopq \to \topq \lsp$, in which $\lsp$ is the LSP.
The second decay scenario considered here, denoted by ``T2tb,'' involves two $\stopq$ decay modes, $\stopq \to \topq \lsp$ (as in T2tt) and $\stopq \to \bq \chipmone$, each with a 50\% branching fraction.
In the latter case, the lightest chargino $\chipmone$ decays with 100\% branching fraction to a virtual $\W$~boson and a $\lsp$.
A natural simplified SUSY spectrum is assumed in which the $\chipmone$ is 5\GeV heavier than the $\lsp$~\cite{Papucci:2011wy,Brust:2011tb,Feng:2013pwa}.
As a result of the mixed decay modes, the T2tb scenario consists of three different final states containing either two $\PQb$ quarks and no top quarks (25\%), one $\PQb$ quark and one top quark (50\%), or two top quarks and no $\PQb$ quarks (25\%).
Figure~\ref{fig:sms_stop} shows the diagrams representing these two simplified models.

\begin{figure}[tb]
  \centering
    \includegraphics[width=0.48\textwidth]{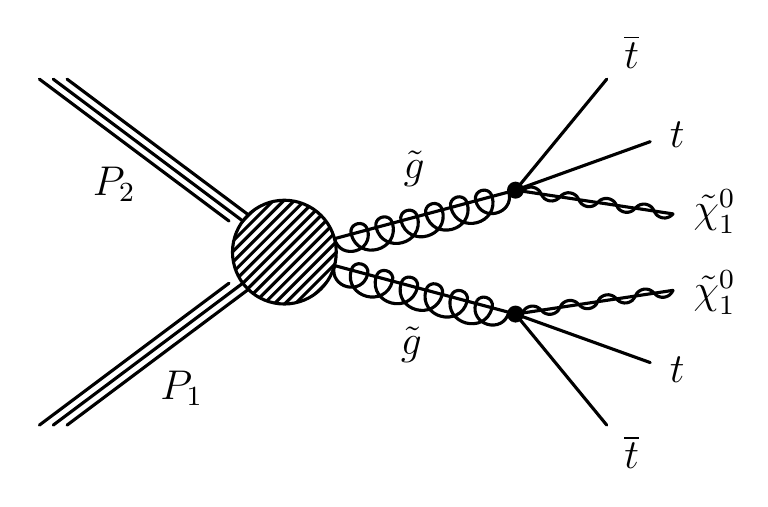}
    \includegraphics[width=0.48\textwidth]{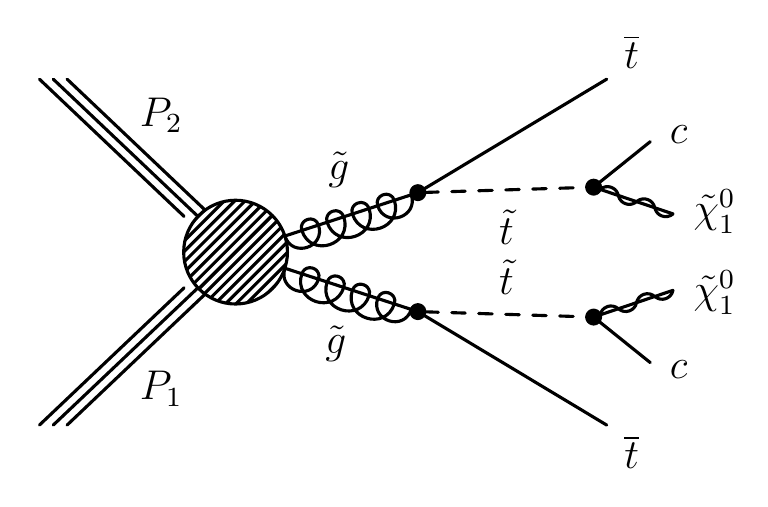}
   \caption{Diagrams representing the simplified models of gluino-mediated top squark production considered in this study: the T1tttt model (\cmsLeft) where the gluino decays to top quarks and the LSP via an off-shell top squark, and the T5ttcc model (\cmsRight) where the gluino decays to an on-shell top squark, which decays to a charm quark and the LSP.}
  \label{fig:sms_gluino}
\end{figure}

Two scenarios are considered for gluino-mediated top squark production, as shown in Fig.~\ref{fig:sms_gluino}.
In the main model, denoted by ``T1tttt,'' the gluino decays to top quarks via an off-shell top squark: $\gluino \to \PQt \cPaqt \lsp$.
This model is complementary to the direct top squark production because it gives sensitivity to the scenario where the gluino is kinematically accessible but the top squark is too heavy for direct production.
The second scenario, denoted by ``T5ttcc,'' features on-shell top squarks in the decay chain with a mass difference between top squark and LSP assumed to be $\Delta m(\stopq,\lsp) = 20\GeV$.
For this model, the gluino decays to a top quark and a top squark, $\gluino \to \cPaqt \stopq$, and the top squark decays to a charm quark and the LSP, $\stopq \to \cPqc \lsp$.
This model again serves as a complement to the direct search by providing sensitivity to very light top squarks, which would not decay to on-shell top quarks.

All scenarios described above share similar final states, containing two neutralinos and up to four top quarks.
Given that the $\lsp$ is stable and only interacts weakly, it does not produce a signal in the detector. Therefore, $\MET$ is one of the most important discriminators between signal and SM background, especially for models with large mass differences between the top squark or gluino and the $\lsp$.
Since top quarks decay almost exclusively to a $\PQb$ quark and a $\PW$ boson, each hadronically decaying top quark can result in up to three identified jets, depending on the top quark \pt and jet size.
For certain signal scenarios, there may be additional bottom, charm, or light-flavor quarks, which increase the expected jet and $\PQb$-tagged jet multiplicities.

\subsection{Top quark reconstruction and identification}
\label{sec:methodAlpha_topTagger}

The procedure to reconstruct and identify the hadronically decaying top quarks (top quark tagging or ``$\PQt$ tagging'') presented here is similar to the one used in Ref.~\cite{stop8TeV}, where reconstruction of the hadronically decaying top quarks from resolved jets is performed as described in Refs.~\cite{Kaplan:2008ie,Plehn:2010st,Kaplan:2012gd}.
The $\PQt$ tagging algorithm is improved in this work, to be more sensitive to boosted scenarios in which decay products from the $\PW$ boson or top quark are merged into a single jet.
Additionally, the algorithm is expanded to allow the reconstruction of multiple top quarks in each event.

The top quark tagging algorithm takes as input all reconstructed AK4 jets that satisfy $\pt>30\GeV$ and $\abs{\eta}<5$.
These jets are clustered into three categories of top quark candidates: trijet, dijet, and monojet.
Trijet candidates, representing the three jets coming from the $\PQb$ quark and the hadronic decay of the \W~boson, are subject to the following conditions:
(i) All jets lie within a cone of radius $\Delta R = 1.5$, centered at the direction defined by the vector sum of the momentum of the three jets.
The radius requirement implies a moderate Lorentz boost of the top quark, as is expected for the vast majority of signal parameter space $(m_{\sTop/\gluino}, m_{\PSGczDo})$ targeted in this search.
(ii) To reduce combinatoric backgrounds, one of the ratios of dijet to trijet masses must be consistent with the $m_{\W}/m_{\PQt}$ ratio~\cite{Plehn:2010st}.
The trijet system must satisfy one of the following three (overlapping) criteria:
\begin{linenomath}
\ifthenelse{\boolean{cms@external}}{
\begin{widetext}\begin{equation} \label{eq:taggerEQ}
\begin{aligned}
     \text{(a)}\quad& 0.2< \arctan\left(\frac{m_{13}}{m_{12}}\right) < 1.3 \quad\text{and}\quad R_\text{min} < \frac{m_{23}}{m_{\text{3-jet}}} < R_\text{max},\\
     \text{(b)}\quad& R_\text{min}^{2}\left[1+\left(\frac{m_{13}}{m_{12}}\right)^2\right]
     < 1-\left(\frac{m_{23}}{m_{\text{3-jet}}}\right)^2
     < R_\text{max}^{2}\left[1+\left(\frac{m_{13}}{m_{12}}\right)^2\right],  \\
     \text{(c)}\quad& R_\text{min}^{2}\left[1+\left(\frac{m_{12}}{m_{13}}\right)^2\right]
     < 1-\left(\frac{m_{23}}{m_{\text{3-jet}}}\right)^2
     < R_\text{max}^{2}\left[1+\left(\frac{m_{12}}{m_{13}}\right)^2\right].
\end{aligned}
\end{equation}
\end{widetext}
}
{
\begin{equation} \label{eq:taggerEQ}
\begin{aligned}
     \text{(a)}\quad& 0.2< \arctan\left(\frac{m_{13}}{m_{12}}\right) < 1.3 \quad\text{and}\quad R_\text{min} < \frac{m_{23}}{m_{\text{3-jet}}} < R_\text{max},\\
     \text{(b)}\quad& R_\text{min}^{2}\left[1+\left(\frac{m_{13}}{m_{12}}\right)^2\right]
     < 1-\left(\frac{m_{23}}{m_{\text{3-jet}}}\right)^2
     < R_\text{max}^{2}\left[1+\left(\frac{m_{13}}{m_{12}}\right)^2\right],  \\
     \text{(c)}\quad& R_\text{min}^{2}\left[1+\left(\frac{m_{12}}{m_{13}}\right)^2\right]
     < 1-\left(\frac{m_{23}}{m_{\text{3-jet}}}\right)^2
     < R_\text{max}^{2}\left[1+\left(\frac{m_{12}}{m_{13}}\right)^2\right].
\end{aligned}
\end{equation}
}
\end{linenomath}
Here, $m_{12}$, $m_{13}$, and $m_{23}$ are the dijet masses, where the jet indices 1, 2, and 3 reflect a decreasing order in \pt.
The numerical constants have values
$R_\text{min} = 0.85 \, (m_{\W}/m_{\PQt})$ and $R_\text{max} = 1.25 \, (m_{\W}/m_{\PQt})$, with $m_{\W} = 80.4\GeV$ and $m_{\PQt} = 173.4\GeV$~\cite{PDG}.
Assuming massless input jets and trijet mass $m_{\text{3-jet}} = m_{\PQt}$, each of the three criteria can be reduced to the condition that the respective ratio of
$m_{23}/m_{\text{3-jet}}$, $m_{12}/m_{\text{3-jet}}$ or $m_{13}/m_{\text{3-jet}}$ is within the range of $[R_{\text{min}}, R_{\text{max}}]$.

The second category of top quark candidates is clustered from just two jets and is designed to tag top quark decays in which the $\PW$ boson decay products
are merged into a single jet ($\PW$ jet).
The jet mass is used to determine if a jet represents a $\PW$ jet with a required mass window of $70$--$110\GeV$.
Additionally, the dijet system is required to pass the requirement:
\begin{linenomath}
\begin{equation}
    \label{eq:taggerDiJReq}
    R_\text{min} < \frac{m_{\PW\text{jet}}}{m_{\text{dijet}}} < R_\text{max},
\end{equation}
\end{linenomath}
where $m_{\PW\text{jet}}$ is the mass of the candidate $\PW$ jet and $m_{\text{dijet}}$ is the mass of the dijet system.
$R_\text{min}$ and $R_\text{max}$ are the same as for the trijet requirements.
The final category of candidates, monojets, are constructed from single jets which have a jet mass consistent with $m_{\PQt}$, \ie, in the range of 110--220\GeV.

After all possible top quark candidates are constructed, the final list of reconstructed top quark objects is determined by making requirements on the total mass of the object and the number of $\PQb$ jets.
Any top quark candidate with more than one $\PQb$ jet is rejected because the probability of having two genuine $\PQb$ jets, or having a second light-flavor jet tagged as a $\PQb$ jet, in a single top quark candidate is negligible.
All candidates with a mass outside the range $100$--$250\GeV$ are rejected.
The list of candidates is pruned to remove candidates that share a jet with another candidate, in favor of the candidate with the mass closer to the true top quark mass.
However, if there is only one $\PQb$ jet in the event, the top quark candidate with the best match to the true top mass may be pruned if it contains the $\PQb$ jet to ensure that there are two objects for the \MTTwo calculation (described below).

\begin{figure}[htb]
  \centering
    \includegraphics[width=\cmsFigWidth]{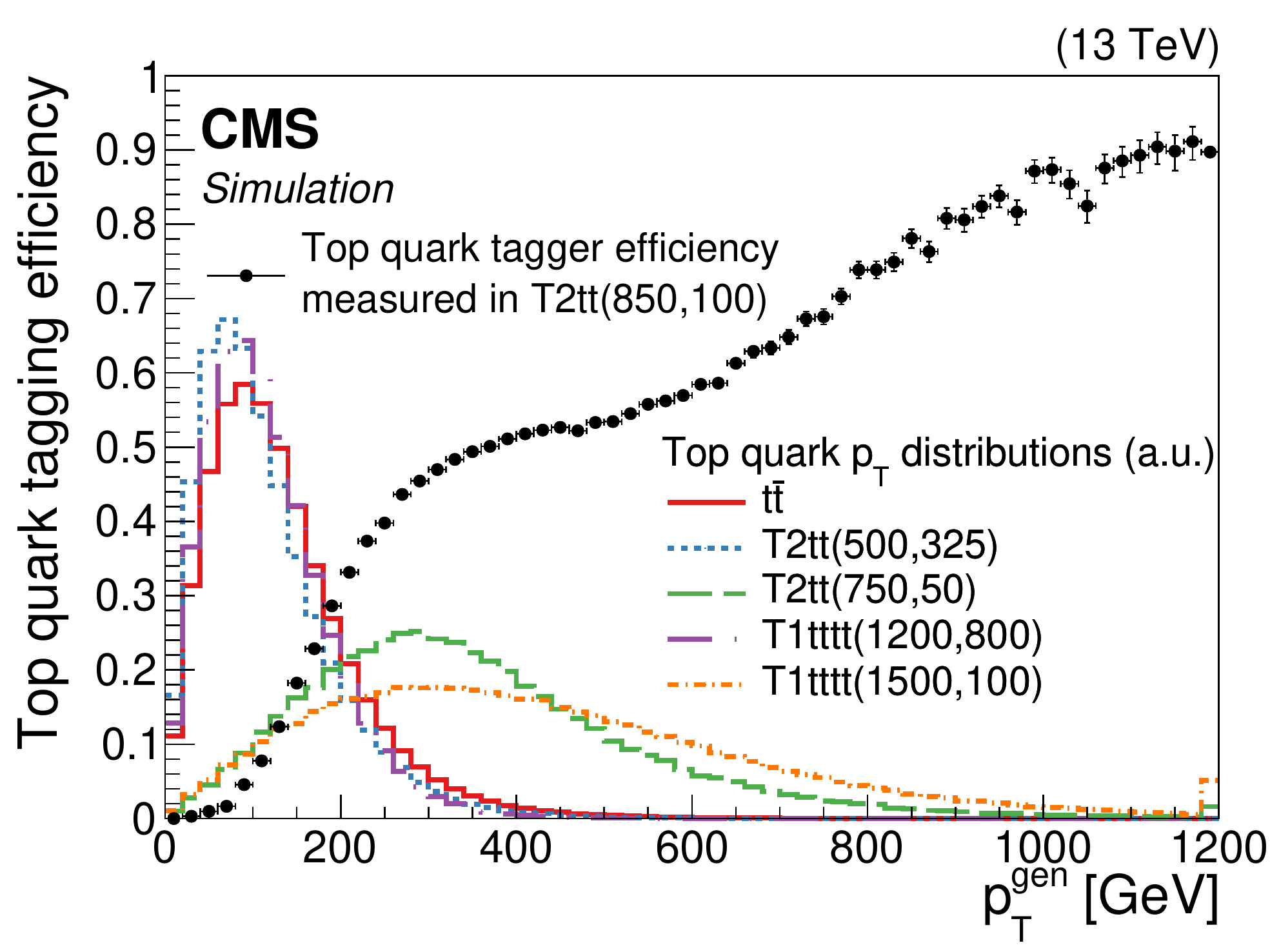}
    \caption{The tagging efficiency of the top quark tagger as a
    function of the generator-level hadronically decaying top quark $\pt$ (black points).
    The efficiency was computed using the T2tt signal model with $m_{\sTop} = 850\GeV$ and $m_{\lsp} = 100\GeV$, and it is similar for \ttbar events.
    The vertical bars depict the statistical uncertainty. The colored lines show the
    expected hadronically decaying top quark $\pt$ distribution from \ttbar (red solid line),
    the T2tt signal model with $m_{\sTop} = 500\GeV$ and $m_{\lsp} = 325\GeV$ (blue short-dashed line),
    the T2tt signal model with $m_{\sTop} = 750\GeV$ and $m_{\lsp} = 50\GeV$ (green long-dashed line),
    the T1tttt signal model with $m_{\gluino} = 1200\GeV$ and $m_{\lsp} = 800\GeV$ (purple long-dash-dotted line),
    and the T1tttt signal model with $m_{\gluino} = 1500\GeV$ and $m_{\lsp} = 100\GeV$ (orange short-dash-dotted line).
    The last bin contains the overflow entries and the top quark $\pt$ distributions are normalized to unit area.}
    \label{fig:MethodAlpha_toptagger_efficiency}
\end{figure}

By considering not only fully resolved (trijet) top quark decays, but also decays from boosted top quarks, manifesting themselves as dijet or monojet topologies,
this $\PQt$ tagger achieves a high efficiency for tagging top quarks over a wide range of top quark \pt values,
from ${\sim}30\%$ at $200\GeV$ to close to 85\% at $1\TeV$.
The tagging efficiency is determined using the T2tt signal model with $m_{\sTop} = 850\GeV$ and $m_{\lsp} = 100\GeV$ since it has a wide top quark $\pt$ spectrum.
The tagging efficiency was also measured using SM \ttbar background and other signal models, and was found to agree with the T2tt measurement within statistical uncertainties.
The event sample used to measure the tagging efficiency was selected by requiring the presence of at least four jets with $\pt > 30\GeV$ and $\abs{\eta}<2.4$.
The $\PQt$-tagged object must be matched to a hadronically decaying generator-level top quark within a
cone of radius 0.4 in ($\eta,\phi$) space.
The $\PQt$ tagging efficiency as a function of top quark \pt is shown in Fig.~\ref{fig:MethodAlpha_toptagger_efficiency}, which also includes
the expected \pt distributions for the hadronically decaying top quark in SM \ttbar events, as well as in various signal models.
Since the top quark \pt spectrum for signal events depends strongly on $m_{\sTop / \gluino}$ and $\Delta m(\sTop / \gluino,\lsp)$, the good tagging efficiency across the top quark \pt spectrum ensures high acceptance for a wide range of signal models.
The tagging efficiency for a previous algorithm, described in Ref.~\cite{stop8TeV}, as evaluated from simulation, is about 20\% at top quark $\pt=600\GeV$ and drops quickly to close to 0 for higher top quark \pt.
Figure~\ref{fig:MethodAlpha_toptagger_efficiency} shows that the top quark tagger performance has substantially improved with respect to that used in Ref.~\cite{stop8TeV}: the efficiency is about 55\% at $\pt=600\GeV$, and it rises with increasing \pt.

The purity of the $\PQt$ tagger, computed as the percentage of $\PQt$-tagged objects that can be matched to a hadronically decaying generator-level top quark within a cone of radius 0.4 in ($\eta,\phi$) space, is 70--90\% in \ttbar events that satisfy $\MET>200\GeV$ and contain at least four jets, at least one of which is \PQb-tagged.
The probability that an event that does not contain hadronically decaying top quarks will be found to contain one or more $\PQt$-tagged objects is about 30--40\% for events passing the selection used for the efficiency calculation. Further details on the $\PQt$ tagger performance are presented in \suppMaterial.
The event yields of these processes, as well as from the \ttbar process, are further reduced by placing requirements on the ``stransverse mass'' variable, \MTTwo, discussed below, as a complement to the top quark tagging requirements.
The top quark tagging efficiency agrees well between data and the \GEANTfour-based simulation as shown in \suppMaterial. However, a correction factor of up to 5\% is needed to account for discrepancies between the fast simulation and the \GEANTfour-based simulation. It is derived using the same T2tt signal model mentioned above and is parametrized as a function of top quark candidate \pt.

The \MTTwo variable~\cite{Lester:1999tx,Barr:2003rg} is an extension of the transverse mass variable that is sensitive to the pair production of heavy particles, \eg, gluinos or top squarks, each of which decays to an invisible particle.
For direct top squark production, \MTTwo has a kinematic upper limit at the \sTop mass, whereas for \ttbar production the kinematic upper limit is the top quark mass.
For gluino pair production, the interpretation of \MTTwo depends on the decay scenario. However, the values of \MTTwo are consistently larger than those for \ttbar or other SM backgrounds due to the larger values of \MET and the high \pt of the top quarks produced in gluino decays.
The \MTTwo variable is defined for two heavy particles, denoted with subscripts $1$ and $2$, decaying to some visible particles and an invisible particle ($\lsp$) as:
\begin{linenomath}
\ifthenelse{\boolean{cms@external}}{
\begin{multline} \label{eq:MT2}
     \MTTwo \equiv \min_{\vec{q}_\text{T,1}+\vec{q}_\text{T,2} = \ptvecmiss} \Bigl\{ \max \bigl[m_\mathrm{T} ^2(\vec{p}_\text{T,1}; m_\text{p,1}, \vec{q}_\text{T,1}; m_{\lsp}), \\
     m_\mathrm{T} ^2(\vec{p}_\text{T,2}; m_\text{p,2}, \vec{q}_\text{T,2}; m_{\lsp})\bigr]\Bigr\} ,
\end{multline}
}{
\begin{equation} \label{eq:MT2}
     \MTTwo \equiv \min_{\vec{q}_\text{T,1}+\vec{q}_\text{T,2} = \ptvecmiss} \left\{ \max \left[m_\mathrm{T} ^2(\vec{p}_\text{T,1}; m_\text{p,1}, \vec{q}_\text{T,1}; m_{\lsp}), m_\mathrm{T} ^2(\vec{p}_\text{T,2}; m_\text{p,2}, \vec{q}_\text{T,2}; m_{\lsp})\right]\right\} ,
\end{equation}
}
\end{linenomath}
where $\vec{p}_\mathrm{T,i} $ and $m_\mathrm{p,i}$ are the transverse momentum and mass of the visible daughters of each heavy particle,
and $\vec{q}_\mathrm{T,i} $ and $m_{\lsp}$ represent the unknown transverse momentum and mass of the invisible $\lsp$ from each heavy particle decay.
The transverse mass squared, $m_\mathrm{T} ^2$, is defined as
\begin{linenomath}
\begin{equation} \label{eq:MTdef}
        m_\mathrm{T} ^2(\vec{p}_\mathrm{T} ; m_\mathrm{p}, \vec{q}_\mathrm{T} ; m_{\lsp}) \equiv m_\mathrm{p}^{2} + m_{\lsp}^2 + 2 \left(|\vec{p}_\mathrm{T} | |\vec{q}_\mathrm{T} | - \vec{p}_\mathrm{T}  \cdot \vec{q}_\mathrm{T} \right).
\end{equation}
\end{linenomath}
The \MTTwo variable is the minimum~\cite{Lester:1999tx} of two transverse masses with the constraint that the sum of the transverse momenta of both neutralinos is equal to the $\ptvecmiss$ in the event, i.e., $\vec{q}_\text{T,1}+\vec{q}_\text{T,2} = \ptvecmiss$.
The invisible particle is assumed to be massless, in order to be consistent with the use of the neutrino as the invisible particle in the \MTTwo calculation for the SM backgrounds, therefore $m_{\lsp}$ equals zero in Eqs.~(\ref{eq:MT2}) and (\ref{eq:MTdef}).

We construct the visible decay products of each heavy particle ($1$ and $2$) from the list of $\PQt$-tagged objects.
The selection requirements used in the analysis ensure that every event has at least one reconstructed $\PQt$-tagged object.
In the case where two $\PQt$-tagged objects are identified, each is used as one visible component in the \MTTwo calculation.
If more than two $\PQt$-tagged objects are found, \MTTwo is calculated for all combinations and the lowest \MTTwo value is used.
In the case where only one $\PQt$-tagged object is identified, the visible component of the second system is taken from the remaining jets not included in the $\PQt$-tagged object, using a $\PQb$-tagged jet as a seed to partially reconstruct a top quark.
The $\PQb$-tagged jet is combined with the closest jet that yields an invariant mass between 50\GeV and $m_{\PQt}$.
The combined ``dijet'' is used as the second visible system.
In case no jet combination satisfies that invariant mass requirement, the $\PQb$-tagged jet is used as the only remnant of the second visible system.

\subsection{Event selection and categorization}
\label{sec:alpha_event_selection}

Events in the search regions are collected with a trigger that applies a lower threshold of $350\GeV$ on \HT in coincidence with a threshold of 100\GeV on \MET.
This trigger is fully efficient at selecting events satisfying the requirements $\HT > 500\GeV$ and $\MET > 175\GeV$, both at the full event reconstruction level.

All events must pass filters designed to remove detector- and beam-related noise.
All jets considered in this analysis are required to have $\pt>30\GeV$, and must pass a set of jet identification criteria as described in Ref.~\cite{CMS-PAS-JME-10-003}.
The minimum number of such jets with $\abs{\eta}<2.4$ in an event must be $\njets\geq4$, with the leading two jets required to have $\pt>50\GeV$.
Events must satisfy $\MET > 200\GeV$ and $\HT > 500\GeV$, where the thresholds are chosen to exceed the trigger efficiency turn-on and to allow a low $175 < \MET < 200\GeV$ sideband for background studies.
A requirement on the angle between \MET and the first three leading jets, $\Delta\phi(\MET, j_{1,2,3})> 0.5$, 0.5, 0.3, is applied to reduce the number of events from QCD multijet processes.
High-\MET QCD multijet events are usually the result of an undermeasurement of the \pt of one of the leading jets, which results in \MET being aligned with that jet and $\Delta\phi(\MET, j_{1,2,3})$  being small.
The undermeasurement can occur because of detector effects or, in the case of semileptonic $\PQb$ or $\cPqc$ quark decays, because a neutrino carries away unmeasured energy.
Finally, requirements that $\ntops \ge 1$, $\nbjets \ge 1$, and $\MTTwo>200$\GeV are applied, after which we observe 288 events in the data.

\begin{figure*}[tbp]
  \centering
    \includegraphics[width=0.48\linewidth]{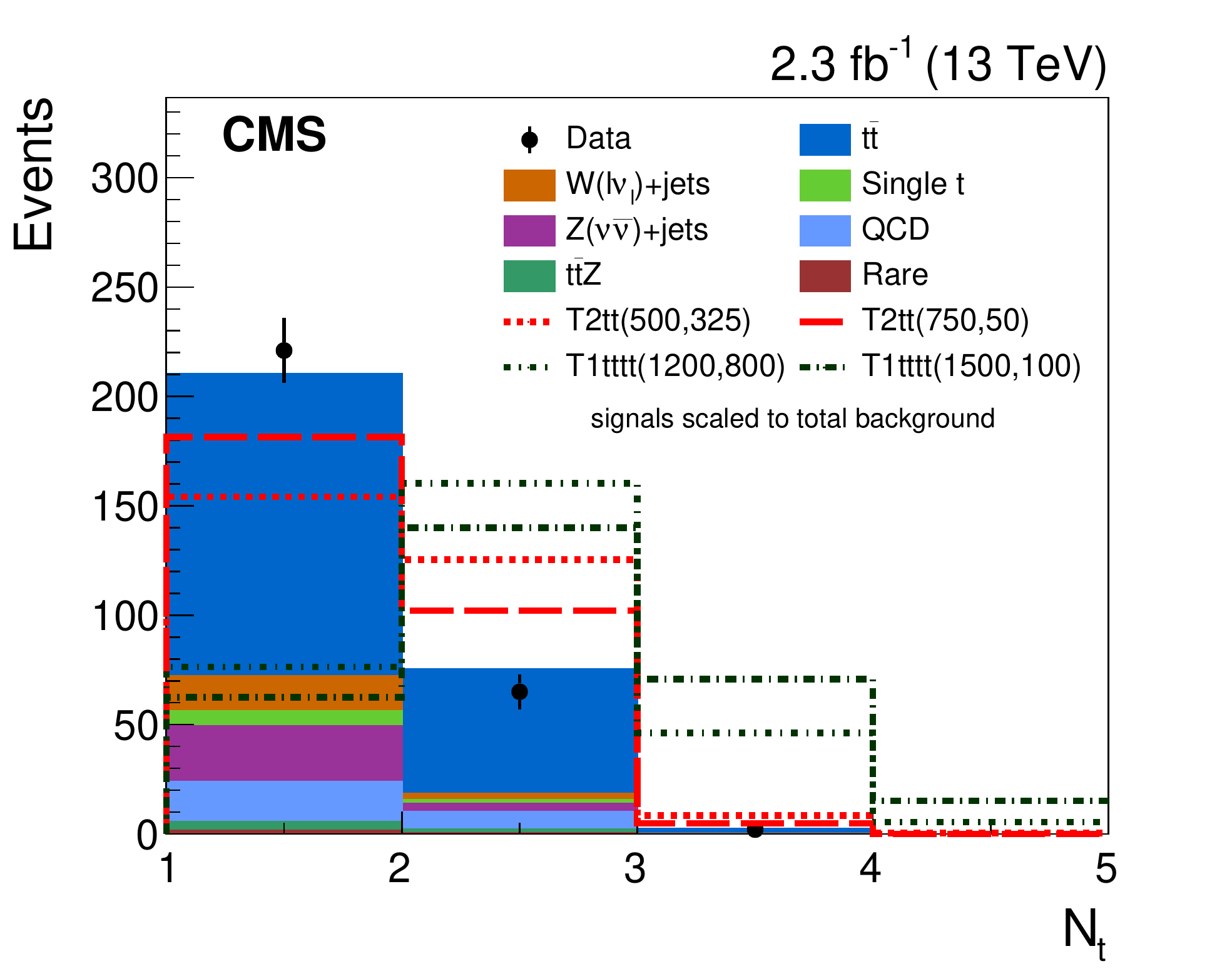}
    \includegraphics[width=0.48\linewidth]{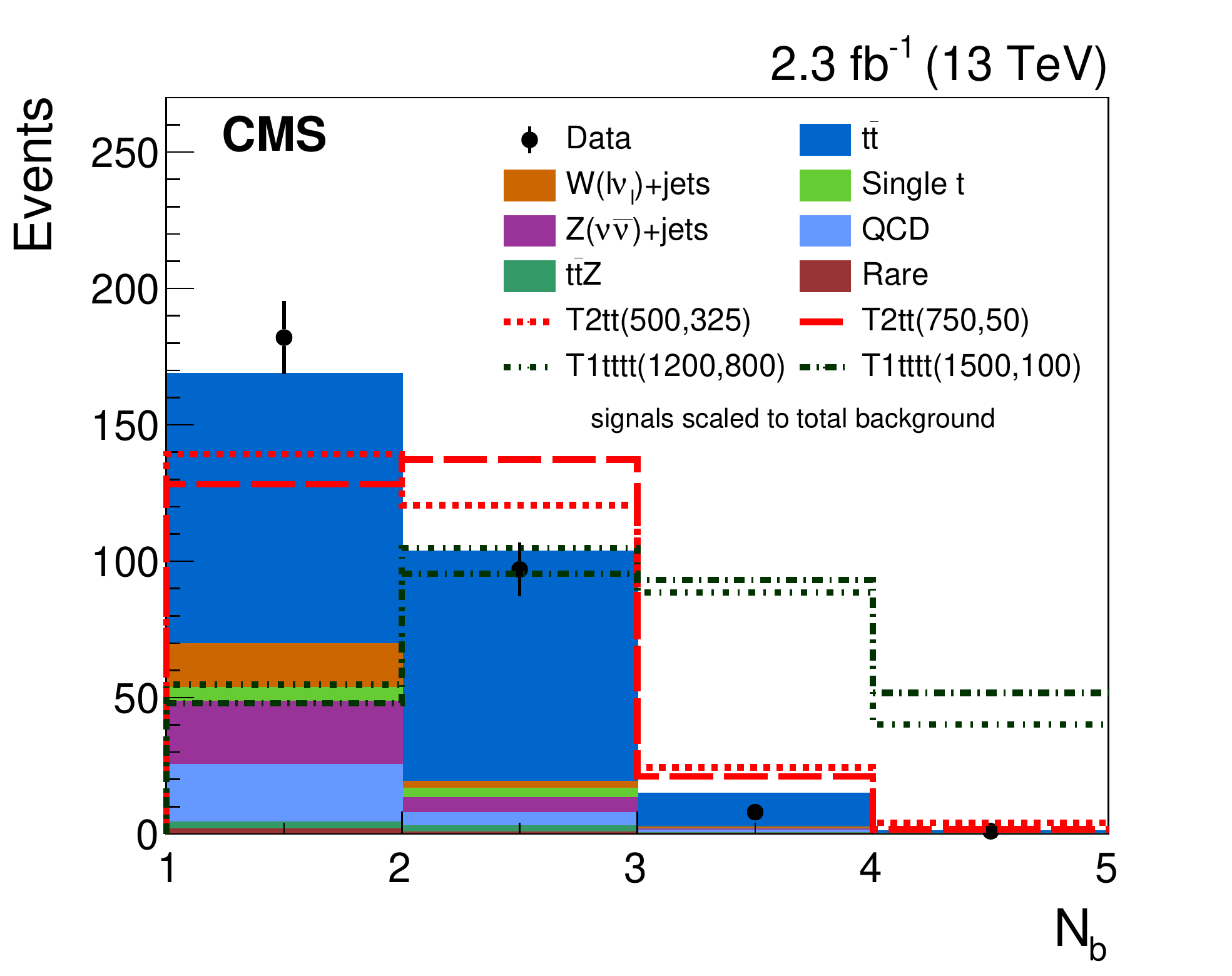} \\
    \includegraphics[width=0.48\linewidth]{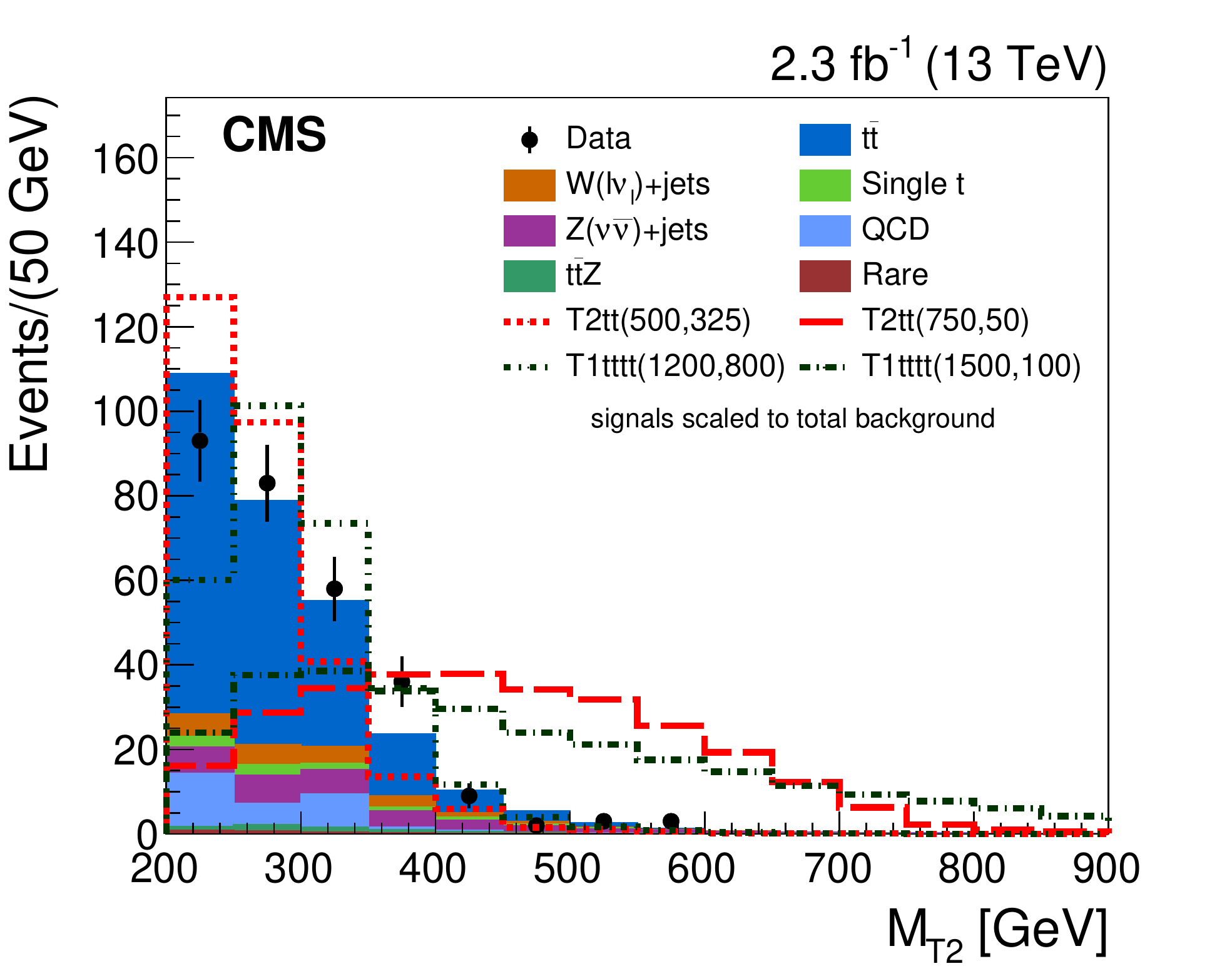}
    \includegraphics[width=0.48\linewidth]{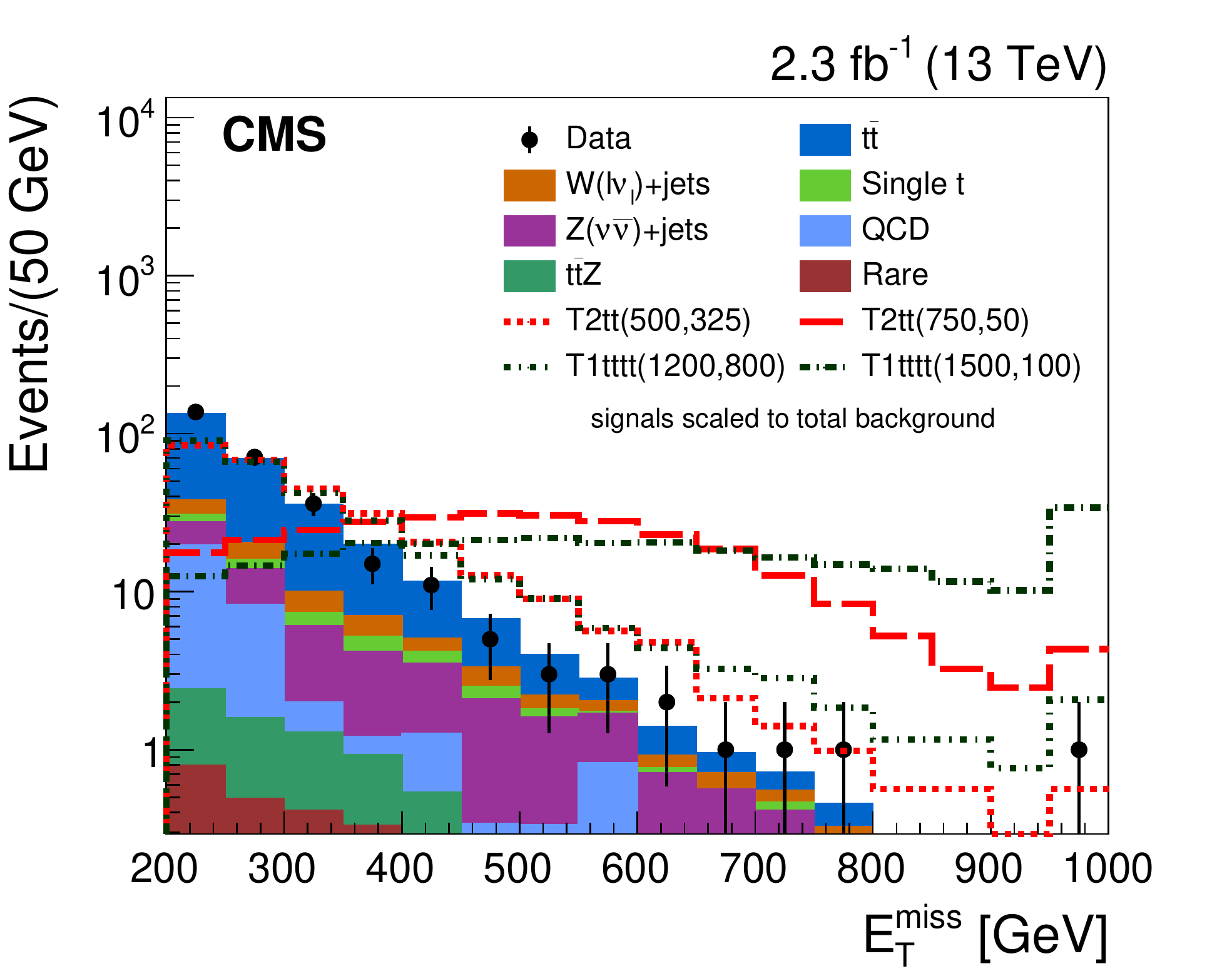}
    \caption{ Comparison of the distributions in data (black points), simulated SM backgrounds (filled stacked histograms) and several signal models in \ntops (top left), \nbjets (top right), \MTTwo (bottom left), and \MET (bottom right), after the preselection requirements have been applied.
    The T2tt signal model with $m_{\sTop} = 500 \,(750)\GeV$ and $m_{\lsp} = 325 \,(50)\GeV$ is shown with a red short-dashed (long-dashed) line,
    and the T1tttt signal model with $m_{\gluino} = 1200 \,(1500)\GeV$ and $m_{\lsp} = 800 \,(100)\GeV$ with a dark green short-dash-dotted (long-dash-dotted) line.
    The distributions for the signal events have been normalized to the same area as the total background distribution, and the last bin contains the overflow events.
    }
    \label{fig:compSBvars}
\end{figure*}

\begin{figure*}[tbhp]
  \centering
    \includegraphics[width=0.32\linewidth]{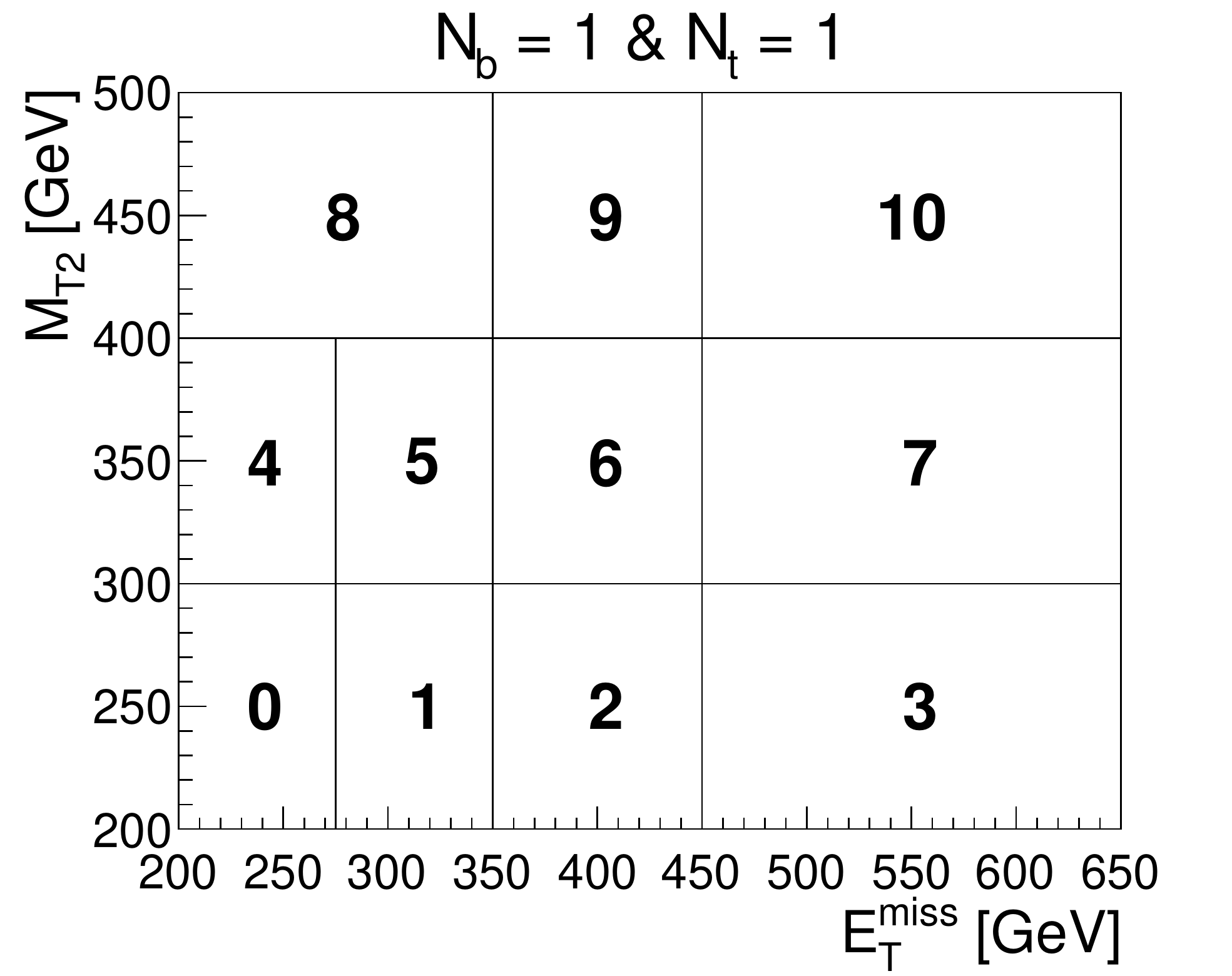}
    \includegraphics[width=0.32\linewidth]{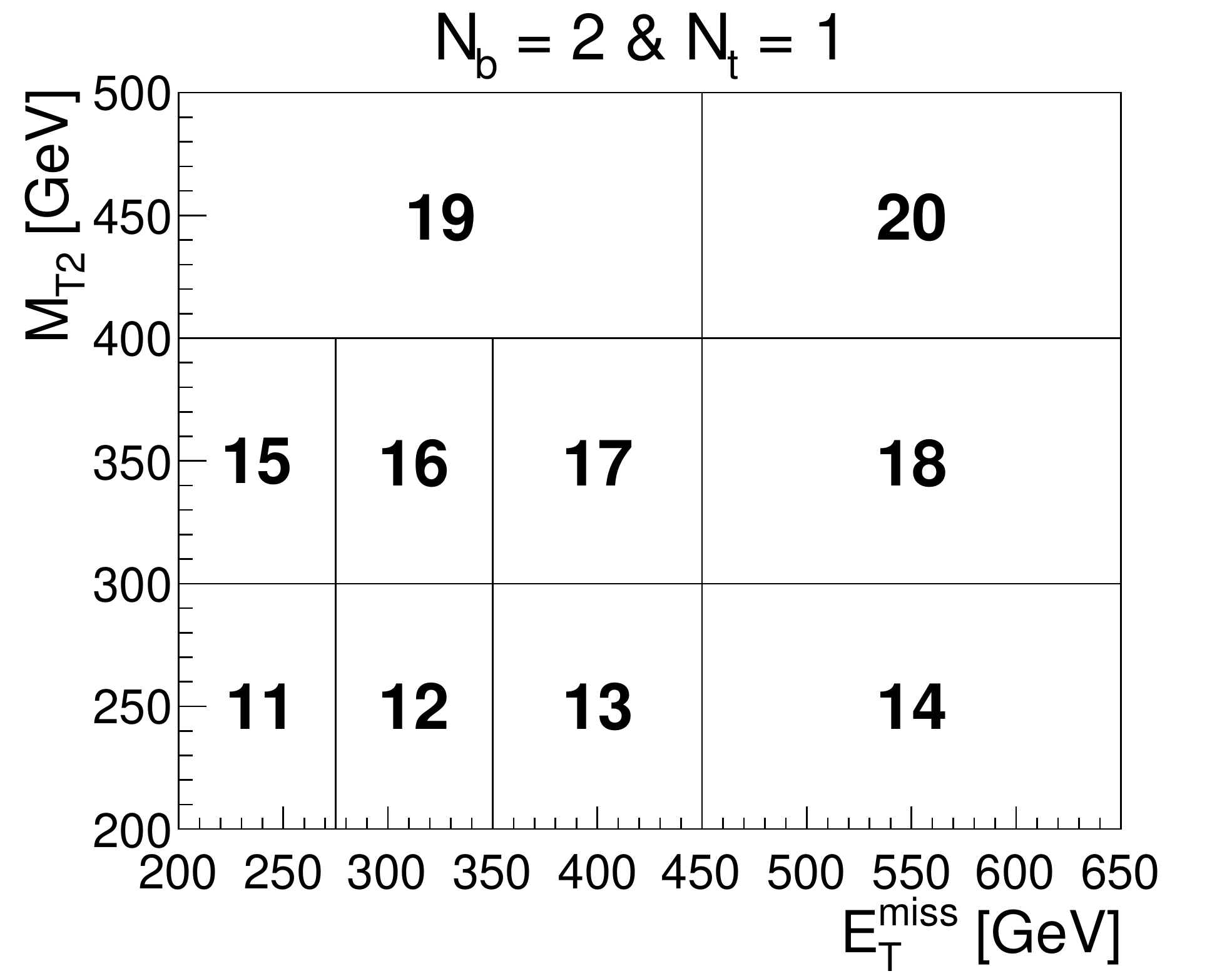}
    \includegraphics[width=0.32\linewidth]{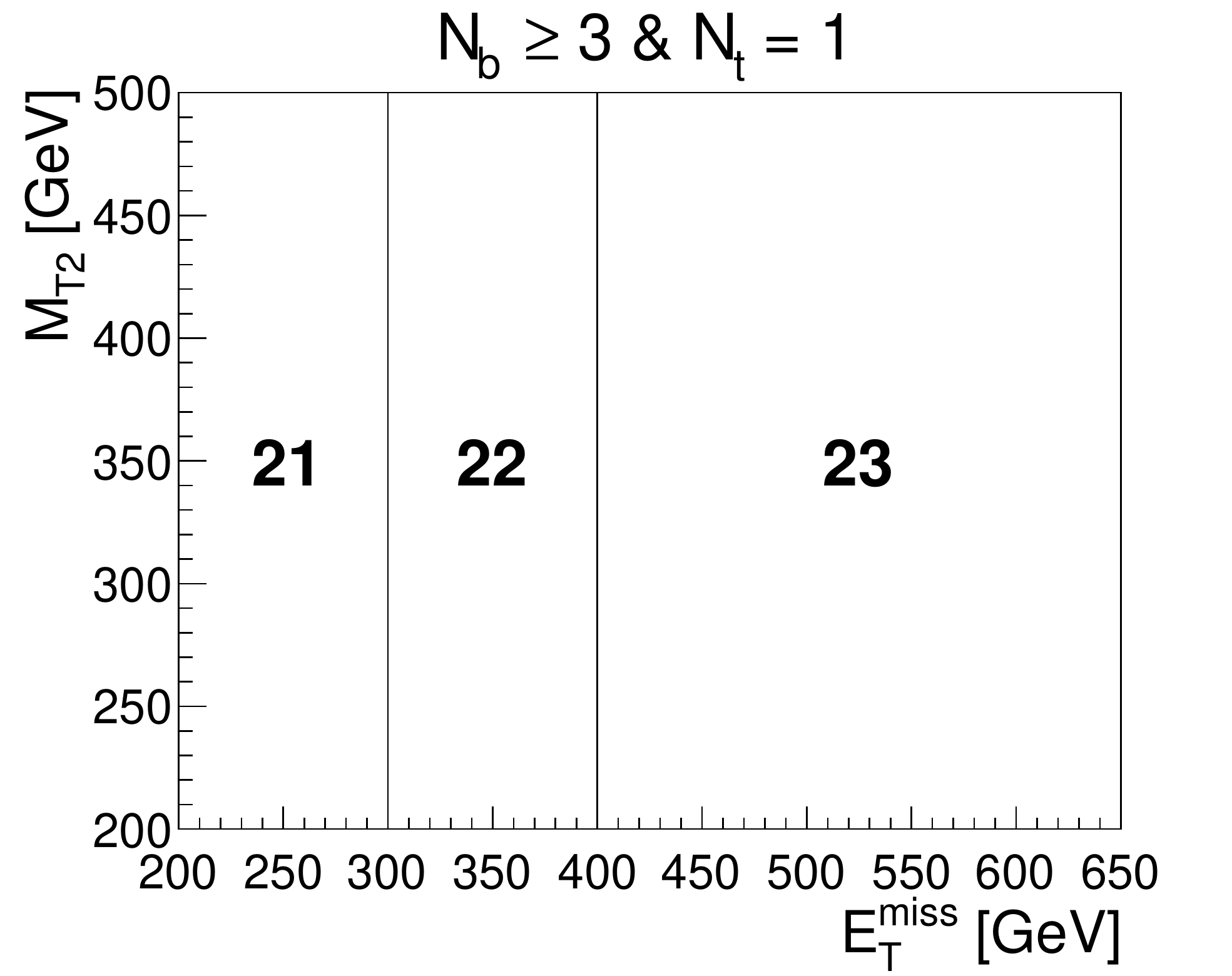}
    \includegraphics[width=0.32\linewidth]{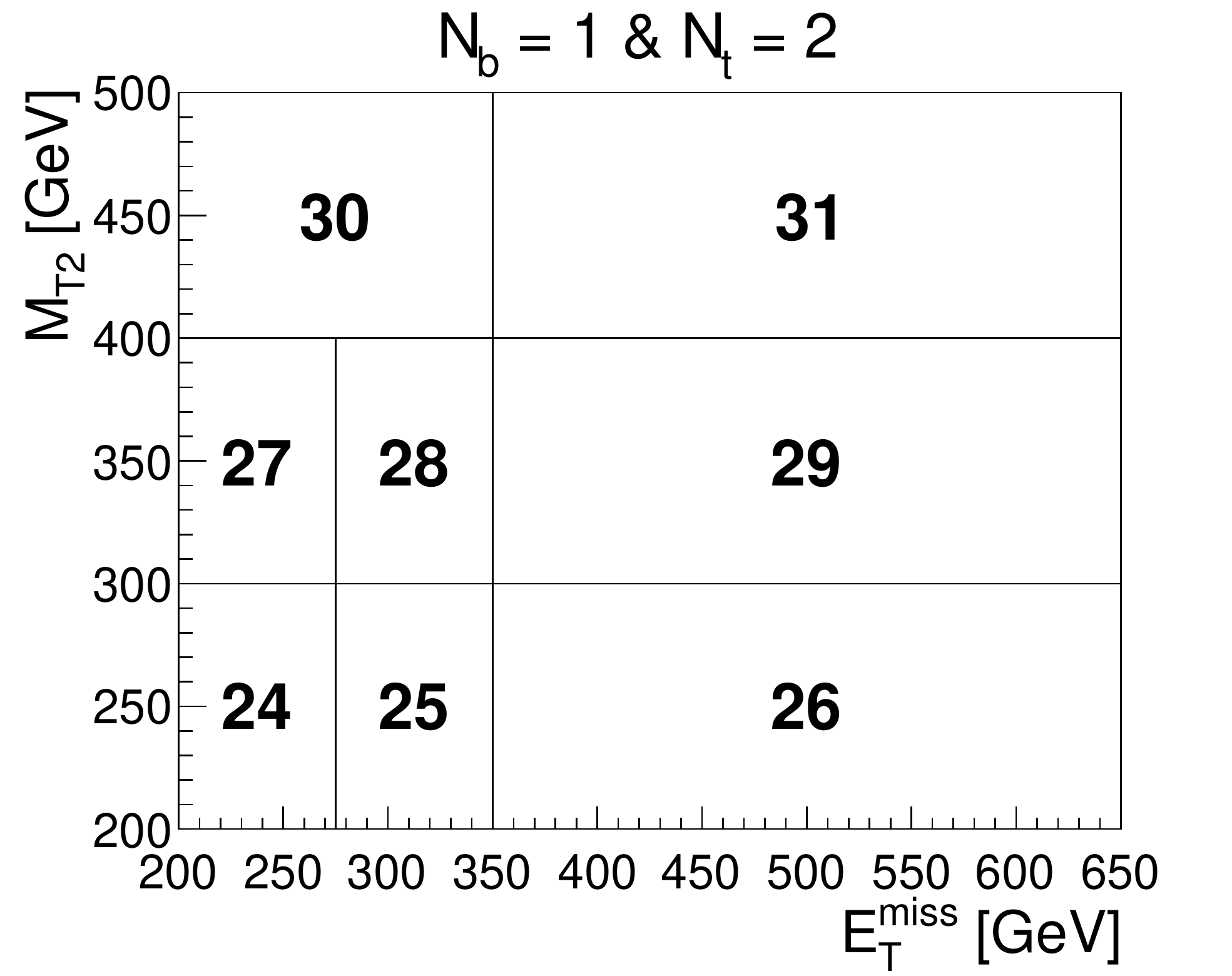}
    \includegraphics[width=0.32\linewidth]{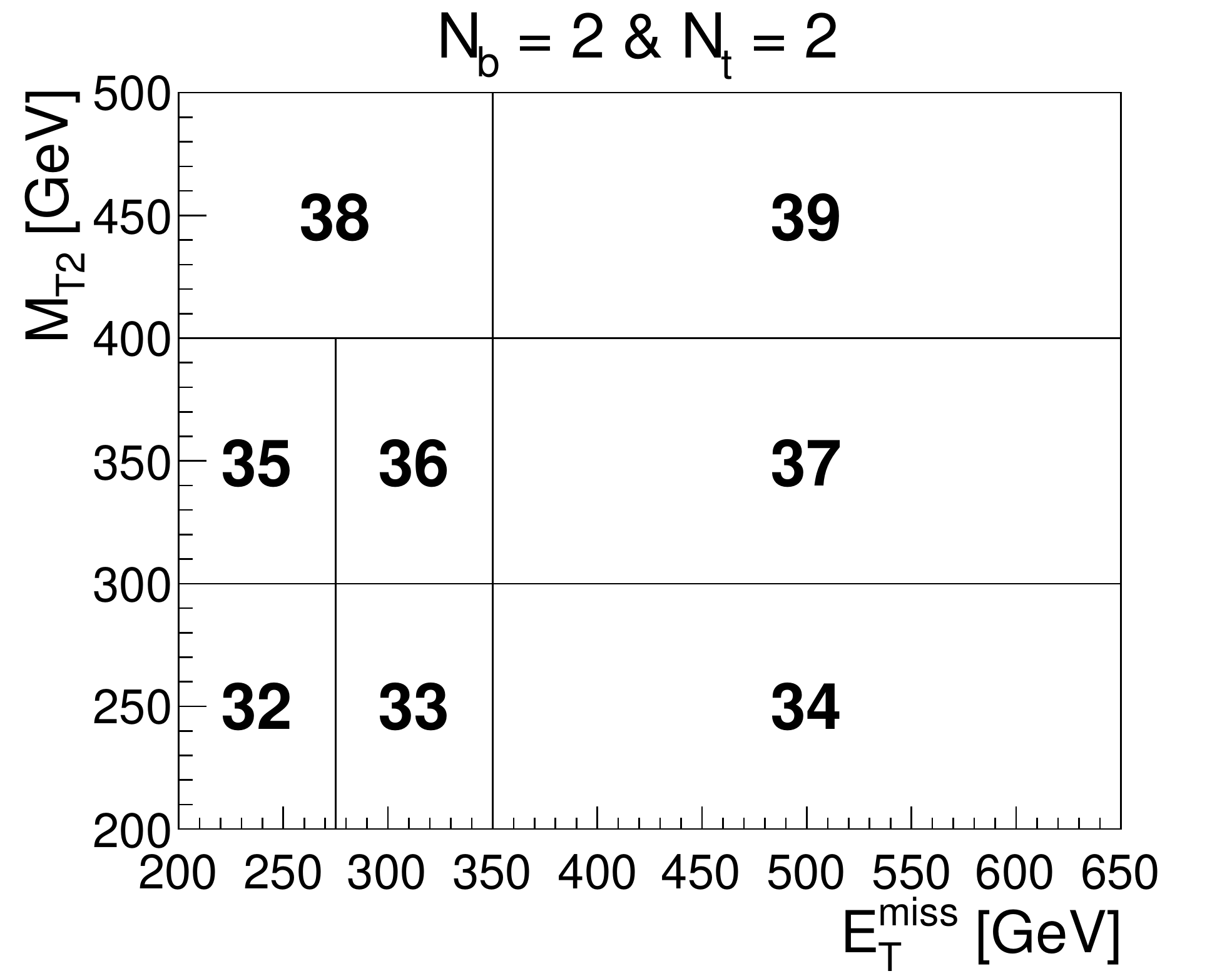}
    \includegraphics[width=0.32\linewidth]{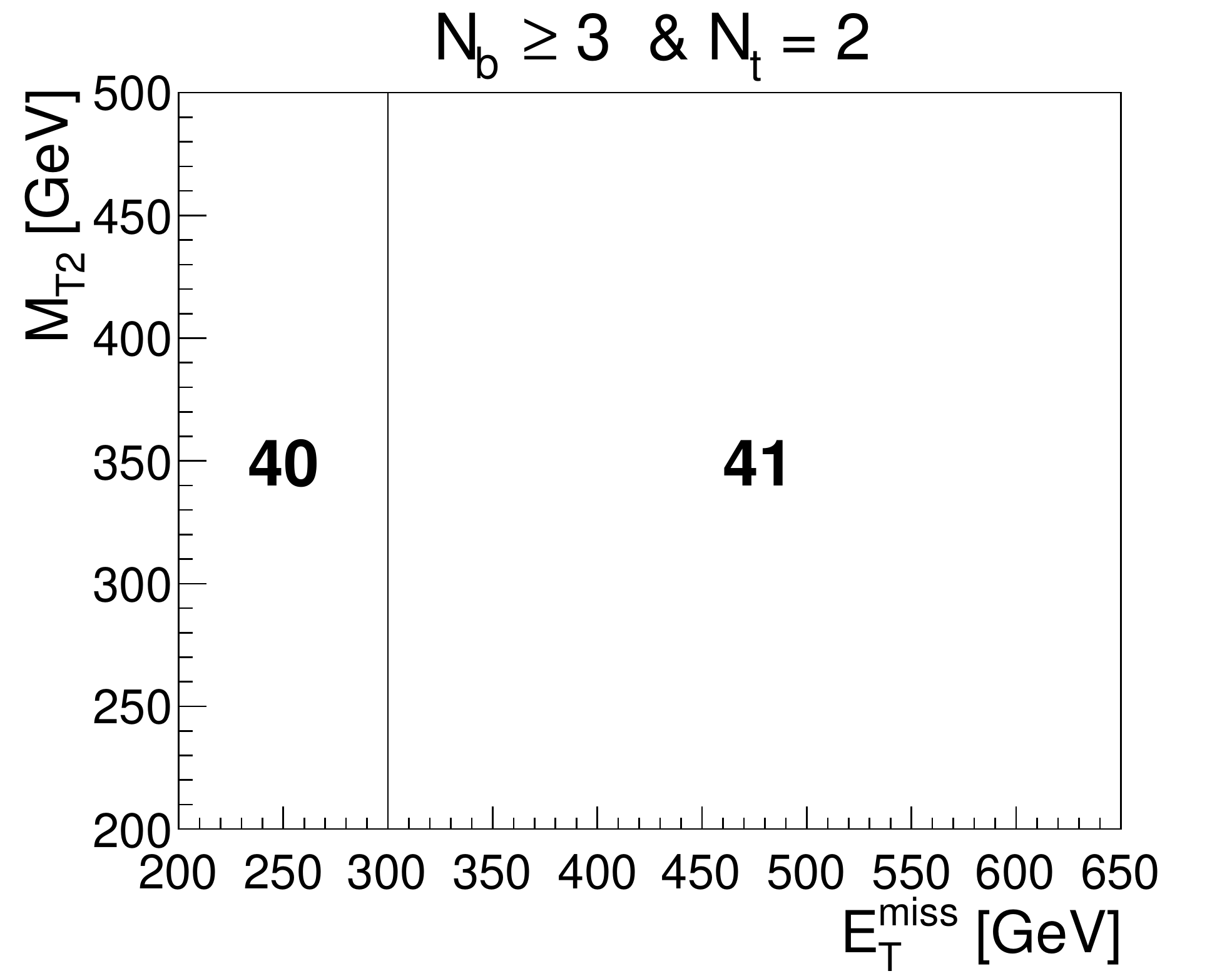}
    \caption{ Search region definitions for bin numbers 0--41 of the gluino-mediated production optimization.
The highest \MET and \MTTwo bins are open-ended, \eg, bin 10 requires $\MET>450\GeV$ and $\MTTwo > 400\GeV$.
In addition to the search bins shown in this figure, there are three bins (42--44) with $\ntops\ge 3$, one for each $\nbjets$ bin, that contain no further binning in \MET or \MTTwo beyond baseline selection requirements.
}
    \label{fig:SBXX}
\end{figure*}

\begin{figure}[htbp]
  \centering
    \includegraphics[width=0.32\textwidth]{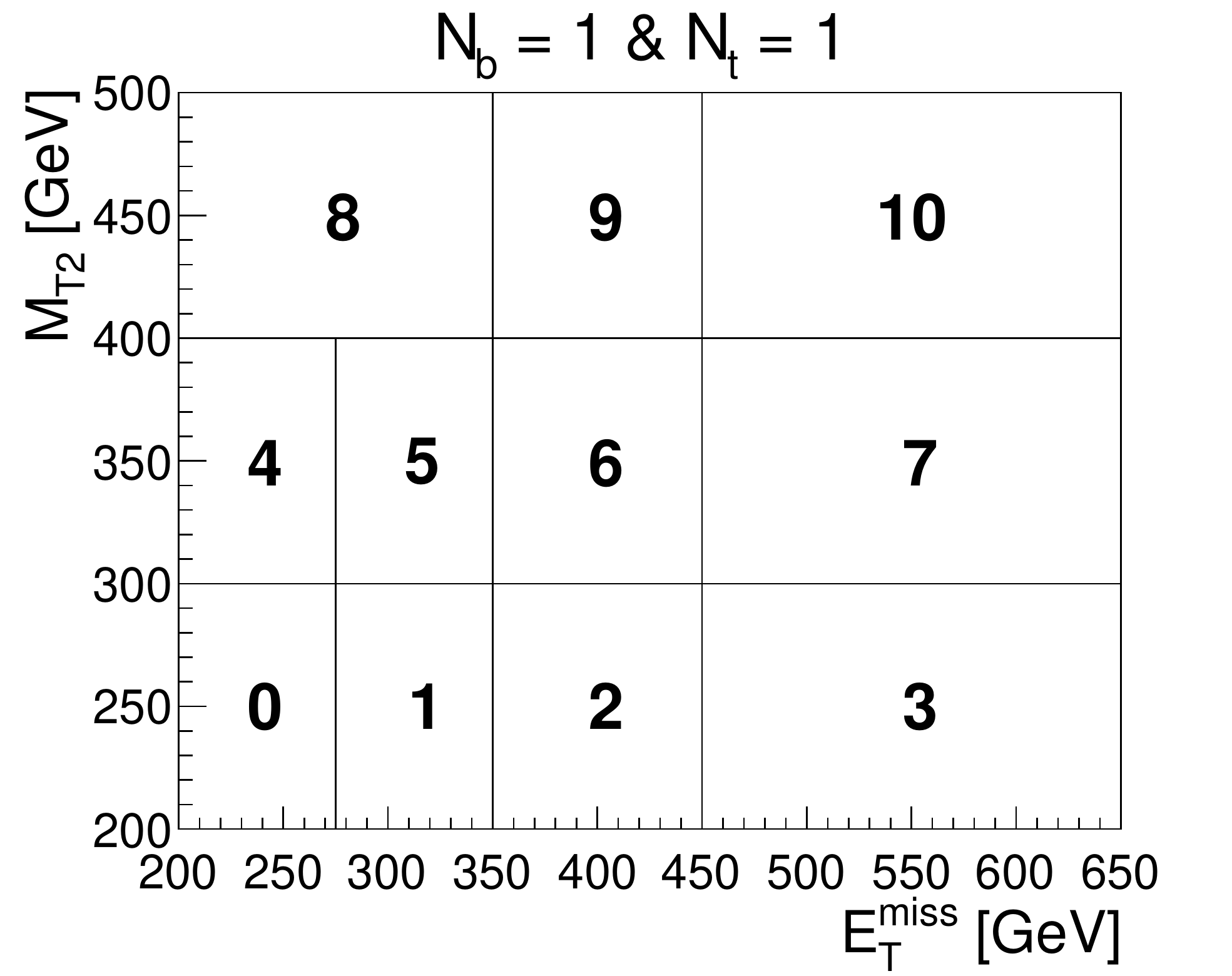}
    \includegraphics[width=0.32\textwidth]{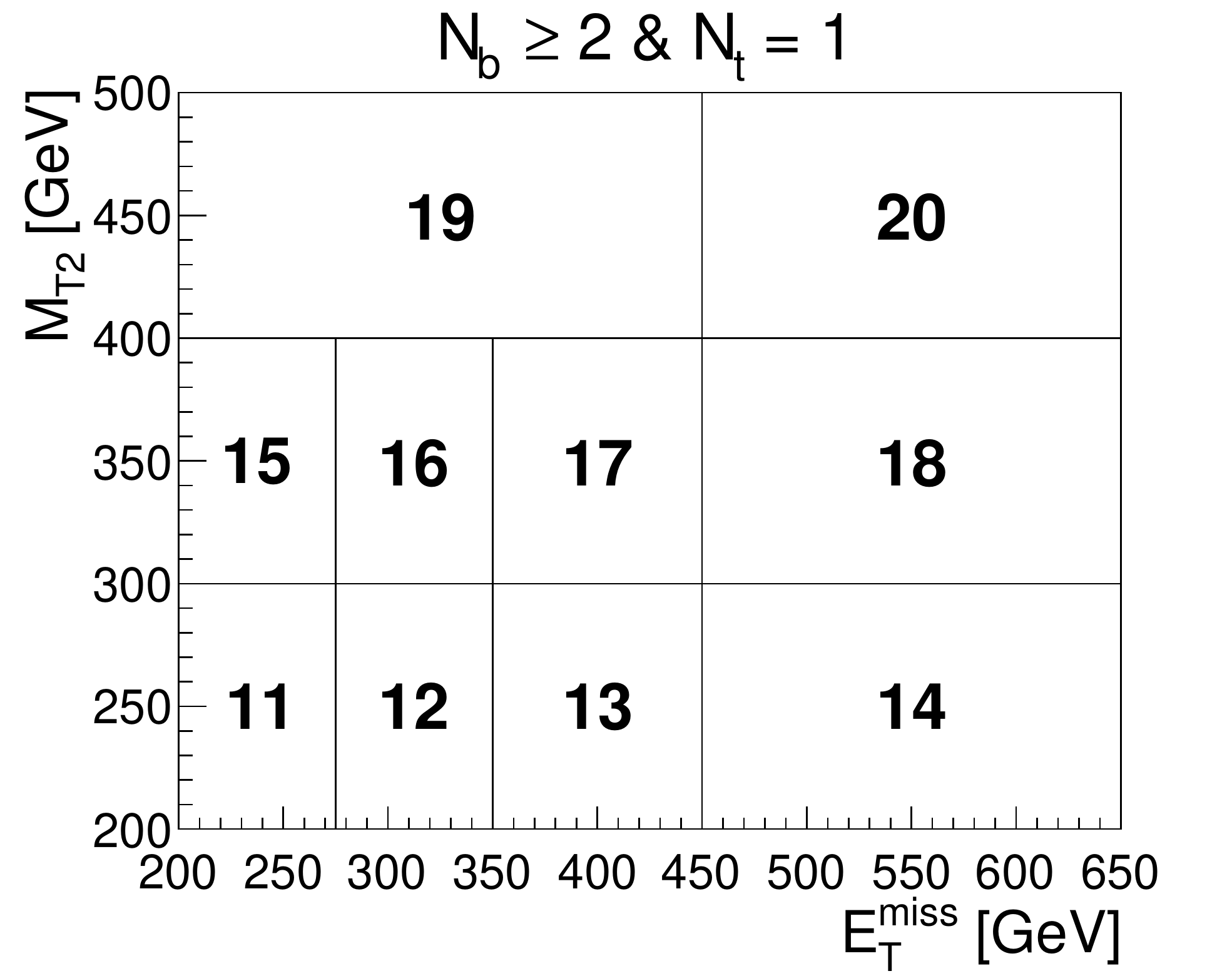}\\
    \includegraphics[width=0.32\textwidth]{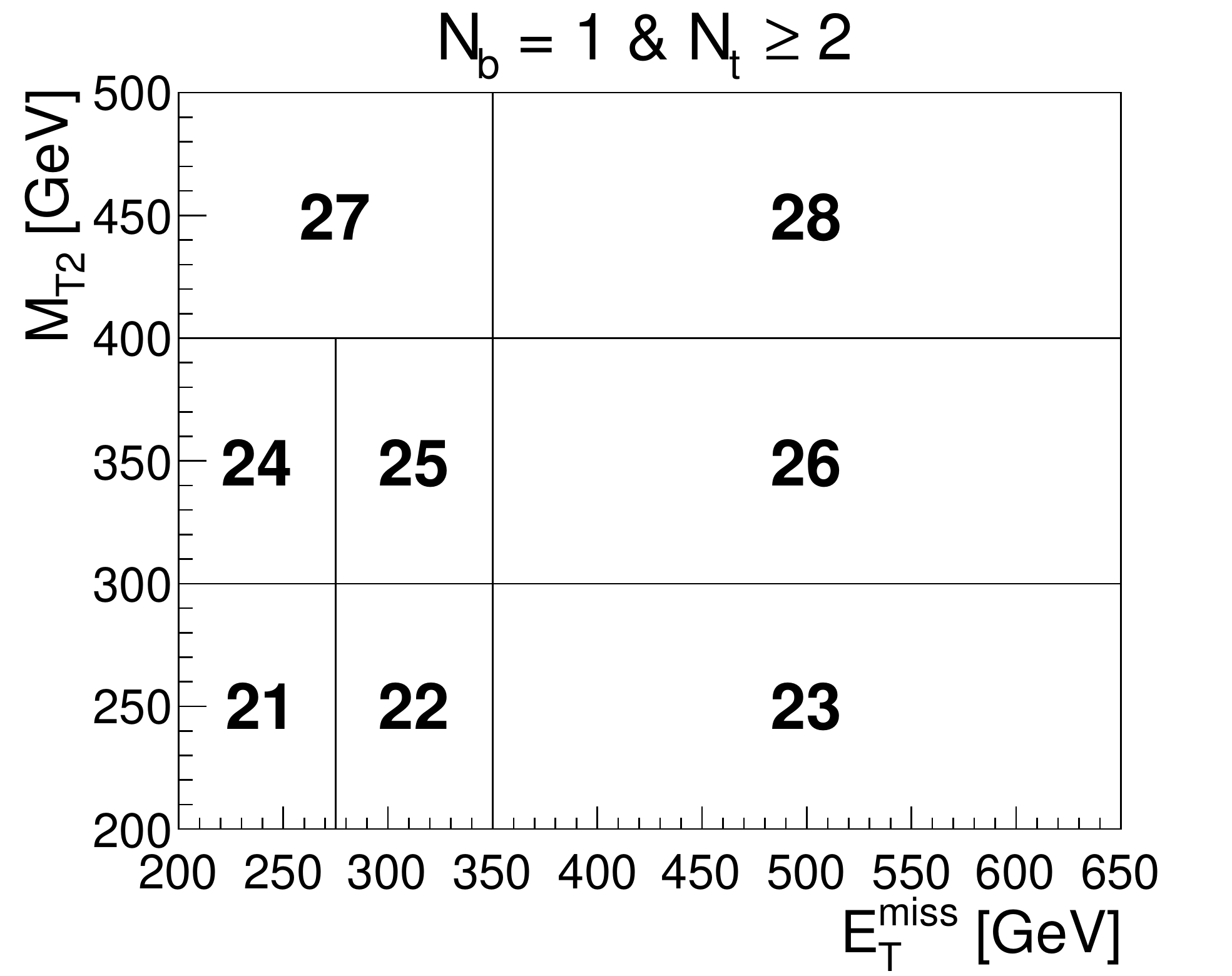}
    \includegraphics[width=0.32\textwidth]{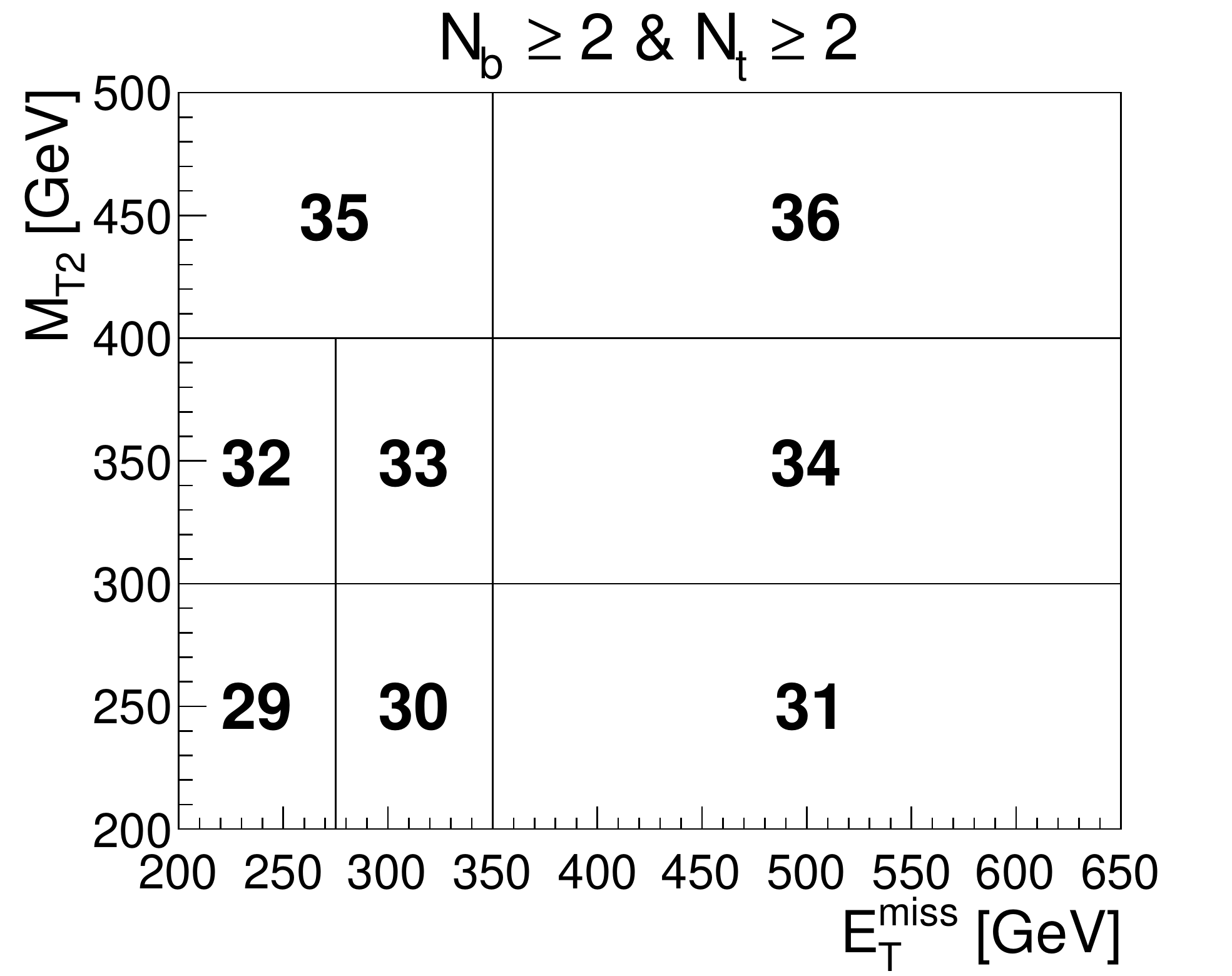}
    \caption{ Search region definitions for bin numbers 0--36 for the direct top squark production optimization.
The highest \MET and \MTTwo bins are open-ended, \eg, bin 10 requires $\MET>450\GeV$ and $\MTTwo > 400\GeV$. }
    \label{fig:SB37}
\end{figure}

After this preselection, we define nonoverlapping search regions in terms of \ntops, \nbjets, \MET, and \MTTwo.
Figure~\ref{fig:compSBvars} displays the background composition, as computed from simulation, following the preselection as a function of each of these four variables.
Note that the $\PQt$-tagged object definition does not require the presence of $\PQb$-tagged jets, nor are $\PQb$-tagged jets inside $\PQt$-tagged objects rejected from the $\PQb$-tagged jet counting.
Thus there is not a one-to-one correspondence between the numbers of $\PQt$-tagged objects and $\PQb$-tagged jets in an event.
Two different analysis optimizations are used to get the best sensitivity for direct top squark production models (T2tt and T2tb) versus gluino-mediated production models (T1tttt and T5ttcc).
For direct top squark production models, the multiplicities of $\PQb$-tagged jets and $\PQt$-tagged objects are binned as $\nbjets=1$, $\nbjets\ge 2$ and $\ntops=1$, $\ntops\ge 2$.
Due to the possibility of having more than two top quarks in the decay chain, the gluino-mediated production models are interpreted using bins with $\nbjets=1$, $\nbjets=2$, $\nbjets\ge 3$ and $\ntops=1$, $\ntops=2$, $\ntops\ge 3$.
To improve background suppression, in particular of the \ttbar contribution, and to improve the sensitivity to the various signal topologies, each (\nbjets, \ntops) bin is further subdivided by placing requirements on the \MET and \MTTwo variables, as shown in Figs.~\ref{fig:SBXX} and \ref{fig:SB37}. These figures also list the search region bin numbers used throughout the paper.
The subdivision of any given (\nbjets, \ntops) bin according to the \MET and \MTTwo variables is the same for both the direct top squark and the gluino-mediated production optimizations.


\section{Background estimation}
\label{sec:backgroundestimation}

About 70\% of the expected SM background (integrated over all search bins) comes from \ttbar, \wjets, and single top quark events with leptonic $\PW$ boson decays.
If the $\PW$ boson decays to a $\tau$ lepton that decays hadronically, this $\tau$ lepton is reconstructed as a jet and passes the lepton vetoes.
If, on the other hand, the $\PW$ boson decays to an electron or muon, events can survive the lepton vetoes when the electron or muon is ``lost,'' \ie, is not isolated, not identified/reconstructed, or out of the acceptance region.
The remaining SM background contributions, in order of decreasing importance, originate from the \znunu+jets, QCD multijet, $\ttbarZ$ and other rare processes such as triboson and $\ttbarW$ production.
The \ttbar, \wjets, single top quark, and QCD multijet backgrounds are determined using data-driven methods and are validated with closure tests in the simulation.
The \znunu+jets background is estimated using simulated events that are weighted to match the data in control regions.
Small contributions from $\ttbarZ$ and other rare processes are estimated directly from simulated events.
The background estimation methods are presented in the following subsections.

\subsection{Estimation of the lost-lepton background}
\label{sec:wtop}

The contribution to the background from events with lost leptons (LL) is determined from a data control sample (CS) that consists mainly of $\ttbar$ events.
This CS is collected using the search trigger and is defined to match the preselection, but the muon veto is replaced by the requirement that there be exactly one well-identified and isolated muon with $\pt>10\GeV$ and $\abs{\eta}<2.4$, and the isolated track veto is removed.
To reduce possible signal contamination in this CS, only events with $\mt$ less than 100\GeV are considered, with $\mt$ reconstructed from the muon \pt and $\MET$ as described for tracks in Eq.~(\ref{eq:mt}).
For \ttbar, \wjets, and single top quark events with one $\PW\to\mu\nu$ decay, $\MET$ originates from the produced neutrino. This means that the \mt distribution represents the transverse $\PW$ mass and falls off sharply above $80\GeV$; however, this is not the case for signal events.

The predicted number of events with lost leptons, $N_\text{LL}$, originating from the \ttbar, \wjets, and single top quark processes contributing to each search region bin is calculated as
\begin{linenomath}
\begin{equation}
N_\text{LL}= \sum_{\mathrm{CS}} ({F_\text{iso}}+{F_\text{ID}}+{F_\text{acc}})  F_\text{dilepton}  \frac{\epsilon_\text{isotrack}}{\epsilon^{\mu}_{\mt}},
\label{eq:lostleptonequation}
\end{equation}
\end{linenomath}
where $\sum_\mathrm{CS}$ is the sum over the events measured directly in the corresponding bin of the single muon CS defined above.
The factors ${F_\text{iso}}$, ${F_\text{ID}}$, and ${F_\text{acc}}$ convert the number of events in the CS to the number of LL events due to isolation, reconstruction and identification, and acceptance criteria (typical values are, respectively, around $0.1$, $0.1$, and $0.3$).
These scale factors are determined from isolation and reconstruction efficiencies, as well as the acceptance, which are obtained for each search region bin using simulated \ttbar events.
The contribution to the signal region from dilepton \ttbar events where both leptons are lost is corrected with the term $F_\text{dilepton}$ ($0.99$ for muons and $0.97$ for electrons).
The CS is normalized by the factor $\epsilon^{\mu}_{\mt}$ (around $0.9$) to compensate for the efficiency of the $\mt<100\GeV$ requirement.
Finally, the isolated track veto efficiency factor, $\epsilon_\text{isotrack}$, is applied to get the final number of predicted LL background events. The isolated track veto efficiency, \ie, the fraction of events surviving the isolated track veto, is around 60\%.

The main systematic uncertainty for the LL background prediction is derived from a closure test, which assesses whether the method can correctly predict the background yield in simulated event samples.
The test is performed by comparing the LL background in the search regions, as predicted by applying the LL background determination procedure to the simulated muon CS, to the expectation obtained directly from \ttbar, single top quark, and \wjets simulation.
The result of the closure test for the 45 search bins optimized for gluino-mediated production is shown in the top plot of Fig.~\ref{fig:LL-hadtau-closure}.
The closure test uncertainty (up to 26\%, depending on the search bin) is dominated by statistical fluctuations and included as a systematic uncertainty in the LL background prediction. The closure uncertainties for the 37 search bins optimized for direct top squark production are of similar size.
The following other sources of systematic uncertainty are also included: lepton isolation efficiency (effect on prediction is between $2$ and $7\%$), lepton reconstruction and identification efficiency ($3$ to $8\%$), lepton acceptance from uncertainty in the PDFs (about $10\%$), control sample purity ($2\%$), corrections due to the presence of dilepton events (around $1\%$), efficiency of the $\mt$ selection (less than $1\%$), and isolated-track veto ($3$ to $11\%$).

\subsection{Estimation of the hadronically decaying \texorpdfstring{$\tau$}{tau} lepton background}

Events from \ttbar, \wjets, and single top quark processes in which a $\tau$ lepton decays hadronically ($\tauh$) are one of the largest components of the SM background contributing to the search regions.
When a $\PW$ boson decays to a neutrino and a $\tauh$, the presence of neutrinos in the final state results in \ptvecmiss, and the event passes the lepton veto because the hadronically decaying $\tau$ lepton is reconstructed as a jet.
A veto on isolated tracks is used in the preselection to reduce the $\tauh$ background with a minimal impact on signal efficiency.

The estimate of the remaining $\tauh$ background is based on a CS of $\mu$+jets events selected from data using a trigger with requirements on both muon \pt and $\HT$, and a requirement of exactly one muon with $\pt>20\GeV$ and $\abs{\eta}<2.4$.
An upper threshold on the transverse mass reconstructed from the muon and~\MET, $\mt < 100\GeV$, is required to select events containing a $\PW\to\mu\nu$ decay and to suppress signal events contaminating the $\mu$+jets sample.
Since both $\mu$+jets and \tauh{}+jets production arise from the same underlying process, the hadronic component of the events is expected to be the same, aside from the
response of the detector to a muon or $\tauh$.
The muon \pt is smeared by response template distributions derived for a hadronically decaying $\tau$ lepton to correct the leptonic part of the event.
The response templates are derived using \ttbar, \wjets, and single top quark simulated samples by comparing the true $\tau$ lepton \pt with the reconstructed \tauh jet \pt.
The kinematic variables of the event are recalculated with this $\tauh$ jet, and the search selections are applied to predict the \tauh background.

The probability to mistag a $\tauh$ jet as a $\PQb$ jet is significant (about 0.1) and affects the \nbjets distribution of $\tauh$ background events.
The dependence of the mistag rate on the $\tauh$ jet \pt is larger for \ttbar events than for \wjets events,
because the $\PQb$ quark from the top quark decay can overlap with the $\tauh$ jet.
This mistag rate is taken into account in the $\mu$+jets CS by randomly selecting a simulated $\tauh$ jet and counting it as a $\PQb$ jet with the probability obtained from MC simulation in \wjets events for the corresponding $\tauh$ jet \pt.

The $\tauh$ background prediction is calculated as follows:
\begin{linenomath}
\ifthenelse{\boolean{cms@external}}{
\begin{multline}
N_{\tauh} = \sum_\mathrm{CS} \Bigl(
\sum_\text{template bins} P_{\tauh}^\text{resp}
\frac{1}{\epsilon^{\mu}_\text{trigger}\,\epsilon^{\mu}_\text{reco}\,\epsilon^{\mu}_\text{iso}\,\epsilon^{\mu}_\text{acc}\,\epsilon^{\mu}_{\mt}}
\\
\times\frac{ \mathcal{B}(\PW \to \tauh)}{\mathcal{B}(\PW \to \mu)}
\,\epsilon_\text{isotrack}
\,F_{\tau \to \mu}
\,F_\text{dilepton}\Bigr),
\label{eqn:tauh}
\end{multline}
}{
\begin{equation}
N_{\tauh} = \sum_\mathrm{CS} \left(\sum_\text{template bins} P_{\tauh}^\text{resp}
\frac{1}{\epsilon^{\mu}_\text{trigger}\,\epsilon^{\mu}_\text{reco}\,\epsilon^{\mu}_\text{iso}\,\epsilon^{\mu}_\text{acc}\,\epsilon^{\mu}_{\mt}}
\frac{ \mathcal{B}(\PW \to \tauh)}{\mathcal{B}(\PW \to \mu)}
\,\epsilon_\text{isotrack}
\,F_{\tau \to \mu}
\,F_\text{dilepton}\right),
\label{eqn:tauh}
\end{equation}
}
\end{linenomath}
where the first summation is over the events in the $\mu$+jets CS, the second is over the bins of the $\tauh$ response template, and $P_{\tauh}^\text{resp}$ is the probability of the $\tauh$ response from each bin.
The various correction factors applied to convert $\mu$+jets events into $\tauh$+jets events to construct the final $\tauh$ sample are:
\begin{itemize}
\item the branching fraction ratio $\mathcal{B}(\W \to \tauh)/\mathcal{B}(\W \to \mu)$ = 0.65;
\item the muon reconstruction and identification efficiency $\epsilon^{\mu}_\text{reco}$ (0.94--0.98) and the muon isolation efficiency $\epsilon^{\mu}_\text{iso}$ (0.5--0.95 depending on the muon $\pt$ and the $\sum\pt$ of PF candidates within an annulus with outer radius of $\Delta R=0.4$ and inner radius equal to the isolation cone);
\item the muon acceptance $\epsilon^{\mu}_\text{acc}$ (typically around 0.8--0.9);
\item  the $\mt$ selection efficiency $\epsilon_{\mt}$ ($>0.9$);
\item the correction to account for the contamination in the CS from muons from $\tau$ decays, $F_{\tau \to \mu}$ (around 0.8 depending on \njets and \MET);
\item the isolated track veto efficiency for $\tauh$, $\epsilon_\text{isotrack}$ (around 0.7), as determined from simulated \ttbar, \wjets and single top quark events by matching isolated tracks to $\tauh$ jets;
\item the $\tauh$ contribution that overlaps with the LL background prediction due to contamination of dileptonic events in the CS, $F_\text{dilepton}$, to avoid double counting (0.98);
\item and a correction for the $\mu$ trigger efficiency, $\epsilon^{\mu}_\text{trigger}$ (0.95).
\end{itemize}
The muon reconstruction, identification, and isolation efficiency are the same as those used for the LL background determination.

\begin{figure}[htp]
  \centering
  \includegraphics[width=\cmsFigWidth]{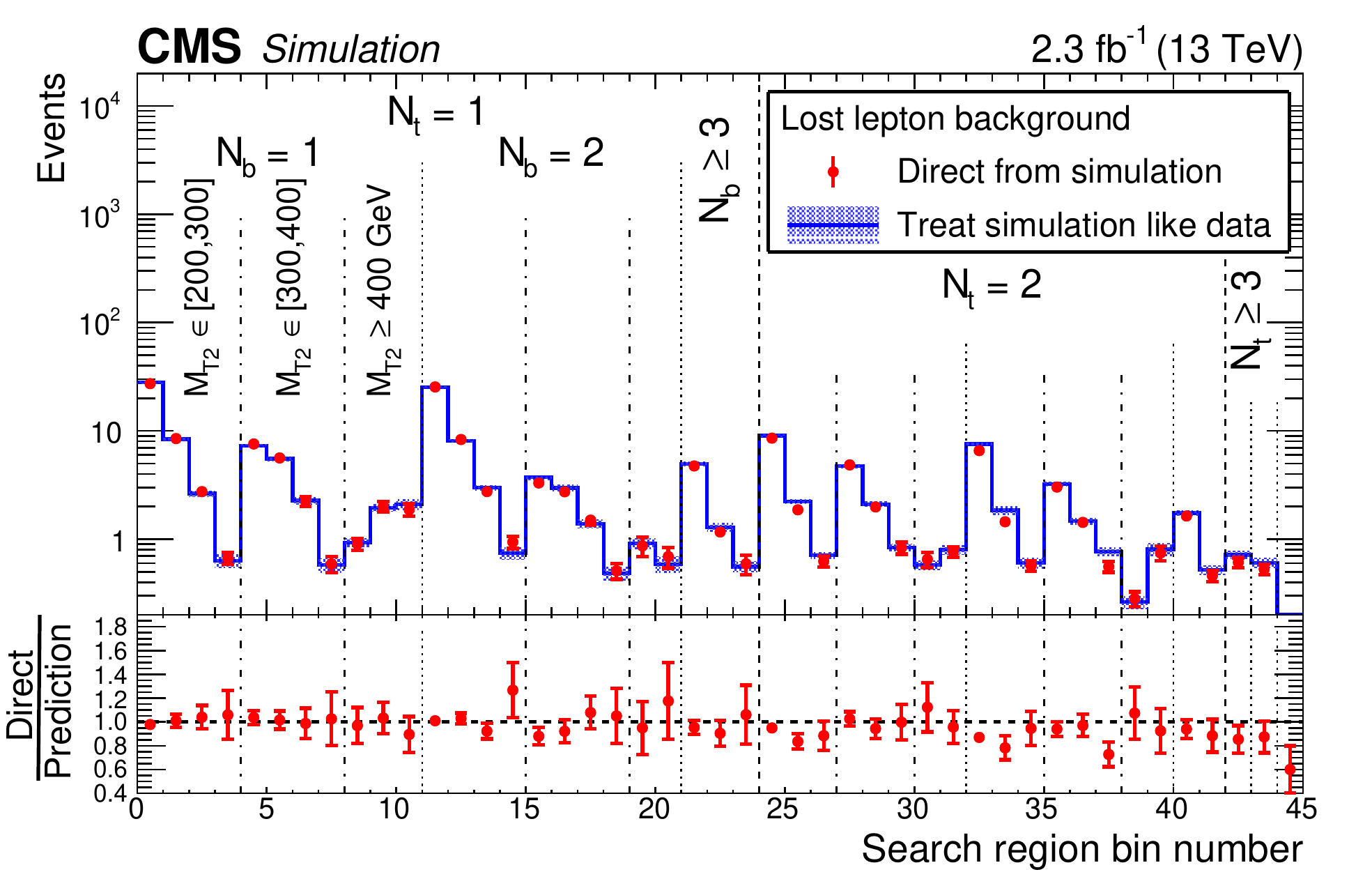}\\
  \includegraphics[width=\cmsFigWidth]{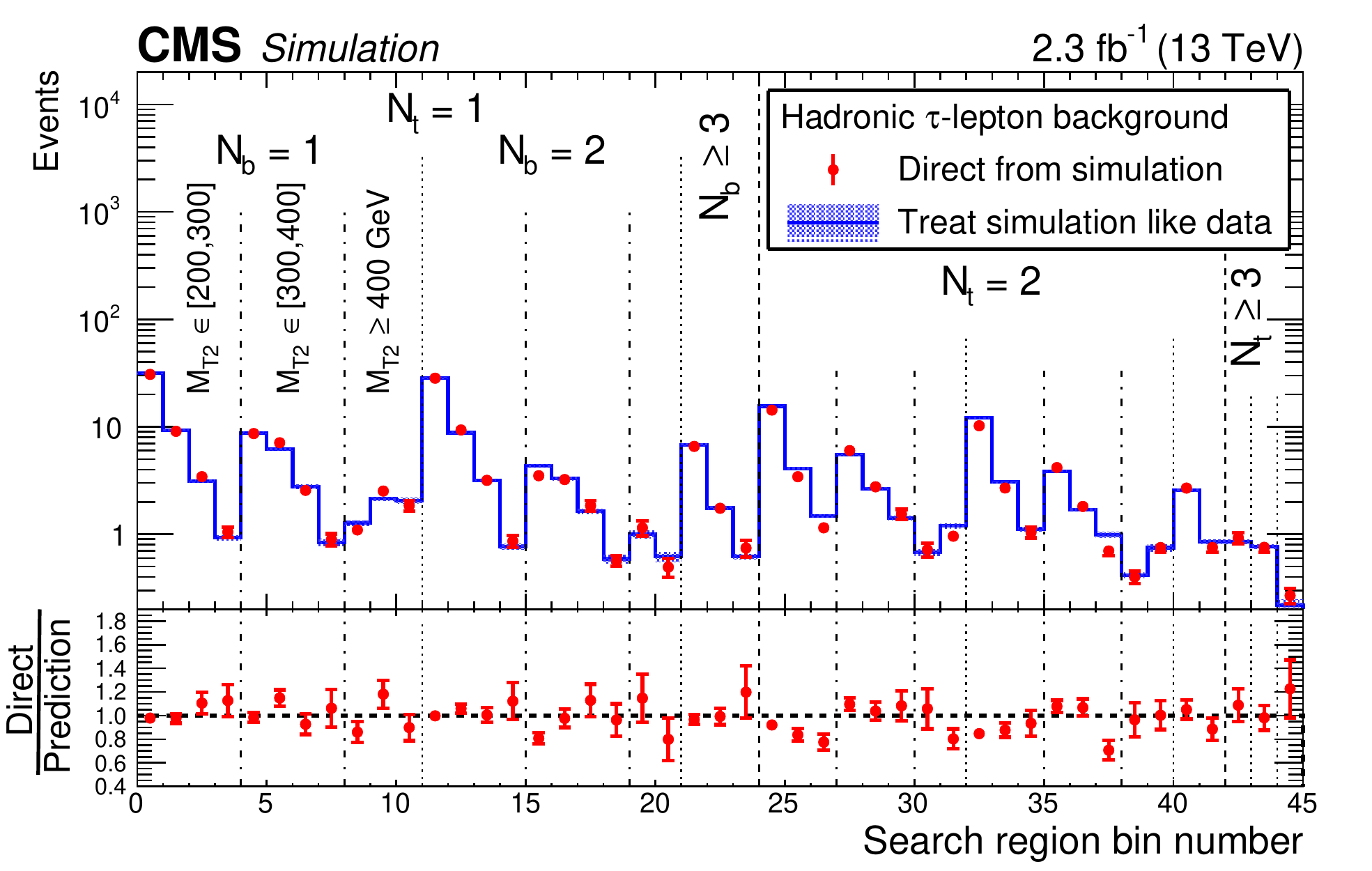}
  \caption{(Top) The lost-lepton background in the 45 search regions
    optimized for gluino-mediated production as determined directly from \ttbar,
    single top quark, and \wjets simulation
    (points) and as predicted by applying the
    lost-lepton background determination procedure to the simulated
    muon control sample (histograms).
    The lower panel shows the same results after dividing
    by the predicted value.
    (Bottom) The corresponding simulated results for the background from
    hadronically decaying $\tau$ leptons.
    For both plots, vertical lines indicate search regions with different
    \ntops, \nbjets, and \MTTwo values.
    Within each (\ntops, \nbjets, \MTTwo) region, the bins indicate the
    different \MET selections, as defined in Fig.~\ref{fig:SBXX}.
    Only statistical uncertainties are shown.
  }
  \label{fig:LL-hadtau-closure}
\end{figure}

A closure test is performed comparing the \tauh background in the search regions as predicted by applying the \tauh background determination procedure to the simulated muon CS to the expectation obtained directly from simulation.
The result of the closure test for the 45 search bins optimized for gluino-mediated production is shown in the lower plot of Fig.~\ref{fig:LL-hadtau-closure}.
The closure uncertainty for each search bin (between 2\% and 28\%) is dominated by statistical fluctuations and is included as a systematic uncertainty in the \tauh background prediction.
The closure uncertainties for the 37 search bins optimized for direct top squark production are of similar size.
In addition, systematic uncertainties are evaluated for each of the ingredients in the prediction, which arise from uncertainties in the following sources:
the $\tauh$ response template (2\%),
the muon reconstruction and isolation efficiency (1\%),
the acceptance due to uncertainties in the PDFs (up to 5\%),
the $\PQb$ mistag rate of the $\tauh$ jet (up to 15\%),
$\epsilon_{\mt}$ due to uncertainties in the \MET scale ($<1\%$),
the efficiency of the isolated track veto ($4$--$6.5\%$),
contamination from lost leptons (2.4\%),
and the trigger efficiency (1\%).

\subsection{\texorpdfstring{Estimation of the \znunu background}{Estimation of the Znunu background}}
\label{sec:zinv}

The \znunu background is derived using simulated events that have been corrected for observed differences between data and simulation.
A $\cPZ\to\mu\mu$ control sample is used to validate the \znunu MC and residual differences in both shape of the jet multiplicity ($N_\mathrm{j}$) distribution and overall normalization present therein are corrected for.
The central value of the \znunu background prediction for each search bin $B$ can be written as
\begin{linenomath}
\begin{equation}
\widehat{N}_B = R_\text{norm} \sum_{\textrm{events}\in B} S_\mathrm{DY}(N_\mathrm{j}) w_\textrm{MC},
\label{eq:zinv_pred}
\end{equation}
\end{linenomath}
where $\widehat{N}_B$ is the predicted number of \znunu background events in search bin $B$.
The sum runs over all simulated \znunu events that fall in search bin $B$,  and $w_\textrm{MC}$ is a standard event weight including the assumed \znunu cross section, the integrated luminosity, the $\PQb$ tagging efficiency scale factors, and the measured trigger efficiency.
Each simulated event is additionally weighted using two scale factors, $R_\text{norm}$ and $S_\mathrm{DY}(N_\mathrm{j})$, that correct the normalization of the simulation and the shape of the simulated $N_\mathrm{j}$ distribution, respectively.
Both scale factors are calculated in a dimuon CS that has events with two muons, with $81 < m_{\mu\mu} < 101\GeV$, and no muon or isolated track vetoes.
In this region the two muons are treated as if they were neutrinos.

The first scale factor, $R_\text{norm}$,
is derived using a tight dimuon CS in data.
This control region has the same selection as the search region preselection, apart from the muon requirement and without any requirements on $\PQb$-tagged jets.
This region is selected for its kinematic similarity to the signal region, but lacks the statistical precision required for shape comparison.
The scale factor is computed by comparing the expected event yield in the tight region in the DY simulation with the observed event yield in data after subtraction of the other SM processes.

\begin{figure}[tbp]
 \centering
 \includegraphics[width=0.48\textwidth]{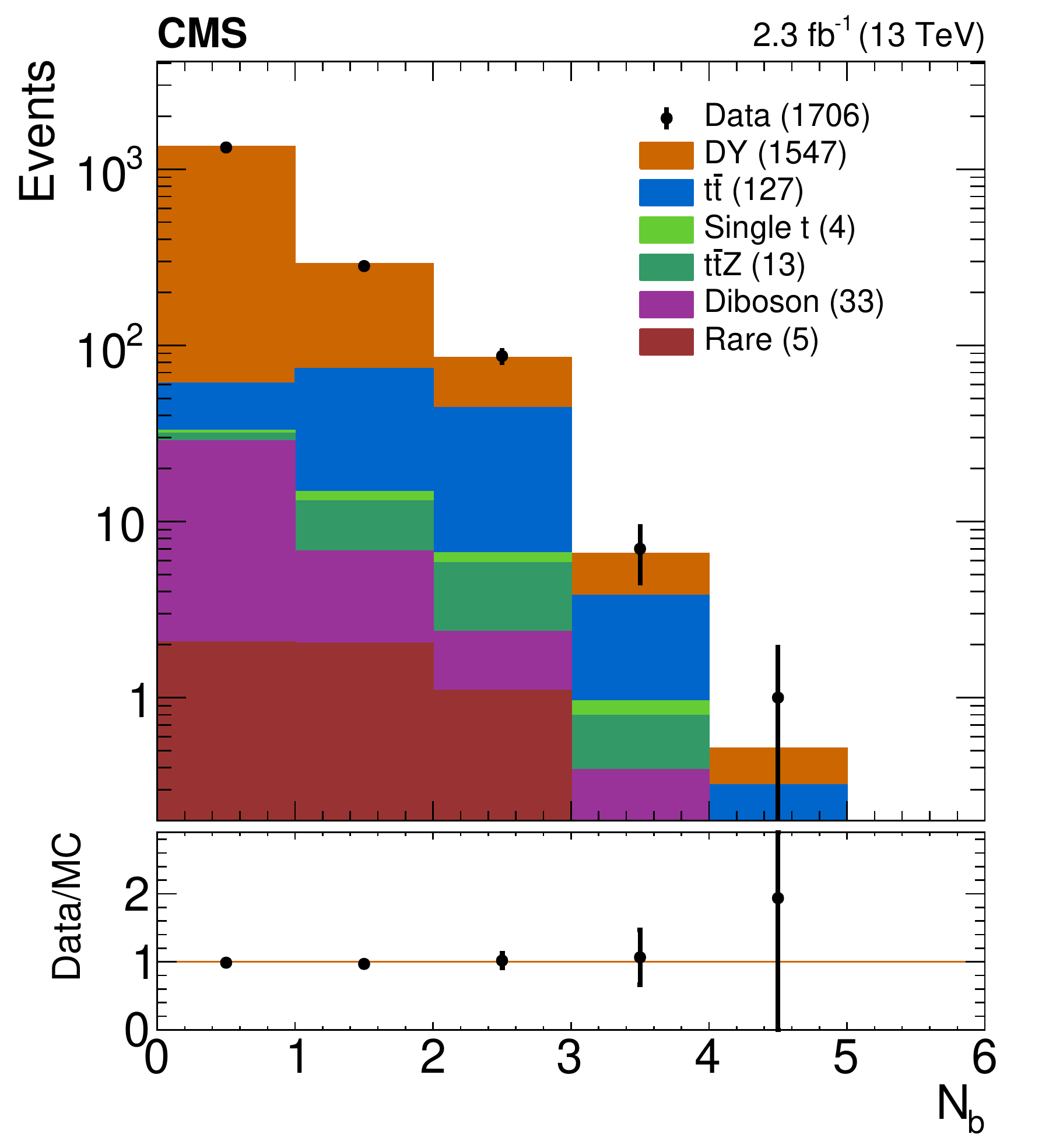}
 \includegraphics[width=0.48\textwidth]{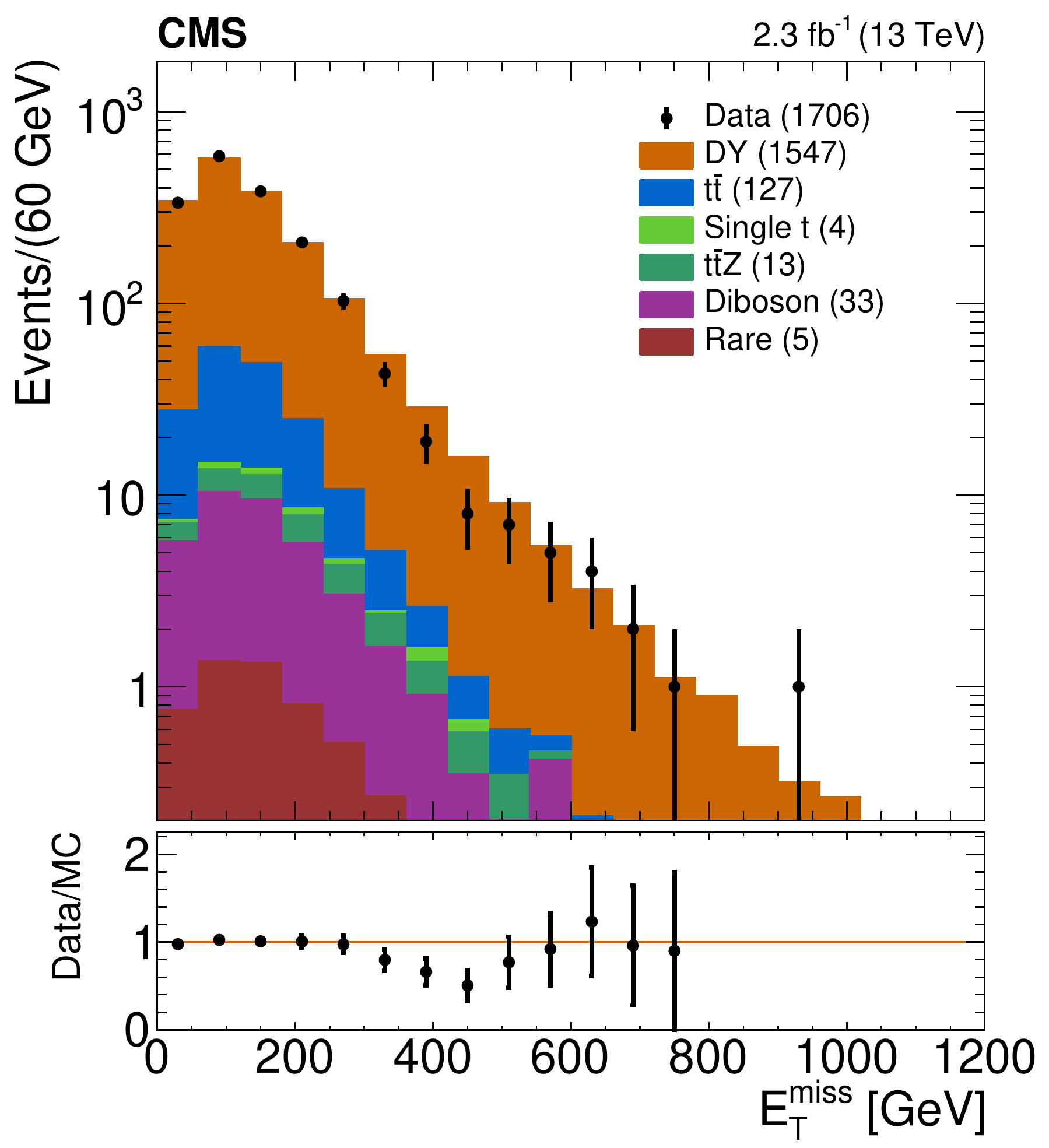}
\caption{The \nbjets (\cmsLeft) and \MET (\cmsRight) distributions in data and simulation in the loose dimuon control region, after applying the $S_\mathrm{DY}(N_\mathrm{j})$ scale factor to the simulation.
The lower panels show the ratio between data and simulation.
Only statistical uncertainties are shown. The values in parentheses in the legend indicate the integrated yield for each given process.
The ``rare'' category includes background processes such as triboson and $\ttbar\W$ production.}
\label{fig:MethodAlpha_zinv_loose}
\end{figure}

The second scale factor, $S_\mathrm{DY}$, depends on the number of jets $N_\mathrm{j}$ in the event and is designed to correct the mismodeling of the jet multiplicity distribution in simulation.
The scale factor is derived in a loose dimuon control region in which the signal region requirements on \MET, \ntops, and \MTTwo are removed,
and the \HT requirement is relaxed to $\HT > 200\GeV$.
The $S_\mathrm{DY}$ scale factor is derived for each $(N_\mathrm{j})$ bin as the ratio between the data, with non-DY backgrounds subtracted, and the DY simulation.
Due to \ttbar contributions similar to the DY processes for greater jet and $\PQb$-tagged jet multiplicities, the \ttbar MC events are similarly reweighted using a CS selected to have an electron and a muon with $81 < m_{\Pe\mu} < 101\GeV$ before subtraction from the dimuon data.
The \nbjets and \MET distributions in the loose dimuon CS after applying the $S_\mathrm{DY}(N_\mathrm{j})$ scale factor are shown in Fig.~\ref{fig:MethodAlpha_zinv_loose}. The \nbjets distribution agrees well between data and simulation, whereas the \MET distribution has some disagreement between 300 and 600$\GeV$.  The disagreement is taken into account with a shape uncertainty equal to the magnitude of the disagreement and has a negligible effect on the final results.

The systematic uncertainties for the \znunu background prediction are divided into two broad categories: uncertainties associated with the use of MC simulation and uncertainties specifically associated with the background prediction method.
The first category includes systematic uncertainties in the PDFs and renormalization/factorization scale choices, jet and \MET energy scale uncertainties, $\PQb$ tagging efficiency scale factor uncertainties, and trigger efficiency uncertainties.
The second category includes uncertainties from the method used to determine $R_\text{norm}$ and the $S_\mathrm{DY}(N_\mathrm{j})$ scale factors, and uncertainties based on the residual shape disagreement between data and DY+jets simulation in the loose dimuon CS.
The uncertainty in $R_\text{norm}$, derived from the statistical uncertainties on data and MC in the tight CS, results in a 19\% uncertainty in the predicted \znunu event yield for each search bin.
The uncertainties associated with $S_\mathrm{DY}$ are the dominant uncertainties and are related to residual shape uncertainties (after applying the $S_\mathrm{DY}$ scale factor) in the search region variables \MET, \MTTwo, $\nbjets$, and $\ntops$.
These uncertainties are evaluated in the loose CS with the additional requirement that $\ntops \ge 1$ so that \MTTwo is well defined.
The resulting shift of the central value of the search bin predictions is used as the systematic uncertainty from the residual shape disagreements. Depending on the search bin, this uncertainty ranges between 10 and 82\%.
The statistical uncertainties in the ratios between data and simulation, as well as in $S_\mathrm{DY}$,  are also included as a 15--75\% systematic uncertainty in the prediction.

\subsection{Estimation of the QCD multijet background}
\label{sec:qcd}

The procedure to predict the QCD multijet background consists of selecting a signal-depleted data CS, rich in QCD multijet events, from which significant contributions of other SM backgrounds, such as \ttbar, \wjets, and \zjets, are subtracted.
Following that, a translation factor, partly determined from data and partly from simulation, is used to convert the number of events measured in the data CS into a prediction for each search region bin.

The CS is defined by applying the full set of preselection requirements described in Sec.~\ref{sec:alpha_event_selection}, except that the $\Delta\phi(\MET, j_{1,2,3})$ requirements are inverted, requiring that the \MET be aligned with one of the leading three jets.
The estimated number of QCD multijet events in the inverted-$\Delta\phi$ CS is computed by subtracting the contributions from LL, hadronically decaying $\tau$ leptons, and \zjets processes from the number of data events observed in that region.
The same methods as described in the previous sections are used to estimate the contributions from LL and $\tauh$ processes, but applied to this QCD multijet-rich CS.
Simulation is used to estimate the contribution from \znunu events, since it is expected to be small.

The translation factor between the QCD multijet-rich CS and the search region bins is computed in data, using a sideband of the preselection region, defined by the requirement $175 < \MET < 200 \GeV$ and without an \nbjets requirement,  where the amount of data is sufficiently large to make an accurate measurement.
The contributions from processes other than QCD multijet are subtracted from the observed number of events in this low-\MET data sideband, following the procedure outlined above.
The dependence of the translation factor as a function of \MET is accounted for by using a linear approximation derived from simulation.
To take into account the dependence as a function of \MTTwo, the translation factor is computed separately for \MTTwo values below and above $300 \GeV$.
The translation factor ranges from 0.01 to 0.14 depending on \MET and \MTTwo.

\begin{figure}[!htb]
\centering
\includegraphics[width=\cmsFigWidth]{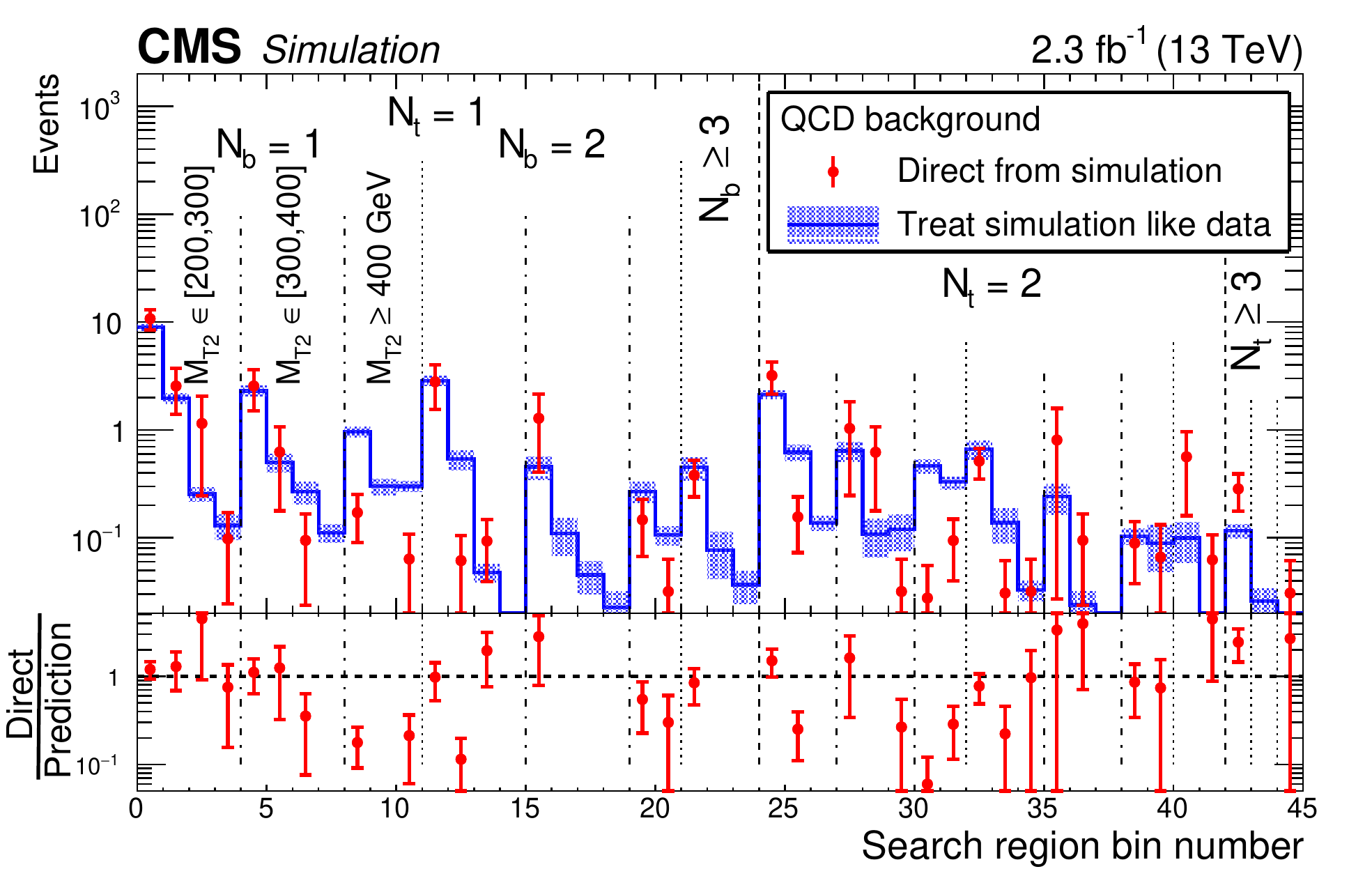}
\caption{
The QCD multijet background in the 45 search regions optimized for gluino-mediated production
as determined directly from simulation (points)
and as predicted by applying the QCD multijet background determination
procedure to simulated event samples in the inverted-$\Delta\phi$ control region (histograms).
The lower panel shows the same results after dividing by the predicted value.
Only statistical uncertainties are shown.
The labeling of the search regions is the same as in Fig.~\ref{fig:LL-hadtau-closure}.
}
\label{fig:qcd-closure}
\end{figure}

The main systematic uncertainty in the QCD multijet prediction is obtained from a closure test in which the expectation for the signal region event yields, as obtained directly from the QCD multijet simulation, is compared to the prediction obtained by applying the QCD multijet background prediction procedure to simulated event samples.
The result for the 45 search bins optimized for gluino-mediated production is shown in Fig.~\ref{fig:qcd-closure}, and any observed nonclosure from the relaxed  \MET and \nbjets requirements is taken into account as the systematic uncertainty.
If there is insufficient simulation to populate a bin in the closure prediction, the uncertainty from the next lowest \MET bin is used.  This uncertainty ranges from 5\% to 500\% depending on the search bin.
The closure uncertainties for the 37 search bins optimized for direct top squark production are of similar size.
The high closure uncertainties for some search bins are due to statistical limitations of the simulation, but have a small effect on the final results because the QCD multijet yields are very low in these search bins compared to other backgrounds.
In addition, another major source of systematic uncertainty in the QCD multijet prediction is the uncertainty in the $T_\text{QCD}$ factors.

\subsection{\texorpdfstring{Backgrounds from $\ttbarZ$ and other SM rare processes}
{Backgrounds from ttZ and other SM rare processes}}
\label{sec:ttZ}

Similar to the $\znunu$ background, $\ttbarZ$ is an irreducible background when $\PZ$ bosons decay to neutrinos and both top quarks decay hadronically.
The $\ttbarZ$ cross section at 13\TeV is only 783\unit{fb} (computed at NLO using \MADGRAPH5\_a\MCATNLO) and the predicted yield of $\ttbarZ$ events in the search bins is less than 10\% of the total background.
Given the presence of genuine \MET and $\PQb$ jets in $\ttbarZ$ events, and given the small cross section associated with this process, we rely on simulation to predict its contribution to each search region bin.
The $\ttbarZ$ simulation is validated using a trilepton control sample in data, and the 30\% statistical uncertainty in this data measurement is propagated to the $\ttbarZ$ prediction.

The contribution of the $\ttbarW$ process to the signal region is covered by the LL and $\tauh$ background estimation methods.
The signal region yields for the diboson and multiboson processes are fully determined by simulation and are combined into a single rare background prediction.

\section{Results and interpretation}
\label{sec:interpretation}

\begin{figure*}[htbp]
  \centering
  \includegraphics[width=\textwidth]{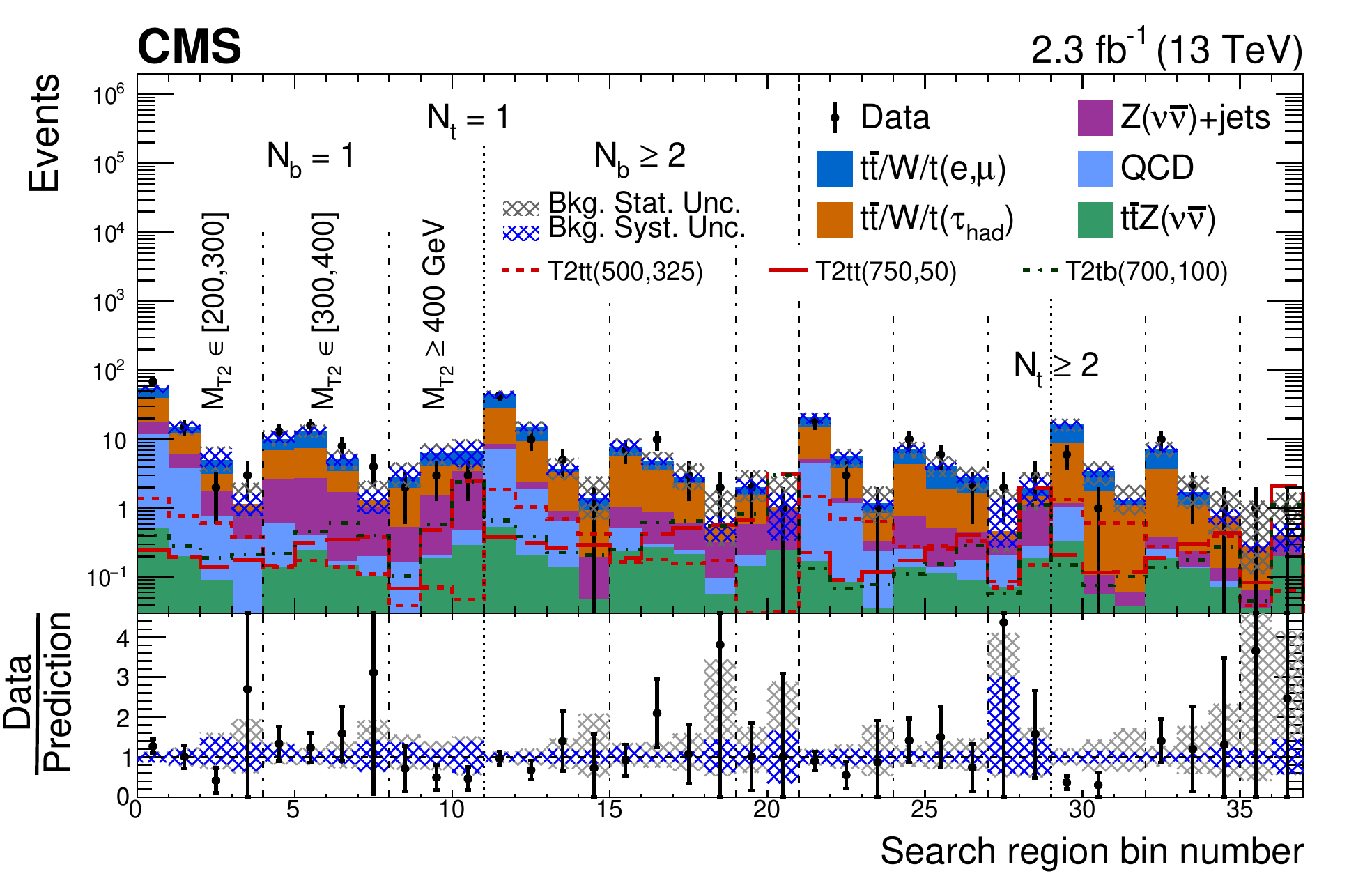}
  \caption{Observed event yields in data (black points) and predicted SM background (filled solid area) for the 37 search bins optimized for direct top squark production.
  The red and dark green lines indicate various signal models: the T2tt model with $m_{\sTop} = 500\GeV$ and $m_{\lsp} = 325\GeV$ (red short-dashed line),
  the T2tt model with $m_{\sTop} = 750\GeV$ and $m_{\lsp} = 50\GeV$ (red long-dashed line),
  and the T2tb model with $m_{\sTop} = 700\GeV$ and $m_{\lsp} = 100\GeV$ (dark green dashed-dotted line).
  The lower panel shows the ratio of data over total background prediction in each search bin.
  For both panels, the error bars show the statistical uncertainty associated with the observed data counts,
  and the grey (blue) hatched bands indicate the statistical (systematic) uncertainties in the total predicted background.}
  \label{fig:baseline_SR}
\end{figure*}

\begin{figure*}[htbp]
  \centering
  \includegraphics[width=\textwidth]{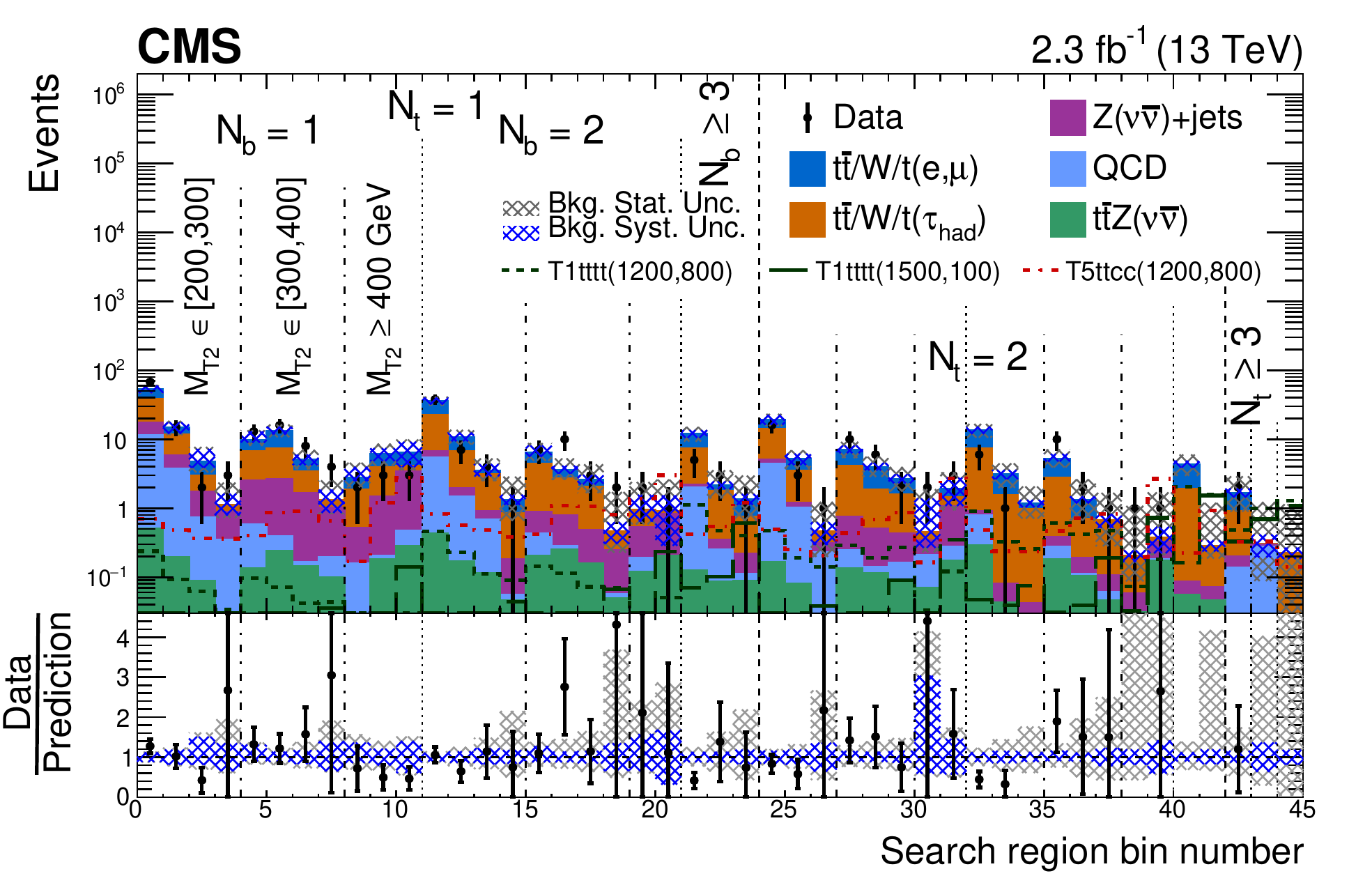}
  \caption{Observed event yields in data (black points) and predicted SM background (filled solid area) for the 45 search bins optimized for gluino models.
  The red and dark green lines indicate various signal models: the T1tttt model with $m_{\gluino} = 1200\GeV$ and $m_{\lsp} = 800\GeV$ (dark green short-dashed line),
  the T1tttt model with $m_{\gluino} = 1500\GeV$ and $m_{\lsp} = 100\GeV$ (dark green long-dashed line),
  and the T5ttcc model with $m_{\gluino} = 1200\GeV$ and $m_{\lsp} = 800\GeV$ (red dashed-dotted line).
  The lower panel shows the ratio of data over total background prediction in each search bin.
  For both panels, the error bars show the statistical uncertainty associated with the observed data counts,
  and the grey (blue) hatched bands indicate the statistical (systematic) uncertainties in the total predicted background.}
    \label{fig:baseline_SR_45}
\end{figure*}

The predicted number of SM background events and the number of events observed in data for each of the search regions defined in Sec.~\ref{sec:alpha_event_selection} are summarized in Fig.~\ref{fig:baseline_SR} and Tables~\ref{tab:obs_vs_pred_common} and \ref{tab:obs_vs_pred_37Bins} for the binning optimized for direct top squark production,
and in Fig.~\ref{fig:baseline_SR_45} and Tables~\ref{tab:obs_vs_pred_common} and \ref{tab:obs_vs_pred_45Bins} for the binning optimized for gluino-mediated production models.
Typically, the most significant background across the search regions comes from SM \ttbar or $\PW$ boson production, where the $\PW$ boson decay contains genuine \MET from a neutrino.
Generally, the next largest contribution comes from \znunu production in association with jets (including heavy-flavor jets) in which the neutrino pair gives rise to large \MET and the top quark conditions are satisfied by an accidental combination of the jets.
For search regions with very high \MET requirements, the \znunu background can become dominant.
The QCD multijet contribution and the contribution from other rare SM processes are subdominant across all bins. The largest rare SM process contribution (though still small) comes from $\ttbarZ$ with the $\cPZ$ boson decaying into a pair of neutrinos.
No statistically significant deviation between the observed data events and the SM background prediction is found.

\begin{table*}[tbp]
\centering
\topcaption{Observed yields from the data compared to the total background predictions for the search bins that are common between the direct top squark and gluino-mediated production  optimizations.
The quoted uncertainties on the predicted background yields are statistical and systematic, respectively.}
\label{tab:obs_vs_pred_common}
\renewcommand{\arraystretch}{1.25}
\cmsTable{
\begin{scotch}{ccccc  cc}
Bin number & $\ntops$ & $\nbjets$ & $\MTTwo$ [\GeVns{}] & \MET [\GeVns{}] & Data & Predicted background \\
\hline
0 & 1 & 1 & 200\,--\,300 & 200\,--\,275 &         68 &     54 $^{+   4}_{-   4}$ $^{+   6}_{-   6}$ \\
1 & 1 & 1 & 200\,--\,300 & 275\,--\,350 &         15 &     15 $^{+   2}_{-   2}$ $^{+   3}_{-   3}$ \\
2 & 1 & 1 & 200\,--\,300 & 350\,--\,450 &          2 &    4.9 $^{+ 1.6}_{- 1.2}$ $^{+ 2.4}_{- 0.9}$ \\
3 & 1 & 1 & 200\,--\,300 & $>$450     &          3 &    1.2 $^{+ 1.1}_{- 0.2}$ $^{+ 0.4}_{- 0.4}$ \\
4 & 1 & 1 & 300\,--\,400 & 200\,--\,275 &         13 &    9.8 $^{+ 1.8}_{- 1.5}$ $^{+ 3.1}_{- 1.0}$ \\
5 & 1 & 1 & 300\,--\,400 & 275\,--\,350 &         16 &     13 $^{+   2}_{-   2}$ $^{+   2}_{-   1}$ \\
6 & 1 & 1 & 300\,--\,400 & 350\,--\,450 &          8 &    5.0 $^{+ 1.7}_{- 1.1}$ $^{+ 0.9}_{- 0.9}$ \\
7 & 1 & 1 & 300\,--\,400 & $>$450     &          4 &    1.3 $^{+ 1.1}_{- 0.1}$ $^{+ 0.5}_{- 0.5}$ \\
8 & 1 & 1 & $>$400 & 200\,--\,350 &          2 &    2.9 $^{+ 1.3}_{- 0.8}$ $^{+ 1.1}_{- 0.4}$ \\
9 & 1 & 1 & $>$400 & 350\,--\,450 &          3 &      6 $^{+   2}_{-   2}$ $^{+   1}_{-   1}$ \\
10& 1 & 1 & $>$400 & $>$450     &          3 &      7 $^{+   2}_{-   1}$ $^{+   3}_{-   3}$ \\
\end{scotch}
}
\end{table*}

\begin{table*}[tpb]
\centering
\topcaption{Observed yields from the data compared to the total background predictions for the search bins that are specific to the direct top squark production optimization.
The quoted uncertainties on the predicted background yields are statistical and systematic, respectively.}
\label{tab:obs_vs_pred_37Bins}
\renewcommand{\arraystretch}{1.25}
\cmsTable{
\begin{scotch}{ccccc  cc}
Bin number & $\ntops$ & $\nbjets$ & $\MTTwo$ [\GeVns{}] & \MET [\GeVns{}] & Data & Predicted background \\
\hline
11 & 1 & $\ge$\,2 & 200\,--\,300 & 200\,--\,275 &         43 &     44 $^{+   4}_{-   4}$ $^{+   5}_{-   5}$ \\
12 & 1 & $\ge$\,2 & 200\,--\,300 & 275\,--\,350 &         10 &     15 $^{+   3}_{-   2}$ $^{+   2}_{-   2}$ \\
13 & 1 & $\ge$\,2 & 200\,--\,300 & 350\,--\,450 &          5 &    3.6 $^{+ 1.5}_{- 0.9}$ $^{+ 0.7}_{- 0.6}$ \\
14 & 1 & $\ge$\,2 & 200\,--\,300 & $>$450     &          1 &    1.4 $^{+ 1.5}_{- 0.7}$ $^{+ 0.2}_{- 0.2}$ \\
15 & 1 & $\ge$\,2 & 300\,--\,400 & 200\,--\,275 &          7 &    7.6 $^{+ 1.7}_{- 1.4}$ $^{+ 2.0}_{- 0.9}$ \\
16 & 1 & $\ge$\,2 & 300\,--\,400 & 275\,--\,350 &         10 &    4.8 $^{+ 1.7}_{- 1.1}$ $^{+ 0.6}_{- 0.5}$ \\
17 & 1 & $\ge$\,2 & 300\,--\,400 & 350\,--\,450 &          3 &    2.8 $^{+ 1.6}_{- 0.9}$ $^{+ 0.4}_{- 0.4}$ \\
18 & 1 & $\ge$\,2 & 300\,--\,400 & $>$450     &          2 &    0.5 $^{+ 1.3}_{- 0.1}$ $^{+ 0.2}_{- 0.2}$ \\
19 & 1 & $\ge$\,2 & $>$400 & 200\,--\,450 &          2 &    2.0 $^{+ 1.4}_{- 0.7}$ $^{+ 0.6}_{- 0.4}$ \\
20 & 1 & $\ge$\,2 & $>$400 & $>$45      &          1 &   0.99 $^{+1.77}_{-0.06}$ $^{+0.65}_{-0.65}$ \\[1ex]
21 & $\ge$\,2 & 1 & 200\,--\,300 & 200\,--\,275 &         18 &     20 $^{+   2}_{-   2}$ $^{+   3}_{-   3}$ \\
22 & $\ge$\,2 & 1 & 200\,--\,300 & 275\,--\,350 &          3 &      5 $^{+   1}_{-   1}$ $^{+   1}_{-   1}$ \\
23 & $\ge$\,2 & 1 & 200\,--\,300 & $>$350     &          1 &    1.1 $^{+ 0.9}_{- 0.5}$ $^{+ 0.2}_{- 0.2}$ \\
24 & $\ge$\,2 & 1 & 300\,--\,400 & 200\,--\,275 &         10 &    7.1 $^{+ 1.8}_{- 1.5}$ $^{+ 1.1}_{- 0.7}$ \\
25 & $\ge$\,2 & 1 & 300\,--\,400 & 275\,--\,350 &          6 &    4.0 $^{+ 1.5}_{- 1.1}$ $^{+ 0.5}_{- 0.5}$ \\
26 & $\ge$\,2 & 1 & 300\,--\,400 & $>$350     &          2 &    2.7 $^{+ 1.2}_{- 0.8}$ $^{+ 0.4}_{- 0.4}$ \\
27 & $\ge$\,2 & 1 & $>$400 & 200\,--\,250 &          2 &    0.5 $^{+ 1.1}_{- 0.1}$ $^{+ 0.9}_{- 0.2}$ \\
28 & $\ge$\,2 & 1 & $>$400 & $>$350     &          3 &    1.9 $^{+ 1.1}_{- 0.5}$ $^{+ 0.9}_{- 0.8}$ \\
29 & $\ge$\,2 & $\ge$\,2 & 200\,--\,300 & 200\,--\,275 &          6 &     16 $^{+   3}_{-   3}$ $^{+   2}_{-   2}$ \\
30 & $\ge$\,2 & $\ge$\,2 & 200\,--\,300 & 275\,--\,350 &          1 &    3.3 $^{+ 1.3}_{- 1.1}$ $^{+ 0.5}_{- 0.5}$ \\
31 & $\ge$\,2 & $\ge$\,2 & 200\,--\,300 & $>$350     &          0 &    1.3 $^{+ 0.9}_{- 0.4}$ $^{+ 0.1}_{- 0.1}$ \\
32 & $\ge$\,2 & $\ge$\,2 & 300\,--\,400 & 200\,--\,275 &         10 &    7.1 $^{+ 1.8}_{- 1.5}$ $^{+ 0.8}_{- 0.7}$ \\
33 & $\ge$\,2 & $\ge$\,2 & 300\,--\,400 & 275\,--\,350 &          2 &    1.7 $^{+ 1.3}_{- 0.7}$ $^{+ 0.2}_{- 0.2}$ \\
34 & $\ge$\,2 & $\ge$\,2 & 300\,--\,400 & $>$350     &          1 &    0.8 $^{+ 1.0}_{- 0.3}$ $^{+ 0.2}_{- 0.2}$ \\
35 & $\ge$\,2 & $\ge$\,2 & $>$400 & 200\,--\,350 &          1 &   0.27 $^{+1.00}_{-0.16}$ $^{+0.05}_{-0.05}$ \\
36 & $\ge$\,2 & $\ge$\,2 & $>$400 & $>$350     &          1 &   0.41 $^{+1.27}_{-0.06}$ $^{+0.19}_{-0.17}$ \\
\end{scotch}
}
\end{table*}

\begin{table*}[tpb]
\centering
\topcaption{Observed yields from the data compared to the total background predictions for the search bins that are specific to the gluino-mediated production optimization.
The quoted uncertainties on the predicted background yields are statistical and systematic, respectively.}
\label{tab:obs_vs_pred_45Bins}
\renewcommand{\arraystretch}{1.25}
\cmsTable{
\begin{scotch}{ ccccc cc}
Bin number & $\ntops$ & $\nbjets$ & $\MTTwo$ [\GeVns{}] & \MET [\GeVns{}] & Data & Predicted background \\
\hline
11 & 1 & 2 & 200\,--\,300 & 200\,--\,275 &         38 &     36 $^{+   4}_{-   3}$ $^{+   4}_{-   4}$ \\
12 & 1 & 2 & 200\,--\,300 & 275\,--\,350 &          7 &     11 $^{+   2}_{-   2}$ $^{+   2}_{-   2}$ \\
13 & 1 & 2 & 200\,--\,300 & 350\,--\,450 &          4 &    3.5 $^{+ 1.5}_{- 0.8}$ $^{+ 0.8}_{- 0.6}$ \\
14 & 1 & 2 & 200\,--\,300 & $>$450     &          1 &    1.3 $^{+ 1.5}_{- 0.6}$ $^{+ 0.2}_{- 0.2}$ \\
15 & 1 & 2 & 300\,--\,400 & 200\,--\,275 &          7 &    6.4 $^{+ 1.6}_{- 1.3}$ $^{+ 1.7}_{- 0.8}$ \\
16 & 1 & 2 & 300\,--\,400 & 275\,--\,350 &         10 &    3.6 $^{+ 1.6}_{- 0.9}$ $^{+ 0.5}_{- 0.5}$ \\
17 & 1 & 2 & 300\,--\,400 & 350\,--\,450 &          3 &    2.6 $^{+ 1.7}_{- 0.9}$ $^{+ 0.4}_{- 0.4}$ \\
18 & 1 & 2 & 300\,--\,400 & $>$450     &          2 &    0.5 $^{+ 1.2}_{- 0.2}$ $^{+ 0.2}_{- 0.2}$ \\
19 & 1 & 2 & $>$400 & 200\,--\,450 &          2 &    1.0 $^{+ 1.3}_{- 0.2}$ $^{+ 0.6}_{- 0.3}$ \\
20 & 1 & 2 & $>$400 & $>$450     &          1 &   0.91 $^{+1.57}_{-0.05}$ $^{+0.62}_{-0.62}$ \\
21 & 1 & $\ge$\,3 & $>$200 & 200\,--\,300 &          5 &     12 $^{+   3}_{-   2}$ $^{+   2}_{-   2}$ \\
22 & 1 & $\ge$\,3 & $>$200 & 300\,--\,400 &          3 &    2.2 $^{+ 1.4}_{- 0.7}$ $^{+ 0.3}_{- 0.3}$ \\
23 & 1 & $\ge$\,3 & $>$200 & $>$400     &          1 &    1.4 $^{+ 1.6}_{- 0.7}$ $^{+ 0.3}_{- 0.2}$ \\[1ex]
24 & 2 & 1 & 200\,--\,300 & 200\,--\,275 &         16 &     19 $^{+   2}_{-   2}$ $^{+   3}_{-   3}$ \\
25 & 2 & 1 & 200\,--\,300 & 275\,--\,350 &          3 &    5.2 $^{+ 1.4}_{- 1.1}$ $^{+ 1.0}_{- 1.0}$ \\
26 & 2 & 1 & 200\,--\,300 & $>$350     &          1 &    0.5 $^{+ 0.8}_{- 0.2}$ $^{+ 0.2}_{- 0.2}$ \\
27 & 2 & 1 & 300\,--\,400 & 200\,--\,275 &         10 &    7.0 $^{+ 1.8}_{- 1.5}$ $^{+ 1.1}_{- 0.8}$ \\
28 & 2 & 1 & 300\,--\,400 & 275\,--\,350 &          6 &    4.0 $^{+ 1.5}_{- 1.1}$ $^{+ 0.5}_{- 0.5}$ \\
29 & 2 & 1 & 300\,--\,400 & $>$350     &          2 &    2.7 $^{+ 1.2}_{- 0.8}$ $^{+ 0.4}_{- 0.4}$ \\
30 & 2 & 1 & $>$400 & 200\,--\,350 &          2 &    0.5 $^{+ 1.1}_{- 0.1}$ $^{+ 0.9}_{- 0.2}$ \\
31 & 2 & 1 & $>$400 & $>$350     &          3 &    1.9 $^{+ 1.1}_{- 0.5}$ $^{+ 0.9}_{- 0.8}$ \\
32 & 2 & 2 & 200\,--\,300 & 200\,--\,275 &          6 &     14 $^{+   3}_{-   3}$ $^{+   2}_{-   2}$ \\
33 & 2 & 2 & 200\,--\,300 & 275\,--\,350 &          1 &    3.1 $^{+ 1.3}_{- 1.0}$ $^{+ 0.5}_{- 0.5}$ \\
34 & 2 & 2 & 200\,--\,300 & $>$350     &          0 &    1.2 $^{+ 0.9}_{- 0.4}$ $^{+ 0.1}_{- 0.1}$ \\
35 & 2 & 2 & 300\,--\,400 & 200\,--\,275 &         10 &    5.3 $^{+ 1.6}_{- 1.3}$ $^{+ 0.9}_{- 0.5}$ \\
36 & 2 & 2 & 300\,--\,400 & 275\,--\,350 &          2 &    1.3 $^{+ 1.3}_{- 0.6}$ $^{+ 0.2}_{- 0.1}$ \\
37 & 2 & 2 & 300\,--\,400 & $>$350     &          1 &    0.7 $^{+ 1.0}_{- 0.4}$ $^{+ 0.2}_{- 0.1}$ \\
38 & 2 & 2 & $>$400 & 200\,--\,350 &          1 &   0.20 $^{+0.87}_{-0.11}$ $^{+0.04}_{-0.04}$ \\
39 & 2 & 2 & $>$400 & $>$350     &          1 &   0.38 $^{+1.31}_{-0.07}$ $^{+0.16}_{-0.16}$ \\
40 & 2 & $\ge$\,3 & $>$200 & 200\,--\,300 &          0 &    4.3 $^{+ 1.6}_{- 1.3}$ $^{+ 0.5}_{- 0.5}$ \\
41 & 2 & $\ge$\,3 & $>$200 & $>$300     &          0 &   0.29 $^{+0.91}_{-0.09}$ $^{+0.06}_{-0.05}$ \\[1ex]
42 & $\ge$\,3 & 1 & $>$200 & $>$200 &          2 &    1.7 $^{+ 1.2}_{- 0.7}$ $^{+ 0.3}_{- 0.2}$ \\
43 & $\ge$\,3 & 2 & $>$200 & $>$200 &          0 &    0.3 $^{+ 0.9}_{- 0.2}$ $^{+ 0.1}_{- 0.1}$ \\
44 & $\ge$\,3 & $\ge$\,3 & $>$200 & $>$200 &          0 &   0.23 $^{+0.92}_{-0.21}$ $^{+0.04}_{-0.04}$ \\
\end{scotch}
}
\end{table*}

The statistical interpretation of the results in terms of exclusion limits for the signal models considered is based on a binned likelihood fit to the observed data, taking into account the predicted background and expected signal yields with their uncertainties in each search bin. The extraction of exclusion limits is based on a modified frequentist approach~\cite{CLs,Read:2002hq,Junk:1999kv,Cowan:2010js} using a profile likelihood ratio as the test statistic.
Signal models for which the 95\% confidence level (\CL) upper limit on the production cross section falls below the theoretical cross section (based on NLO+NLL calculations~\cite{Borschensky:2014cia}) are considered to be excluded by the analysis.

The uncertainties in the signal modeling are determined per search region bin and include the following sources:
simulation sample size (up to 50\% for top squark pair production models and up to 10\% for gluino-mediated production models),
luminosity determination ($2.7\%$), lepton and isolated track veto (up to 4\%),
$\PQb$ tagging efficiency corrections used to scale simulation to data (up to 36\%), trigger efficiency ($<1\%$),
renormalization and factorization scale variations (up to 3\%), initial-state radiation (up to 30\%),
jet energy scale corrections (up to 25\%),
and the modeling of the fast simulation compared with the full simulation for top quark reconstruction and mistagging (up to 7\%).
All these uncertainties, apart from those arising from the simulation sample size, are treated as fully correlated between the search bins when computing exclusion limits.
Potential contamination of signal events in the single-lepton control regions is taken into account for each signal model considered in the interpretation. The potential contamination in the dilepton and inverted-$\Delta\phi$ region is negligible.
The uncertainties from the background predictions are also taken into account using a similar method as used for the signal modeling, but evaluated separately for each physics process.

\begin{figure}[htbp]
\centering
\includegraphics[width=0.45\textwidth]{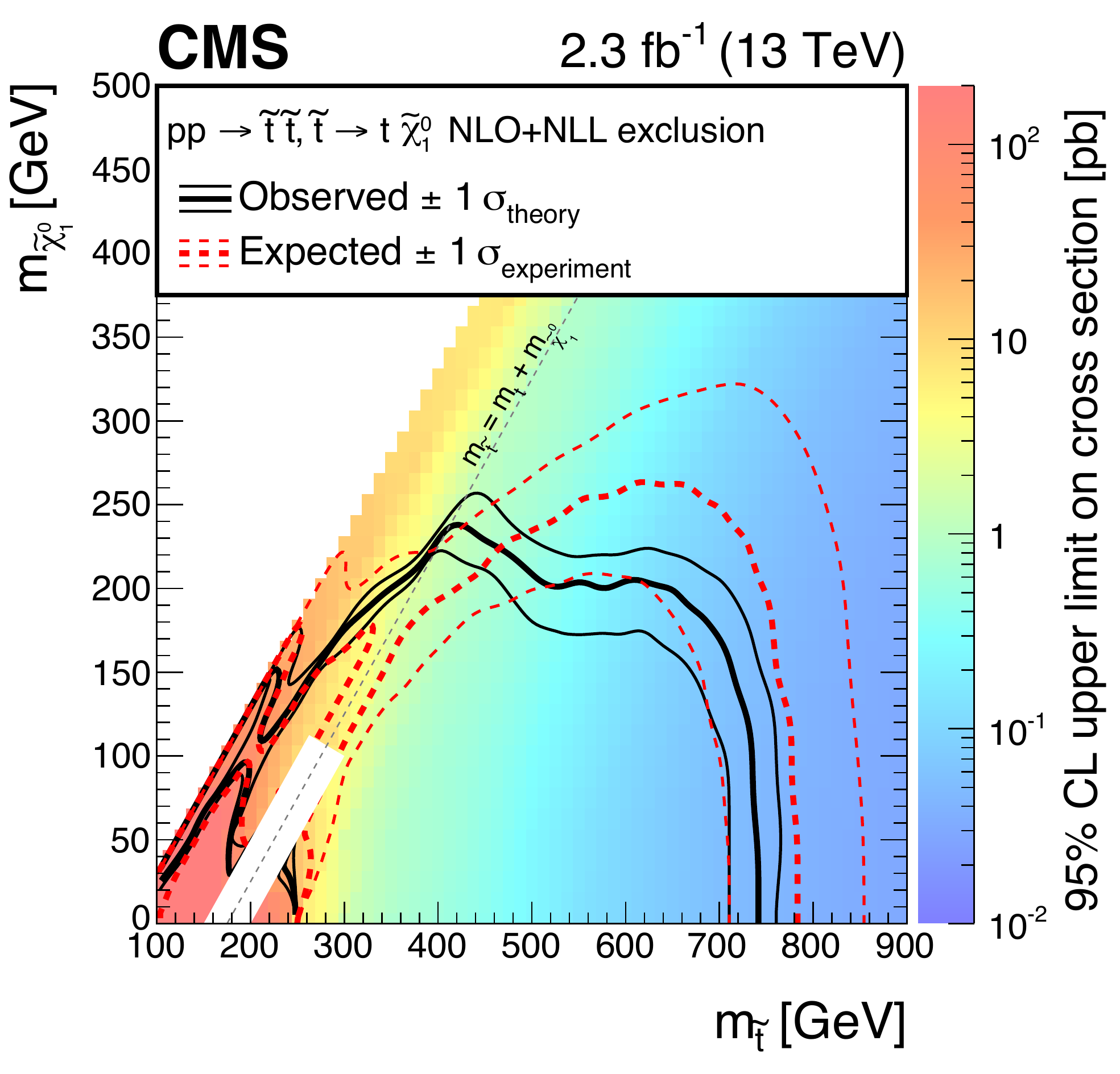}
\includegraphics[width=0.45\textwidth]{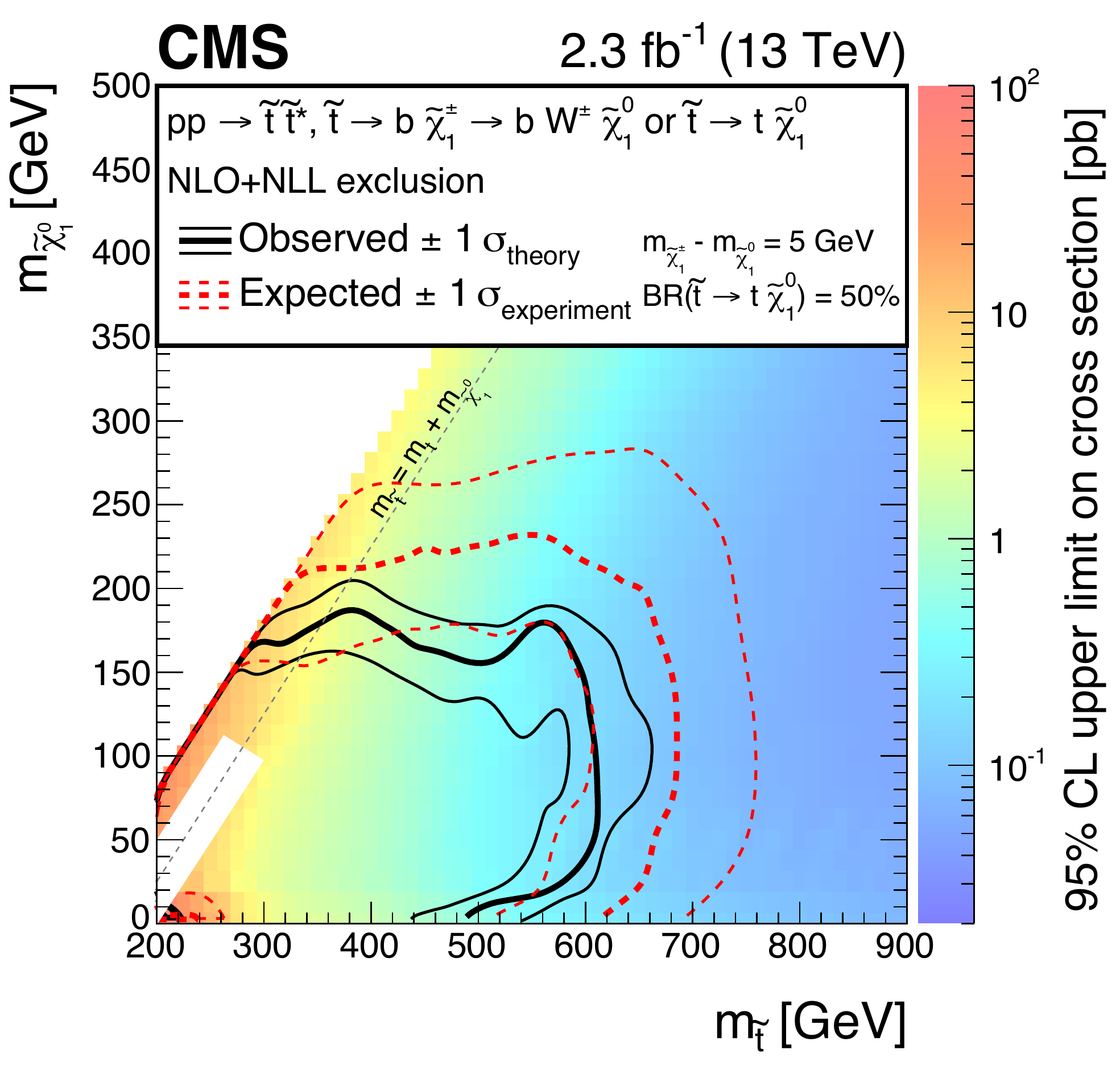}
\caption{\label{fig:T2ttlimits_MethodAlpha} Exclusion limits at 95\% \CL for simplified models of top squark pair production in the T2tt (\cmsLeft) and T2tb (\cmsRight) scenario, assuming a 50\% branching fraction for each of the $\stopq \to \topq \lsp/\stopq \to \bq \chipmone$ modes and a 5\GeV mass difference between the $\chipmone$ and $\lsp$. The solid black curves represent the observed exclusion contour with respect to NLO+NLL cross section calculations~\cite{Borschensky:2014cia} and the corresponding $\pm$1 standard deviation uncertainties. The dashed red curves indicate the expected exclusion contour and the $\pm$1 standard deviation uncertainties including experimental uncertainties.
No interpretation is provided for signal models for which $\abs{m_{\stopq} - m_{\lsp} - m_{\PQt}} \le 25\GeV$ and $m_{\stopq} \leq 275\GeV$
because of significant differences between the fast simulation and the \GEANTfour-based simulation for these low-\MET scenarios.
}
\end{figure}

Figure~\ref{fig:T2ttlimits_MethodAlpha} shows 95\% \CL exclusion limits obtained for simplified models in the pure T2tt scenario, and in the mixed T2tb scenario assuming a 50\% branching fraction for each of the two decay modes ($\stopq \to \topq \lsp/\stopq \to \bq \chipmone$).
In the latter case, the $\chipmone$ and $\lsp$ are assumed to be nearly degenerate in mass, with a 5\GeV difference between their masses.
As a result of this analysis, we exclude top squark masses up to 740\GeV (for zero LSP mass) and LSP masses up to 240\GeV (for top squark mass of 420\GeV) in the T2tt scenario.
In the T2tb scenario, top squark masses up to 610\GeV (for LSP mass of 60\GeV) and LSP masses up to 190\GeV (for top squark mass of 380\GeV) are excluded. These results are comparable to those from the top squark searches at 8\TeV based on an order of magnitude larger data sets.
The improvements of the top quark tagging algorithm, in particular the addition of merged jet scenarios to recover efficiency for boosted top quarks, extends the reach of the analysis to higher top squark masses than would have been possible with the approach used in Ref.~\cite{stop8TeV}.
No interpretation is provided for the T2tt and T2tb signal models for which both $\abs{m_{\stopq} - m_{\lsp} - m_{\PQt}} \le 25\GeV$ and $m_{\stopq} \leq 275\GeV$
because of significant differences between the fast simulation and the \GEANTfour-based simulation for these low-\MET scenarios.

\begin{figure}[htbp]
\centering
\includegraphics[width=0.45\textwidth]{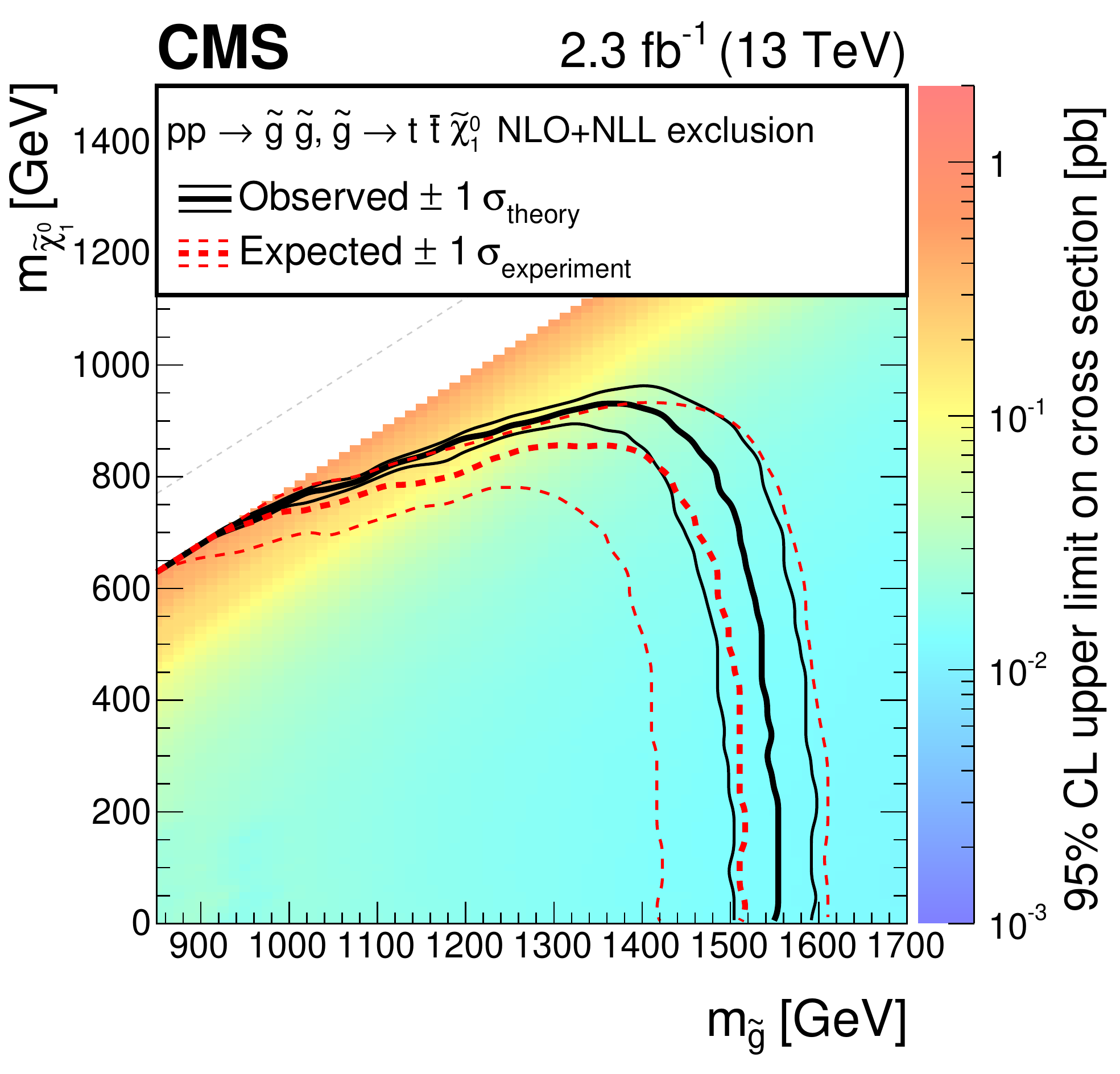}
\includegraphics[width=0.45\textwidth]{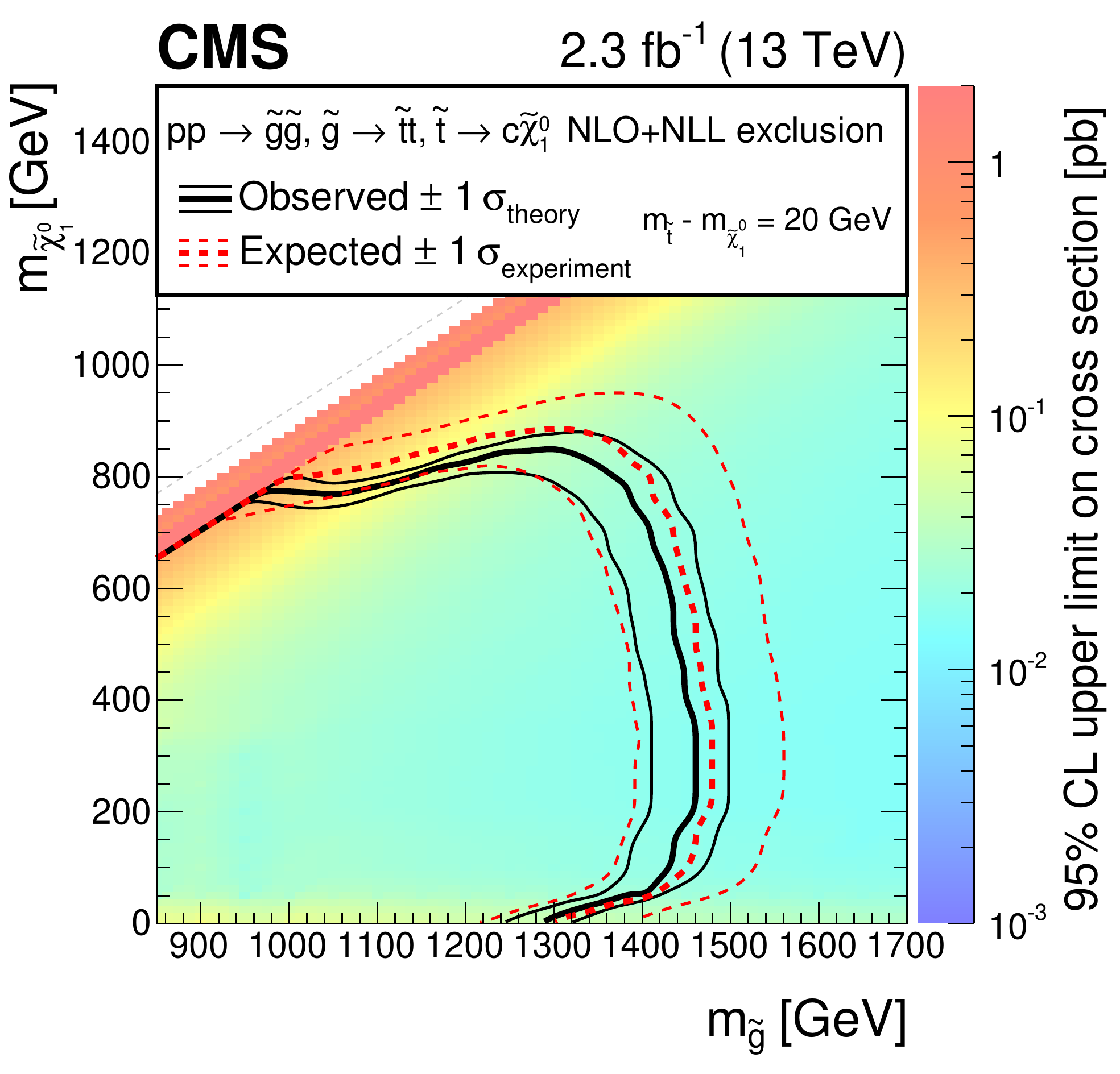}
\caption{\label{fig:T1ttttlimits_MethodAlpha} Exclusion limits at 95\% \CL for simplified models of top squarks produced via decays of gluino pairs in the T1tttt (\cmsLeft) and T5ttcc (\cmsRight) scenarios. The solid black curves represent the observed exclusion contour with respect to NLO+NLL cross section calculations~\cite{Borschensky:2014cia} and the corresponding $\pm$1 standard deviation uncertainties. The dashed red curves indicate the expected exclusion contour and the $\pm$1 standard deviation uncertainties including experimental uncertainties. }
\end{figure}

Figure~\ref{fig:T1ttttlimits_MethodAlpha} shows 95\% \CL exclusion limits obtained for simplified models in the T1tttt and T5ttcc scenarios.
Gluino masses up to 1550\GeV (for zero LSP mass) and LSP masses up to 900\GeV (for top squark mass of 1360\GeV) are excluded for the T1tttt model, whereas gluino masses up to 1450\GeV (for LSP mass of 200-400\GeV) and LSP masses up to 820\GeV (for top squark mass of 1300\GeV) are excluded for the T5ttcc model.
These results significantly extend the mass reach compared to analyses at 8\TeV, which excluded gluino masses up to about 1380 (1340)\GeV and LSP masses up to about 700 (650)\GeV for the T1tttt (T5ttcc) model.
The search bins with $\ntops\ge3$ provide additional sensitivity for T1tttt models with high gluino and LSP masses, since they allow suppression of SM backgrounds while keeping a low \MET threshold.
The decrease in the $m_\gluino$ limit for very small LSP masses for the T5ttcc model can be explained by Lorentz boosts.
For LSP masses near the mass of the charm quark, the LSP and charm quark share the momentum available in the top squark decay about equally.
This results in a softer \MET spectrum, and, therefore, a reduced efficiency, compared to models that have a heavier LSP.

\section{Summary}
\label{sec:summary}

Results have been presented from a search for direct and gluino-mediated top squark production in final states that include tagged top quark decays.
The search uses all-hadronic events with at least four jets and a large imbalance in transverse momentum (\MET), selected from data collected in proton-proton collisions at a center-of-mass energy of 13\TeV with the CMS detector and corresponding to an integrated luminosity of $2.3\fbinv$.
A set of search regions is defined based on \MET, \MTTwo, the number of top quark tagged objects, and the number of $\PQb$-tagged jets.
No statistically significant excess of events is observed above the expected standard model background. Exclusion limits are set at 95\% confidence level for simplified models of direct top squark pair production and of gluino pair production, where the gluinos decay to final states that include top quarks.
For simplified models of pair production of top squarks, which decay to a top quark and a neutralino (T2tt), top squark masses of up to 740\GeV and neutralino masses up to 240\GeV are excluded at 95\% confidence level.
For models that assume 50\% branching fractions for top squark decays to a top quark and a neutralino, or to a bottom quark and a chargino that is nearly degenerate in mass with the neutralino (T2tb), top squark masses of up to 610\GeV and neutralino masses up to 190\GeV are also excluded.
For simplified models of gluino pair production where each gluino decays to a top-antitop quark pair and a neutralino (T1tttt), gluino masses of up to 1550\GeV, and neutralino masses up to 900\GeV are excluded.
Gluino masses of up to 1450\GeV, and neutralino masses up to 820\GeV are excluded for models in which the gluino decays to an on-shell top squark and a top quark, and the top squarks decays to a charm quark and a neutralino (T5ttcc).
These are among the most restrictive currently available limits.

\begin{acknowledgments}
\hyphenation{Bundes-ministerium Forschungs-gemeinschaft Forschungs-zentren} We congratulate our colleagues in the CERN accelerator departments for the excellent performance of the LHC and thank the technical and administrative staffs at CERN and at other CMS institutes for their contributions to the success of the CMS effort. In addition, we gratefully acknowledge the computing centers and personnel of the Worldwide LHC Computing Grid for delivering so effectively the computing infrastructure essential to our analyses. Finally, we acknowledge the enduring support for the construction and operation of the LHC and the CMS detector provided by the following funding agencies: the Austrian Federal Ministry of Science, Research and Economy and the Austrian Science Fund; the Belgian Fonds de la Recherche Scientifique, and Fonds voor Wetenschappelijk Onderzoek; the Brazilian Funding Agencies (CNPq, CAPES, FAPERJ, and FAPESP); the Bulgarian Ministry of Education and Science; CERN; the Chinese Academy of Sciences, Ministry of Science and Technology, and National Natural Science Foundation of China; the Colombian Funding Agency (COLCIENCIAS); the Croatian Ministry of Science, Education and Sport, and the Croatian Science Foundation; the Research Promotion Foundation, Cyprus; the Ministry of Education and Research, Estonian Research Council via IUT23-4 and IUT23-6 and European Regional Development Fund, Estonia; the Academy of Finland, Finnish Ministry of Education and Culture, and Helsinki Institute of Physics; the Institut National de Physique Nucl\'eaire et de Physique des Particules~/~CNRS, and Commissariat \`a l'\'Energie Atomique et aux \'Energies Alternatives~/~CEA, France; the Bundesministerium f\"ur Bildung und Forschung, Deutsche Forschungsgemeinschaft, and Helmholtz-Gemeinschaft Deutscher Forschungszentren, Germany; the General Secretariat for Research and Technology, Greece; the National Scientific Research Foundation, and National Innovation Office, Hungary; the Department of Atomic Energy and the Department of Science and Technology, India; the Institute for Studies in Theoretical Physics and Mathematics, Iran; the Science Foundation, Ireland; the Istituto Nazionale di Fisica Nucleare, Italy; the Ministry of Science, ICT and Future Planning, and National Research Foundation (NRF), Republic of Korea; the Lithuanian Academy of Sciences; the Ministry of Education, and University of Malaya (Malaysia); the Mexican Funding Agencies (BUAP, CINVESTAV, CONACYT, LNS, SEP, and UASLP-FAI); the Ministry of Business, Innovation and Employment, New Zealand; the Pakistan Atomic Energy Commission; the Ministry of Science and Higher Education and the National Science Centre, Poland; the Funda\c{c}\~ao para a Ci\^encia e a Tecnologia, Portugal; JINR, Dubna; the Ministry of Education and Science of the Russian Federation, the Federal Agency of Atomic Energy of the Russian Federation, Russian Academy of Sciences, and the Russian Foundation for Basic Research; the Ministry of Education, Science and Technological Development of Serbia; the Secretar\'{\i}a de Estado de Investigaci\'on, Desarrollo e Innovaci\'on and Programa Consolider-Ingenio 2010, Spain; the Swiss Funding Agencies (ETH Board, ETH Zurich, PSI, SNF, UniZH, Canton Zurich, and SER); the Ministry of Science and Technology, Taipei; the Thailand Center of Excellence in Physics, the Institute for the Promotion of Teaching Science and Technology of Thailand, Special Task Force for Activating Research and the National Science and Technology Development Agency of Thailand; the Scientific and Technical Research Council of Turkey, and Turkish Atomic Energy Authority; the National Academy of Sciences of Ukraine, and State Fund for Fundamental Researches, Ukraine; the Science and Technology Facilities Council, UK; the US Department of Energy, and the US National Science Foundation.

Individuals have received support from the Marie-Curie program and the European Research Council and EPLANET (European Union); the Leventis Foundation; the A. P. Sloan Foundation; the Alexander von Humboldt Foundation; the Belgian Federal Science Policy Office; the Fonds pour la Formation \`a la Recherche dans l'Industrie et dans l'Agriculture (FRIA-Belgium); the Agentschap voor Innovatie door Wetenschap en Technologie (IWT-Belgium); the Ministry of Education, Youth and Sports (MEYS) of the Czech Republic; the Council of Science and Industrial Research, India; the HOMING PLUS program of the Foundation for Polish Science, cofinanced from European Union, Regional Development Fund; the Mobility Plus programme of the Ministry of Science and Higher Education (Poland); the OPUS programme of the National Science Center (Poland); the Thalis and Aristeia programmes cofinanced by EU-ESF and the Greek NSRF; the National Priorities Research Program by Qatar National Research Fund; the Programa Clar\'in-COFUND del Principado de Asturias; the Rachadapisek Sompot Fund for Postdoctoral Fellowship, Chulalongkorn University (Thailand); the Chulalongkorn Academic into Its 2nd Century Project Advancement Project (Thailand); and the Welch Foundation, contract C-1845.
\end{acknowledgments}
\clearpage
\bibliography{auto_generated}
\ifthenelse{\boolean{cms@external}}{}{
\appendix
\clearpage
\section{Additional information on the performance of top quark identification\label{app:suppMat}}
\newcommand{\cPjet}{\ensuremath{\cmsSymbolFace{j}}\xspace}
\ifthenelse{\boolean{cms@external}}{\setlength\cmsFigWidth{0.98\columnwidth}}{\setlength\cmsFigWidth{0.65\textwidth}}

\begin{figure}[h!]
  \begin{center}
    \includegraphics[width=\cmsFigWidth]{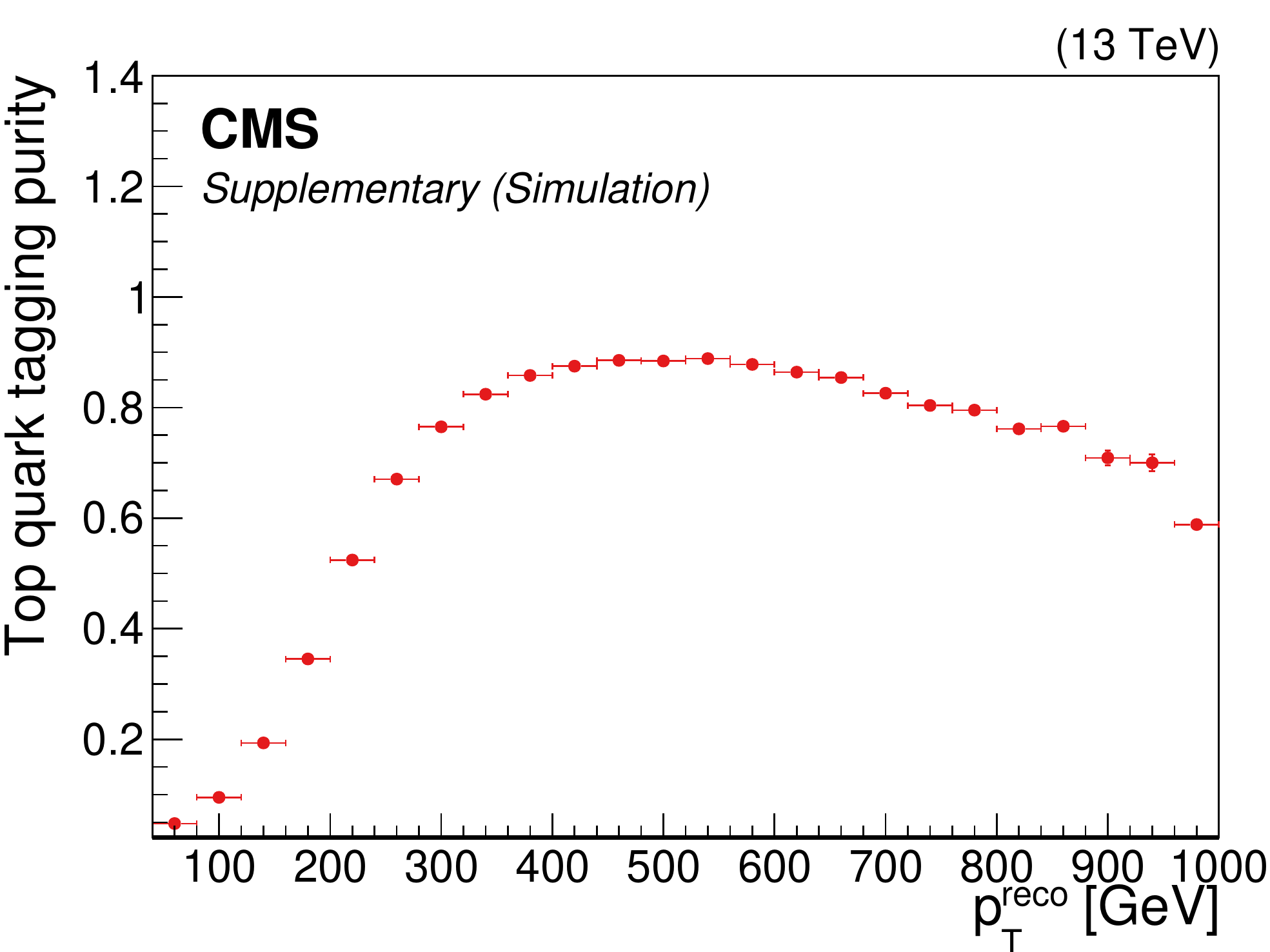}
    \caption{The purity of the top quark tagger as a
    function of the reconstructed top quark $\pt$.
    The purity is defined as the fraction of reconstructed top quarks that are matched to a generator-level hadronically decaying top quark within a cone of $\Delta R=0.4$,
    and was measured in a sample of simulated one-lepton $\cPqt\cPaqt$ events.
    The following event selection requirements were applied:
    $N_{\cPjet} \ge 4$ for $\pt > 30 \GeV$, $\abs{\eta} < 2.4$ and $N_{\cPjet} \ge 2$ for $\pt> 50 \GeV$, $\abs{\eta} <2.4$;
    $N_{\PQb} \ge 1$;
    and $\ETmiss>200\GeV$.
    }
    \label{fig:topPurity}
  \end{center}
\end{figure}

\begin{figure}[b!]
  \begin{center}
    \includegraphics[width=\cmsFigWidth]{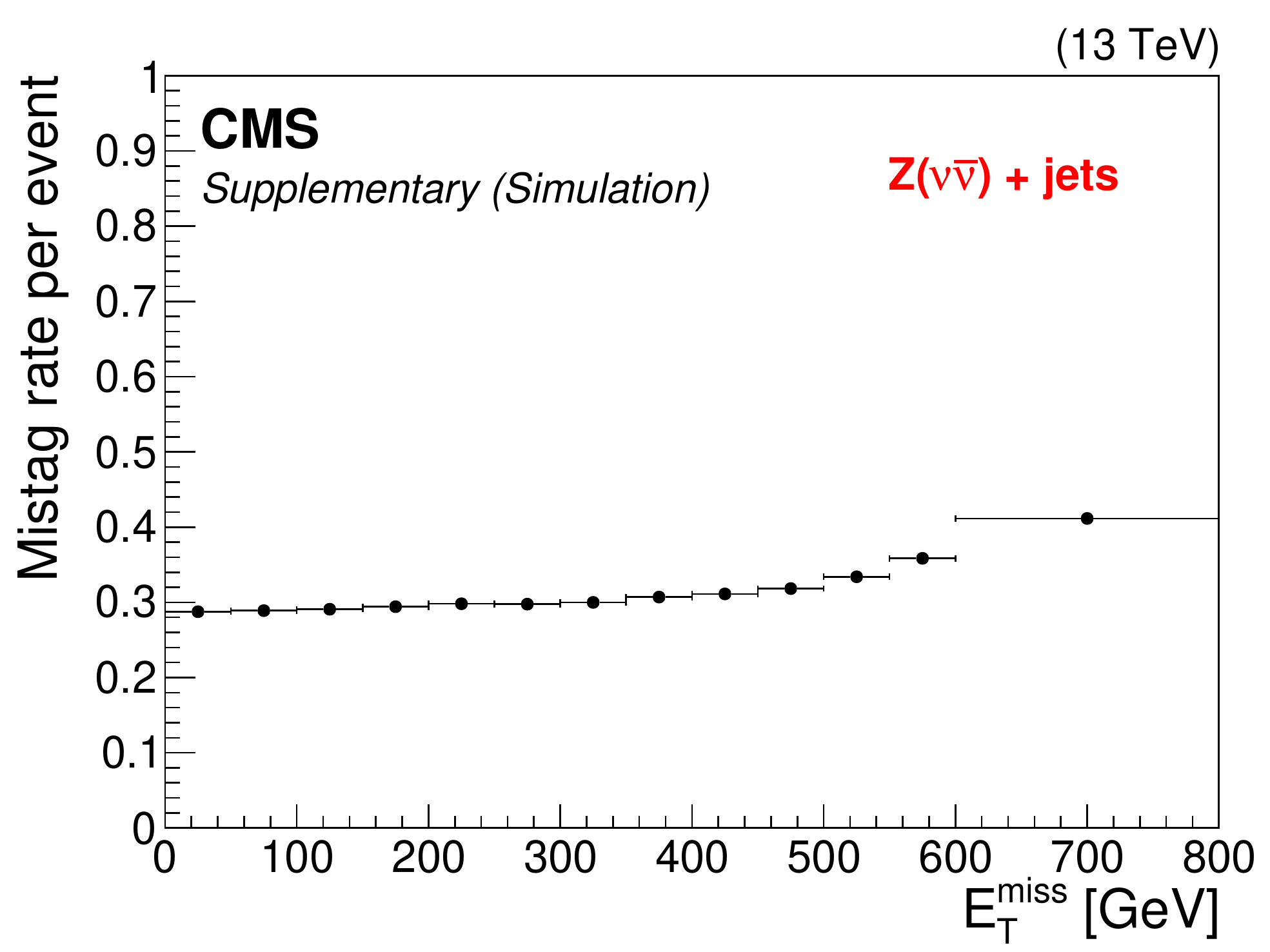}
    \caption{The event mistag rate of the top quark tagger as a
    function of $\ETmiss$ in the $\cPZ\to\nu\nu$ simulated sample, with the following event selection requirements applied:
    $N_{\cPjet} \ge 4$ for $\pt > 30 \GeV$, $\abs{\eta} < 2.4$ and $N_{\cPjet} \ge 2$ for $\pt> 50 \GeV$, $\abs{\eta} <2.4$;
    no electrons, muons, or isolated tracks;
    $\Delta\phi(\ETmiss, \text{jets})$ matching preselection requirements; and $N_{\PQb} \ge 1$.
    }
    \label{fig:topMisTag}
  \end{center}
\end{figure}

\begin{figure}[p]
  \begin{center}
    \includegraphics[width=\cmsFigWidth]{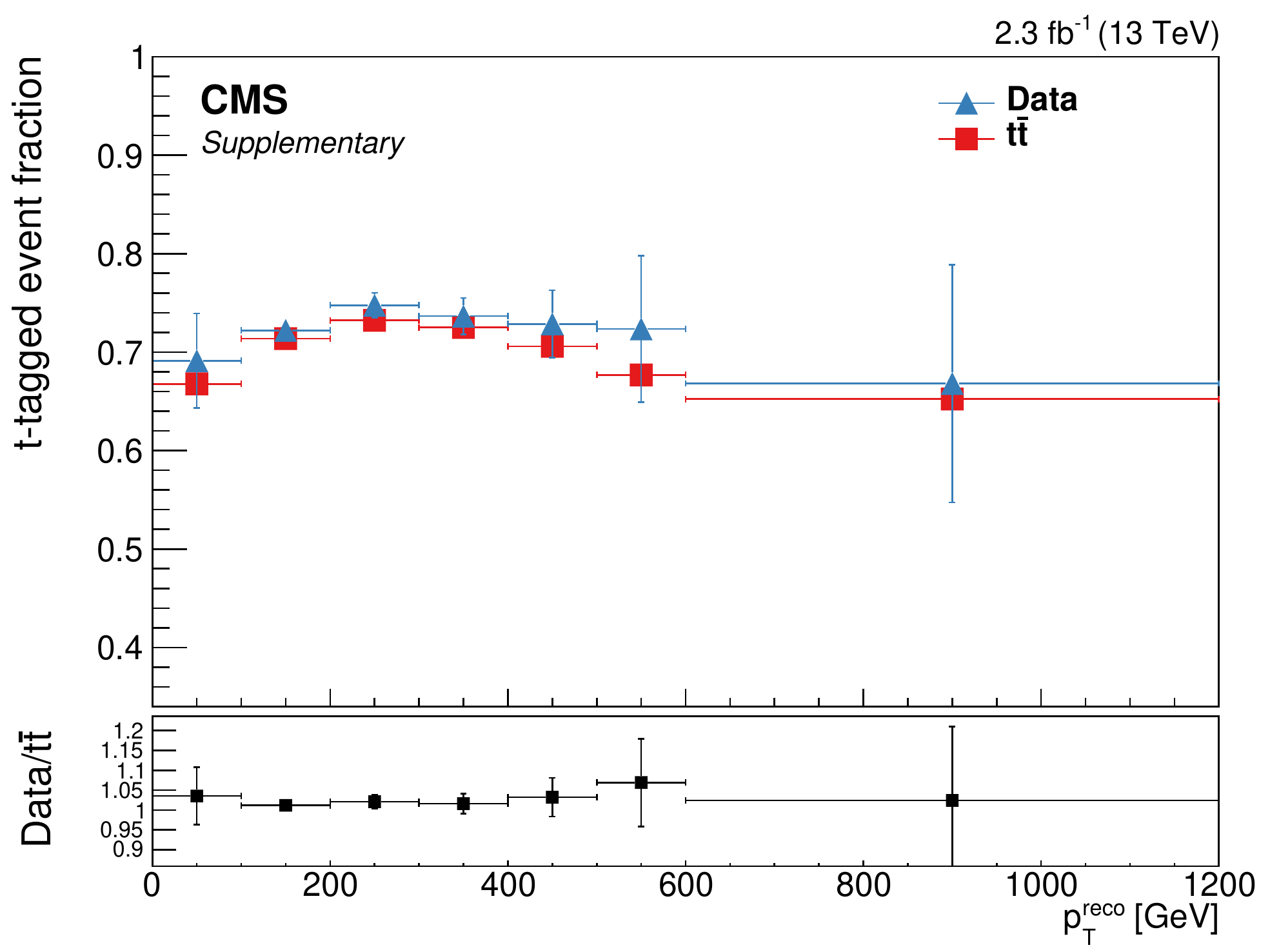}
    \caption{The t-tagged event fraction measured in data and $\cPqt\cPaqt$ simulated
    samples, as a function of the reconstructed top quark candidate \pt.
    The data are selected from a single muon dataset with the following selection applied:
    events pass noise filters;
    $N_{\cPjet} \ge 4$ for $\pt > 30 \GeV$, $\abs{\eta} < 2.4$ and $N_{\cPjet} \ge 2$ for $\pt> 50 \GeV$, $\abs{\eta} <2.4$;
    at least one muon with $\pt > 45 \GeV$, $\abs{\eta} <2.1$;
    no electrons; muon $\pt + \ETmiss > 150\GeV$;
    $\Delta\phi(\ETmiss, \text{jets})$ matching preselection requirements;
    $N_{\PQb} \ge 1$;
    and $\ETmiss > 20 \GeV$.
    We also require the presence of at least one candidate from the t-tagger in the event.
    This candidate can be either: (i) a trijet candidate, composed of three jets with $\pt>30\GeV$ that are within $\Delta R=1.5$ of the candidate four-momentum,
    (ii) a dijet candidate, composed of two jets that are within $\Delta R=1.5$ of the candidate four-momentum,  one of which should have a mass between 70 and 110\GeV,
    or (iii) a monojet candidate, which is simply a single jet with a mass between 110 and 220\GeV.
    The candidate used to compute the t-tagged event fraction is the candidate whose mass is closest to the top quark mass.
    The t-tagged event fraction then is defined as the fraction of events for which this top quark candidate satisfies all requirements of the top quark tagging algorithm.
    The error bar depicts the statistical uncertainty.
    The ratio of the t-tagged event fraction in data and simulated $\ttbar$ is shown in the lower plot, indicating good agreement.}
    \label{fig:DatatopTaggerSF}
  \end{center}
\end{figure}

}
\cleardoublepage \section{The CMS Collaboration \label{app:collab}}\begin{sloppypar}\hyphenpenalty=5000\widowpenalty=500\clubpenalty=5000\textbf{Yerevan Physics Institute,  Yerevan,  Armenia}\\*[0pt]
V.~Khachatryan, A.M.~Sirunyan, A.~Tumasyan
\vskip\cmsinstskip
\textbf{Institut f\"{u}r Hochenergiephysik,  Wien,  Austria}\\*[0pt]
W.~Adam, E.~Asilar, T.~Bergauer, J.~Brandstetter, E.~Brondolin, M.~Dragicevic, J.~Er\"{o}, M.~Flechl, M.~Friedl, R.~Fr\"{u}hwirth\cmsAuthorMark{1}, V.M.~Ghete, C.~Hartl, N.~H\"{o}rmann, J.~Hrubec, M.~Jeitler\cmsAuthorMark{1}, A.~K\"{o}nig, I.~Kr\"{a}tschmer, D.~Liko, T.~Matsushita, I.~Mikulec, D.~Rabady, N.~Rad, B.~Rahbaran, H.~Rohringer, J.~Schieck\cmsAuthorMark{1}, J.~Strauss, W.~Waltenberger, C.-E.~Wulz\cmsAuthorMark{1}
% ...
\vskip\cmsinstskip
% ...
1:~~Also at Vienna University of Technology, Vienna, Austria\\
% ...

\end{sloppypar}
\end{document}